\documentclass[%
 reprint,
 superscriptaddress,
 amsmath,amssymb,
 aps,
 longbibliography,
]{revtex4-2}
\usepackage{braket}
\usepackage{subcaption}
\usepackage{caption}
\usepackage{graphicx}
\usepackage{xcolor}
\usepackage{dcolumn}
\usepackage{bm}
\usepackage[export]{adjustbox} 
\usepackage{multirow}
\usepackage{booktabs}
\usepackage{makecell} 
\usepackage[normalem]{ulem}
\usepackage[
    colorlinks,
    linkcolor=blue,
    anchorcolor=blue,
    citecolor=blue,
    urlcolor=blue
]{hyperref}

\begin{document}

\preprint{APS/123-QED}

\title{Benchmarking quantum trial wavefunctions for phaseless auxiliary-field quantum Monte Carlo}
\author{Rod Rofougaran}
\email{rod.rofougaran@pnnl.gov}
\affiliation{Physical and Computational Sciences, Pacific Northwest National Laboratory, Richland, Washington, USA}

\affiliation{National Energy Research Scientific Computing Center, Lawrence Berkeley National Laboratory, Berkeley, CA 94720,
USA}
\affiliation{Department of Physics, Columbia University, New York, NY 10027, USA}

\author{Neil Mehta}
\affiliation{National Energy Research Scientific Computing Center, Lawrence Berkeley National Laboratory, Berkeley, CA 94720,
USA}

\author{Katherine Klymko}
\affiliation{National Energy Research Scientific Computing Center, Lawrence Berkeley National Laboratory, Berkeley, CA 94720,
USA}

\author{Pooja Rao}
\affiliation{NVIDIA Corporation, Santa Clara, CA 95051, USA}

\author{J. Wayne Mullinax}
\affiliation{KBR, Inc., Intelligent Systems Division, NASA Ames Research Center, Moffet Field, CA 94035, USA}

\author{Samuel Stein}
\affiliation{Physical and Computational Sciences, Pacific Northwest National Laboratory, Richland, Washington, USA}

\author{Norm M. Tubman}
\affiliation{NASA Ames Research Center, Moffet Field, CA 94035, USA}

\author{Ermal Rrapaj}
\affiliation{National Energy Research Scientific Computing Center, Lawrence Berkeley National Laboratory, Berkeley, CA 94720,
USA}
\affiliation{Department of Physics, University of California, Berkeley, CA 94720, USA}
\affiliation{RIKEN iTHEMS, Wako, Saitama 351-0198, Japan}

\begin{abstract}

The phaseless auxiliary-field quantum Monte Carlo (ph-AFQMC) method is a stochastic imaginary-time projection technique for computing ground-state properties of strongly correlated quantum systems, with accuracy that depends critically on the choice of trial wavefunction. We present a comprehensive benchmarking study of trial wavefunctions prepared using parameterized quantum circuits, spanning unitary coupled cluster (UCC), Hamiltonian-informed, Jastrow-inspired, and adaptively constructed ansatzes, and evaluate their accuracy, expressibility, and resource requirements within the quantum-computing-enhanced AFQMC (QC-AFQMC) framework. Trial states are prepared with GPU-accelerated statevector simulations using CUDA-Q and tested on linear hydrogen chains under bond stretching and on nine closed-shell molecules at their equilibrium geometries with varying degrees of static correlation. We find that when ansatzes are compared at similar numbers of variational parameters, ph-AFQMC largely removes the differences between families for weakly and moderately correlated systems: for such molecules every trial yields a ph-AFQMC energy within chemical accuracy despite trial energy errors spanning nearly three orders of magnitude, whereas the choice of ansatz becomes consequential as static correlation grows. Our results indicate that the variational energy of a trial is not always a reliable indicator of its quality for ph-AFQMC, and reveal instances in which adding variational parameters can lower the trial energy while degrading the ph-AFQMC energy. Across all of our benchmark systems, trial fidelity ranks candidate trials more reliably than variational energy, and for spin-contaminated trials, the singlet-projected fidelity is more reliable than unprojected fidelity. The UCCSD ansatz initialized from unrestricted CCSD amplitudes provides a robust and inexpensive-to-optimize baseline across all benchmark systems, with circuit depth as its principal limitation, while adaptive and compact layered ansatzes remain competitive in accuracy in the strongly correlated regime with substantially shallower circuits. Finally, we find that UCCSD trial wavefunctions initialized from unrestricted rather than restricted Hartree–Fock references improve accuracy at stretched geometries but yield non-variational energies in the strongly stretched limit.

\end{abstract}
\maketitle

\section{Introduction}

Determining the ground-state properties of interacting many-electron systems remains a central challenge in quantum chemistry, condensed matter physics, and materials science. Although the Schrödinger equation provides a complete description, in principle, its exact solution becomes intractable at even modest system sizes due to the exponential growth of the Hilbert space. As a result, a variety of approximate methods have been developed to balance accuracy and computational cost. Methods such as Kohn–Sham density functional theory (DFT) and single-reference coupled cluster (CC) theory provide accurate descriptions for weakly correlated systems, but often fail in strongly correlated regimes where multi-reference effects become important. Here, we use strong (or static) correlation to refer to regimes in which no single Slater determinant (SD) provides a dominant description of the exact ground state \cite{ganoe2024notion}.

Auxiliary-field quantum Monte Carlo (AFQMC) has emerged as a highly accurate and scalable approach for treating strongly correlated systems \cite{Lee:2022, motta2018ab}. AFQMC achieves this by applying stochastic imaginary-time evolution to a population of single-SD walkers $\{\ket{\phi}\}$, which collectively project the many-body wavefunction toward the ground state. In its original formulation, referred to as free-projection (fp)-AFQMC \cite{sugiyama1986auxiliary}, walkers evolve in SD space without any constraint, and fermionic antisymmetry leads to fluctuating walker signs/phases, resulting in the Monte Carlo sign/phase problem~\cite{Pan:2024}. To mitigate this issue, constrained variants such as constrained path (cp-) \cite{zhang1995constrained} and phaseless (ph-) \cite{zhang2003quantum} AFQMC methods introduce a controllable bias through a trial wavefunction $\ket{\Psi_T}$, where the overlaps, $\langle \Psi_T | \phi \rangle$, guide the stochastic propagation of the walkers. In this work, we focus on the phaseless approximation, which has attracted significant interest and demonstrated considerable success in \textit{ab initio} electronic structure calculations.

The ph-AFQMC method introduces a controllable approximation to the ground-state wavefunction, whose accuracy depends on the quality of the trial wavefunction $\ket{\Psi_T}$, as reflected in its overlap with the true ground state and the phase constraint it provides \cite{zhang2003quantum}. In the simplest case, a single-SD mean-field solution such as spin-restricted (RHF) or spin-unrestricted Hartree–Fock (UHF) can be used as the trial. However, as systems become more strongly correlated, it becomes necessary to go beyond simple SDs and use linear combinations of determinants or multi-SD (MSD) wavefunctions. Most commonly, trial wavefunctions of this form can be constructed in various ways, such as through truncated \cite{morales2012multideterminant,clark2011computing} or selected \cite{giner2013using,mahajan2021taming} configuration interaction (CI), complete active space self-consistent field (CASSCF) \cite{roos1980complete}, or non-orthogonal MSD approaches \cite{landinez2019non}. Throughout this work, ``U'' denotes \emph{unrestricted} in UHF and \emph{unitary} in acronyms for quantum ansatzes such as unitary coupled cluster (UCC).

More generally, classically constructing trial states that accurately capture many-body
correlations remains a major challenge. The number of determinants in CI-like trial
wavefunctions grows exponentially with system size, while truncating the expansion
compromises size extensivity: the sum of the energies of non-interacting subsystems does
not yield the same energy as treating the combined system. These limitations have
motivated interest in trial states based on compact, nonlinear parameterizations of
correlation, including Slater--Jastrow \cite{mahajan2020efficient}, matrix product state
\cite{jiang2025unbiasing}, and CC~\cite{kjonstad2025systematic} forms, which
encode correlations through the structure of the ansatz rather than through an explicit
linear expansion over determinants. Recent work has also shown that perturbative CC
trial wavefunctions can restore size extensivity in the subsequent ph-AFQMC calculation
\cite{zhang2026size}. In practice, however, evaluating the overlaps and local quantities
required by AFQMC with such correlated trial states can introduce substantial
computational cost \cite{xiao2025implementing}.  Quantum state preparation provides an alternative route to constructing correlated trial states within a systematically improvable circuit representation. Here, we investigate the utility of such quantum-prepared states as
ph-AFQMC trials.

The quantum-computing-enhanced AFQMC (QC-AFQMC) approach was first introduced by Huggins et al.~\cite{huggins2022unbiasing}. In this framework, the trial state is prepared on a quantum computer and subsequently used within ph-AFQMC simulations. In the original formulation, $\ket{\Psi_T}$ is prepared on a quantum device, and the classical shadows protocol \cite{huang2020predicting} is employed to perform a series of randomized Clifford measurements that encode classical information about the quantum state. These measurement results are then used to estimate $\langle \Psi_T | \phi \rangle$ and $\langle \Psi_T | \hat{H} |\phi \rangle$ required during the ph-AFQMC simulation. It was subsequently shown that estimating the local energy, $E_L = \frac{\langle \Psi_T | \hat{H} | \phi \rangle}{\langle \Psi_T | \phi \rangle}$, using classical shadows requires measurement data that scales exponentially with system size $L$ to maintain constant relative precision~\cite{mazzola2022exponential,lee2022response}. 
This bottleneck arises because the amplitudes of correlated quantum trial states, $|\Psi_T(x)|$, are typically exponentially small in system size ($\mathcal{O}(2^{-L})$). To overcome this bottleneck, a tomographic protocol tailored for fermionic systems, known as matchgate shadows, was later introduced~\cite{wan2023matchgate}. 
Subsequent studies investigated the scaling behavior of QC-AFQMC with matchgate shadows and demonstrated that chemical accuracy could be achieved for small systems (\(<10\) qubits), with a post-processing cost scaling polynomially as \(\mathcal{O}(N^{8.5})\), where \(N\) denotes the number of spin orbitals~\cite{huang2024evaluating,kiser2024classical}.
More recently, a reduction in the computational cost to \(\mathcal{O}(N^{5.5})\) through algorithmic optimizations and GPU-accelerated post-processing on a 16-qubit trial wavefunction has been reported~\cite{zhao2025quantum}. This scaling is comparable to that of standard classical ph-AFQMC implementations~\cite{jiang2025unbiasing}, marking an important step towards making QC-AFQMC a practically viable hybrid quantum-classical algorithm. A complementary strategy is to bypass tomographic reconstruction altogether and evaluate the required trial-state overlaps explicitly. A related approach has recently been demonstrated for AFQMC using wavefunctions obtained from sample-based quantum diagonalization~\cite{danilov2025enhancing}.

Despite substantial progress made in reducing the post-processing cost of QC-AFQMC, fundamental questions remain regarding the types of quantum trial states that are most effective in practice. Across various studies of QC-AFQMC to date, several forms of trial wavefunctions have been explored. While the original proposal used a simple non-variational valence-bond wavefunction and a hardware-efficient circuit \cite{huggins2022unbiasing}, since then, the variational quantum eigensolver (VQE) has been used to optimize trial wavefunctions for AFQMC constructed from well-established quantum chemistry ansatzes, including UCC with singles and doubles (UCCSD) \cite{huang2024evaluating,amsler2023classical,khinevich2025enhancing}, paired UCC with doubles (UpCCD) \cite{zhao2025quantum,goings2025molecular}, orbital-optimized UpCCD \cite{goings2025molecular}, as well as the local unitary cluster Jastrow (LUCJ)~\cite{blunt2025quantum} and the quantum number preserving ansatz \cite{kiser2024classical}, along with alternative QC-AFQMC formulations utilizing other trial states \cite{kiser2025contextual,yoshida2025auxiliary}. Our work presents a comprehensive benchmarking study of a wide range of quantum trial ansatzes, including members of the UCC family, hardware-efficient, adaptive, and other variational forms. 

We benchmark our ansatzes using linear hydrogen chains, which are widely regarded as an effective testbed for many-electron methods~\cite{motta2017towards,hachmann2006multireference,huggins2021efficient}, and, as a second benchmark, we study nine closed-shell molecules at their equilibrium geometries. Despite their apparent simplicity, hydrogen chains exhibit a rich range of correlation effects. As the H–H bond length increases, the system transitions from a weakly correlated regime to a strongly correlated regime. At the same time, they avoid additional complexities present in heavier molecules and materials, such as the need to treat core electrons separately or include relativistic effects. Moreover, they provide a natural and systematic scaling with system size while preserving the essential correlation physics, enabling controlled studies of method performance as the Hilbert space grows. Hydrogen chains are, however, chemically homogeneous. To introduce greater chemical diversity, we use the molecular test set introduced in Section~\ref{sec:molecules}. This set comprises first- and second-row hydrides and diatomics with single, double, and triple bonds, and ranges from weakly to strongly correlated at equilibrium geometry \cite{mullinax2025classical}.

The quality of the trial states is assessed using several key metrics: the variational energy, the ph-AFQMC projected energy and its associated error, the ph-AFQMC accuracy as a function of the number of determinants, the unprojected and spin-projected fidelities with respect to the full configuration interaction (FCI) ground state, and the overall resource efficiency of optimizing and preparing the trial state.

Beyond characterizing the behavior and effectiveness of different trial wavefunctions within the QC-AFQMC framework, these results provide broader insight into the performance of modern VQE ansatzes at larger qubit counts under realistic computational constraints. Simulations were performed using GPU-accelerated quantum circuit simulations with CUDA-Q, NVIDIA's open-source hybrid quantum--classical platform \cite{cudaq}, with the workflow distributed across multiple compute nodes on the Perlmutter supercomputer. This enables scalable benchmarking of increasingly complex ansatzes and large-scale parallelization of trial-state optimization. Therefore, the present work complements existing VQE benchmarking studies~\cite{zhao2023orbital} by examining these ansatzes in a hybrid quantum--classical setting leveraging high-performance computing resources. The trial wavefunctions analyzed here may also serve as correlated reference states or benchmarks for other many-body methods.

In addition to exploring various quantum ansatzes, we also explore different initial HF states that precede the variational circuit. While quantum circuits for VQE and QC-AFQMC for spin-symmetric molecules are typically initialized with the RHF state, we also investigate initialization from the UHF state. Under bond stretching and in strongly correlated regimes, UHF can yield spin-symmetry-broken solutions, providing a qualitatively improved description of static correlation. Although such states are spin contaminated and may have reduced overlap with the exact spin-symmetric ground state, previous AFQMC studies have shown that they can nevertheless improve performance and reduce phaseless bias, as the symmetry of the walker ensemble effectively enforces a symmetry-projected description \cite{purwanto2008eliminating,lee2019auxiliary,motta2017towards,al2007bond, shi2014symmetry}. More generally, symmetry-broken single SDs provide a simple and inexpensive means of capturing strong-correlation effects in AFQMC. Generalized Hartree--Fock (GHF) trials, which may break both $S_z$ and $S^2$ symmetries, have also been successfully applied to both molecular and lattice systems \cite{qin2016benchmark,danilov2025capturing, chang2017multi, rosenberg2017ultracold}, providing an important single-determinant baseline for comparison with more elaborate trial forms, including the quantum-prepared states studied here. Here, we examine how the choice of initial state influences AFQMC accuracy and stability across geometries.

The organization of the paper is as follows: in Sec.~\ref{sec:wv} we provide a summary of the various quantum ansatzes considered in this study; in Sec.~\ref{sec:ph}, we provide the main ingredients in the ph-AFQMC approach; in Sec.~\ref{sec:results} we provide the results of both VQE and AFQMC for the hydrogen chain under bond stretching, and compare RHF to UHF; in Sec.~\ref{sec:molecules}, we test the transferability of these findings on a chemically diverse set of molecules at their equilibrium geometries; finally, we discuss our findings and draw conclusions in Sec.~\ref{sec:Discussion} and Sec.~\ref{sec:Conclusion}.

\section{Trial wavefunctions for quantum chemistry}
\label{sec:wv}

The second-quantized representation of the molecular Hamiltonian is:

\begin{equation}
    \hat{H}  = \hat{H}_1 + \hat{H}_2 
    = \sum_{pq}^N T_{pq} a_p^{\dagger} a_q 
    + \frac{1}{2} \sum_{pqrs}^N V_{pqrs} a_p^{\dagger} a_q^{\dagger} a_s a_r ,
\label{eq: ham second quantized}
\end{equation}
where $N$ denotes the number of basis functions in the chosen single-particle basis and 
$a_p^{\dagger}$, $a_q$ are fermionic creation and annihilation operators for orbitals indexed by $p, q$.

Preparing accurate approximations to the ground state of $\hat{H}$ is a central task in 
quantum algorithms for chemistry. The earliest quantum approaches to this problem were
based on adiabatic state preparation and quantum phase estimation~\cite{farhi2000quantum,aspuru2005simulated,abrams1999quantum}. However, these methods
require circuit depths that exceed the capabilities of current noisy quantum devices. Consequently, ground-state estimation on 
near-term hardware has shifted toward the VQE framework, 
which combines a parameterized quantum circuit with a classical optimizer. The 
quantum circuit prepares a compact variational trial state $|\Psi(\boldsymbol{\theta})\rangle = U(\boldsymbol{\theta})|\Phi_0\rangle$ for a given ansatz $U(\boldsymbol{\theta})$, while the parameters $\boldsymbol{\theta}$ are iteratively updated to minimize the cost function, which is the expectation value of the Hamiltonian,
$\langle \Psi(\boldsymbol{\theta}) | \hat{H} | \Psi(\boldsymbol{\theta}) \rangle$. To estimate the energy of a trial state on a quantum computer, $\hat{H}$ is typically mapped to a linear combination of Pauli operators using transformations such as the Jordan--Wigner (JW)~\cite{jordan1928paulische} or Bravyi--Kitaev~\cite{bravyi2002fermionic} mappings, after which the energy is computed as a weighted sum of Pauli expectation values.

For molecular systems described in a second-quantized representation, each qubit typically encodes the occupation of a spin orbital in the RHF basis used to construct $\hat{H}$ in Eq.~\ref{eq: ham second quantized}. The initial state of the variational circuit, $\lvert \Phi_0 \rangle$, is therefore usually chosen as the RHF reference state $\lvert \Psi_{\mathrm{RHF}} \rangle$. However, we also explore initializing the circuit with the UHF reference state, i.e., setting $\lvert \Phi_0 \rangle = \lvert \Psi_{\mathrm{UHF}} \rangle$.

Optimizing the variational parameters that define the trial state, however, is complicated by barren plateaus. These are regimes in which the energy gradients with respect to circuit parameters vanish exponentially with system size~\cite{mcclean2018barren}, causing the cost landscape to become effectively flat and requiring an exponentially large number of measurements to resolve meaningful gradients and maintain desired precision, thus potentially eliminating any exponential advantage offered by variational quantum algorithms. To mitigate barren plateaus, two primary strategies have been proposed: the use of problem-inspired ansatzes and improved parameter initialization schemes~\cite{cerezo2021variational}. Consequently, successful variational circuits in quantum chemistry are typically chemically inspired, with the UCC family of ansatzes serving as a prominent example~\cite{peruzzo2014variational}. The success of the UCC ansatz in VQE arises not only from its close correspondence to classical CC theory, but also from the fact that a significant portion of the correlation energy can be captured by initializing the variational parameters with second-order Møller-Plesset perturbation theory (MP2) \cite{moller1934note, romero2018strategies} or CCSD amplitudes \cite{hirsbrunner2024beyond}. It was shown in Ref.~\cite{hirsbrunner2024beyond} that using CCSD amplitudes to initialize the UCCSD parameters consistently outperforms MP2-based initialization and can serve as good variational starting points even when CCSD returns non-physical energies. Other problem-inspired ansatzes were also explored in this work, including the Hamiltonian Variational Ansatz (HVA) \cite{wecker2015progress, park2024hamiltonian, wiersema2020exploring}, most commonly applied to lattice systems, and the unitary cluster Jastrow (UCJ) ansatz \cite{matsuzawa2020jastrow}.

Due to hardware limitations, it is often necessary to explore informed truncations of these problem-inspired ansatzes. For instance, one can exploit molecular point-group symmetries to eliminate excitation operators in the UCCSD ansatz that do not preserve the symmetry of the Hartree--Fock reference state~\cite{cao2021towards}. Other truncation methods include restricting the ansatz to $k$ repetitions of generalized and paired excitations, leading to the $k$-UpCCGSD family of ansatzes~\cite{lee2018generalized}. Additionally, locality constraints can be imposed on the UCJ ansatz for hardware architectures with local qubit connectivity, resulting in the LUCJ form~\cite{motta2023bridging,motta2024quantum}. Another prominent strategy is adaptive ansatz construction, as in the Adaptive Derivative-Assembled Pseudo-Trotter VQE (ADAPT-VQE) algorithm, where operators from a predefined pool, such as those corresponding to UCCSD excitations, are iteratively selected and added based on their expected contribution to lowering the energy of the trial state~\cite{grimsley2019adaptive}. 
 
These considerations motivate our selection of four representative families of quantum trial ansatzes. All four families have been executed on current hardware, at least in truncated form: UCCSD on trapped-ion and superconducting devices \cite{hempel2018quantum,nam2020ground,guo2024experimental}, paired UCC variants on trapped ions \cite{zhao2023orbital, zhao2025quantum}, LUCJ on superconducting processors \cite{motta2023bridging,robledo2025chemistry}, HVA for lattice models \cite{stanisic2022observing}, and adaptive ansatzes via qubit-ADAPT \cite{tang2021qubit,bauman2025coupled}.

\subsection{Unitary Coupled Cluster}

The UCC ansatz is defined as an exponential of excitation operators acting on a reference determinant, which is usually the RHF state:

\begin{equation}
    \ket{\Psi_{\text{UCC}}} = e^{\hat{T}-\hat{T}^\dagger}\ket{\Phi_{0}},
\label{eq: ucc ansatz }
\end{equation}
where $\hat{T}$ is the excitation operator. In the standard UCC ansatz, excitations are allowed only from occupied to virtual molecular orbitals (MOs) of the reference state. In this case, when including up to $n$-th order excitations, $\hat{T}$ has the form:
\begin{equation}
\hat{T} = \sum_{k=1}^{n} \hat{T}_k, \qquad 
\hat{T}_k = \frac{1}{(k!)^2}
\sum_{i,j,\ldots}^{\text{occ}}
\sum_{a,b,\ldots}^{\text{vir}}
\theta_{ij\ldots}^{ab\ldots}
\hat{a}^\dagger_a \hat{a}^\dagger_b \cdots
\hat{a}_j \hat{a}_i,
\label{eq:excitation operators}
\end{equation}
where occupied and virtual molecular orbitals are indexed by \( i, j, \ldots \) and \( a, b, \ldots \), respectively. 
The variational parameters $\theta_{ij\ldots}^{ab\ldots}$ are associated with the corresponding excitation operators. To map the UCC ansatz to a quantum circuit, we first perform a Trotterization of the exponential operator. 
It has been shown that a single-step Trotter expansion can efficiently approximate the UCC ansatz \cite{barkoutsos2018quantum,chen2021quantum}. 
In this case, the factorized form of the ansatz can be written as:

\begin{equation}
|\Psi_{\mathrm{UCC}}\rangle =
\prod_{i,j,\ldots}^{\text{occ}}
\prod_{a,b,\ldots}^{\text{vir}}
\exp\!\left(
\theta^{ab\ldots}_{ij\ldots} 
\hat{G}^{ab\ldots}_{ij\ldots}
\right)
|\Phi_0\rangle,
\label{eq:ucc_wavefunction}
\end{equation}
where the anti-Hermitian generator $\hat{G}^{ab\ldots}_{ij\ldots}$ is defined as 
$
\hat{a}^{\dagger ab\ldots}_{ij\ldots} - 
\hat{a}^{ij\ldots}_{ab\ldots}.$
By introducing a factor of $i$ to render $\hat{G}$ Hermitian and mapping the resulting second-quantized operators to Pauli operators using the JW transformation, the UCC ansatz can be conveniently expressed on a quantum circuit as a product of exponentials of Pauli strings $e^{i\theta \hat{P}}$. 

Most variants of the UCC ansatz modify Eq.~\eqref{eq:ucc_wavefunction} by restricting the allowed excitation operators.
The letter labels in an ansatz name specify these constraints. For example, ``S'' and ``D'' correspond to single and double excitations, respectively. The label ``G'' denotes generalized excitations in which excitations are permitted between any molecular orbitals rather than being restricted to the occupied-to-virtual transitions of the reference determinant. Additionally, the label ``p'' indicates a paired ansatz, in which double excitations include only electrons in spin-paired units (i.e., $(i_\alpha,i_\beta)\rightarrow(a_\alpha,a_\beta)$). Finally, a prefactor in the ansatz label (e.g., 3-UpCCGSD) denotes the number of circuit repetitions (layers), as is standard for compact paired ansatzes \cite{lee2018generalized}. These constraints reduce the number of excitation operators and variational parameters, while still preserving expressivity. 

\subsection{Local Unitary Cluster Jastrow}

The LUCJ ansatz is constructed by applying locality constraints on the more general UCJ ansatz. By combining chemically motivated structure with hardware-efficient locality, LUCJ aims to provide a low-depth, yet expressive, alternative to UCC.

The general $k$-layer UCJ ansatz has the form:
\begin{equation}
|\Psi_{\text{UCJ}}\rangle =
\prod_{l=0}^{k-1}
\hat{\mathcal{U}}_l\, e^{i\hat{\mathcal{J}}_l}\, \hat{\mathcal{U}}_l^\dagger
|\Phi_{0}\rangle,
\label{eq:LUCJ wavefunction}
\end{equation}
where each $\hat{\mathcal{U}}_l$ is a particle-number-conserving orbital
rotation operator, and each $\hat{\mathcal{J}}_l$ is a diagonal Coulomb
operator of the form:
\begin{equation}
\hat{\mathcal{J}}_l
= \frac{1}{2} \sum_{ij,\sigma\tau}
J^{\sigma\tau}_{ij,l}\, \hat{n}_{i,\sigma} \hat{n}_{j,\tau},
\label{eq:Jastrow operator}
\end{equation}
where $\hat{n}_{i,\sigma} = a^\dagger_{i,\sigma} a_{i,\sigma}$ is the
fermionic number operator. 
Orbital rotations can be decomposed into nearest-neighbor Givens rotations \cite{PhysRevA.92.062318,jiang2018quantum},
enabling efficient implementation on qubit architectures with limited
connectivity. In the spin-balanced case, the Jastrow operator $\hat{\mathcal{J}}_l$ is
fully characterized by two symmetric matrices:
$J^{\alpha\alpha}=J^{\beta\beta}$, which couple same-spin orbitals, and
$J^{\alpha\beta}=J^{\beta\alpha}$, which couple opposite-spin orbitals. The
non-zero entries of these matrices determine which spin orbitals (and hence
qubits) are coupled.

To impose locality, we mask out all interactions except those between nearest
neighbors in the chosen qubit layout. In our simulations, we assume a simple square topology in which the $\alpha$ and $\beta$ qubits occupy two parallel rows adjacent to one another.
Under this layout, the following coupling patterns implement strictly
nearest-neighbor interactions:
\begin{equation}
\begin{aligned}
J^{\alpha\alpha} &: \{(p, p+1), \quad p = 0,\ldots, N-2\},\\
J^{\alpha\beta} &: \{(p, p),   \quad p = 0,\ldots, N-1\},
\label{eq:square topology}
\end{aligned}
\end{equation}
where $N$ denotes the number of MOs. Thus, $k$-LUCJ is obtained by masking $\hat{\mathcal{J}}$ in Eq.~\eqref{eq:LUCJ wavefunction} according to a locally constrained coupling map, such as the one defined in Eq.~\eqref{eq:square topology}.

The variational parameters of the LUCJ ansatz are the non-zero matrix elements
of the corresponding orbital-rotation matrix of operator $\hat{\mathcal{U}}_l$ and the corresponding Jastrow
matrices of operator $\hat{\mathcal{J}}_l$. Analogous to the UCCSD ansatz, the LUCJ ansatz can be
initialized using CCSD $t$-amplitudes by applying a double factorization of the
$T_2$ operator to approximate the UCJ form \cite{motta2023bridging}:
\begin{equation}
e^{\hat{T}_2 - \hat{T}_2^\dagger}
= e^{\, i \sum_{l=0}^{k-1} \hat{\mathcal{U}}_l \hat{\mathcal{J}}_l \hat{\mathcal{U}}_l^\dagger}
\;\approx\;
\prod_{l=0}^{k-1} \hat{\mathcal{U}}_l\, e^{i\hat{\mathcal{J}}_l}\, \hat{\mathcal{U}}_l^\dagger,
\label{eq: double factorization}
\end{equation}
where the final expression follows from a first-order Trotter approximation.
However, enforcing locality by masking out non-nearest-neighbor interactions can
significantly degrade the quality of this factorization. To mitigate this effect, one may
employ a compressed double factorization with appropriate regularization \cite{lin2025improved,ffsim}.

\subsection{ADAPT-VQE}
The ADAPT-VQE algorithm \cite{grimsley2019adaptive} constructs the ansatz dynamically during optimization. This algorithm begins with a reference state $|\Phi_{0}\rangle$ and a Hamiltonian $\hat{H}$, as in standard VQE for molecular systems, but rather than specifying a fixed ansatz, ADAPT-VQE grows the circuit by selecting operators from a predefined operator pool. In our work, this pool consists of the UCCSD excitation operators, referred to as ADAPT-UCCSD.

At the beginning of each ADAPT iteration $n$, the energy gradient with respect
to each operator $\hat{A}_j$ in the pool is evaluated:
\begin{equation}
\frac{\partial E^{(n)}}{\partial \theta_j}
= \left\langle \psi^{(n)} \middle|
\left[ \hat{H}, \hat{A}_j \right]
\middle| \psi^{(n)} \right\rangle,
\label{eq:adapt energy gradient}
\end{equation}
where $|\psi^{(n)}\rangle$ is the state produced by the
current ansatz.

After computing all gradients, the operator with the largest gradient magnitude, corresponding to the steepest expected reduction in energy, is appended to the ansatz together with a new variational parameter. All parameters, including the newly added one, are then re-optimized to obtain the next state $|\psi^{(n+1)}\rangle$.

This cycle of gradient evaluation, operator selection, and parameter
re-optimization is repeated until a convergence criterion is satisfied,
typically when all gradient magnitudes fall below a specified threshold or when
successive energy changes become smaller than a given tolerance.
ADAPT-VQE therefore constructs a compact, chemically inspired ansatz that often
yields significantly shallower circuits than fixed-structure UCC ansatzes, at
the cost of computing energy gradients with respect to the entire operator pool
at each iteration. In practical implementations on quantum hardware, this cost is further compounded by the need to estimate these energy gradients from finite-shot measurements. However, this additional sampling overhead may be more manageable than the circuit depth requirements associated with less compact ansatzes \cite{grimsley2019adaptive}.

\subsection{Hamiltonian Variational Ansatz}

The HVA is motivated by the idea that the structure of the many-body Hamiltonian itself should guide the design of the variational circuit. Rather than employing an excitation-based ansatz, the HVA constructs the variational state using unitary evolutions generated directly by Hamiltonian terms. This approach is inspired by adiabatic state preparation and Trotterized time evolution, in which sequential evolution under non-commuting Hamiltonian components drives the system toward low-energy states. 

After applying the JW transformation, we decompose the resulting qubit Hamiltonian as:
\begin{equation}
\hat{H} = \sum_{p=0}^{P-1} \hat{H}_p,
\label{eq:HVA ham}
\end{equation}
where each term $\hat{H}_p$ consists of a sum of mutually commuting Pauli operators, and different terms do not commute, $[\hat{H}_p, \hat{H}_{p'}] \neq 0$ for $p \neq p'$. Because the Pauli operators within a given $\hat{H}_p$ commute, the corresponding unitary evolution, $\exp(-i \theta \hat{H}_p)$, can be implemented exactly as a product of Pauli rotations, without introducing any Trotter approximation error.

A $k$-layer HVA (labeled as $k$-HVA) is defined as a layered product of such unitary evolutions:
\begin{equation}
\ket{\Psi_{\text{HVA}}} = \prod_{l=0}^{k-1} \prod_{p=0}^{P-1} e^{-i \theta_{p,l} \hat{H}_p}|\Phi_0\rangle,
\label{eq: HVA ansatz }
\end{equation}
where $\theta_{p,l}$ are the variational parameters, and once again $|\Phi_0\rangle = |\Psi_{\text{RHF}}\rangle$. The HVA layers act as a controlled mechanism for systematically entangling and correlating the RHF state in directions dictated by the physical Hamiltonian.

This ansatz can be interpreted as a variationally optimized form of Trotterized time evolution, in which the effective evolution times associated with each Hamiltonian term are treated as independent variational parameters. As the number of layers increases, the expressive power of the HVA systematically improves. 
In the limit of large depth, the ansatz approaches a variationally optimized 
Trotterized time evolution and can approximate the ground state of many local Hamiltonians \cite{wecker2015progress}.

\begin{table}[t]
\centering
\renewcommand{\arraystretch}{1.2}
\begin{tabular}{lc}
\hline
\textbf{Ansatz} & \textbf{Circuit Depth} \\
\hline
UCCSD & $\mathcal{O}(N^3)$ \\
$k$-UpCCGSD & $\mathcal{O}(kN)$ \\
$k$-LUCJ (square topology) & $\mathcal{O}(kN)$ \\
$k$-HVA & $\mathcal{O}(kN^4)$ \\
\hline
\end{tabular}

\caption{
Asymptotic circuit depth for molecular Hamiltonians at half-filling. Here $N$ denotes the number of spatial orbitals. Reported scalings are under standard structured implementations of each ansatz \cite{lee2018generalized,motta2023bridging}. For HVA, the quoted scaling corresponds to a straightforward implementation based on the $O(N^4)$ terms of the electronic Hamiltonian; grouping or factorization can reduce the practical depth.}
\label{tab:depth_scaling}
\end{table}

\section{\lowercase{ph}-AFQMC Overview}
\label{sec:ph}

In AFQMC, as in other projector QMC methods, we approximate the exact ground state, $\ket{\Psi_0}$, via imaginary time evolution:

\begin{equation}
    \ket{\Psi_0} = \lim_{\tau \rightarrow \infty}e^{-\tau \hat{H}}\ket{\Phi_I}.
\label{eq: imag time evolution}
\end{equation}
This expression is formally exact provided that $\langle \Phi_I | \Psi_0 \rangle \neq 0$. Numerically, the continuous imaginary-time evolution is implemented through a discrete-time iteration, $\ket{\Psi^{(n+1)}} = e^{-\Delta \tau \hat{H}} \ket{\Psi^{(n)}}$, where $\Delta\tau$ is a small imaginary-time step.

In our QC-AFQMC simulations, we prepare $|\Psi_T\rangle$ by running GPU-accelerated VQE with a chosen quantum ansatz, extracting the optimized statevector, and expanding it into an MSD wavefunction of the form:
\begin{equation}
    \ket{\Psi_T} = \sum_{n=1}^{N_{\mathrm{dets}}} c_n \ket{D_n}.
\label{eq:MSD}
\end{equation}
Given a set of $N$ MOs, typically obtained from an RHF mean-field solution, each $\ket{D_n}$ denotes a determinant corresponding to a specific occupation configuration of $M$ electrons distributed among $N$ spatial orbitals per spin (i.e., $N$ $\alpha$ and $N$ $\beta$ orbitals). Once the trial state is expressed in MSD form, the remainder of the workflow reduces to standard classical ph-AFQMC.

The specific QMC variant depends on how the imaginary-time evolution is implemented, i.e., the space used to represent the wavefunction and the importance sampling employed to control the fermionic sign/phase problem. In AFQMC, an arbitrary state at imaginary time $\tau$, $\ket{\Psi(\tau)}$, is represented by a weighted ensemble of single-SD walkers $\{\ket{\phi(\tau)}\}$. Each walker is expressed in terms of spin-resolved orbital-coefficient matrices $\Phi^\alpha \in \mathbb{C}^{N \times M_\alpha}$ and $\Phi^\beta \in \mathbb{C}^{N \times M_\beta}$, corresponding to the occupied spin orbitals for the $\alpha$ and $\beta$ spin sectors, respectively.

The population of walkers, each initialized in the RHF occupation, explores the manifold of SDs through the application of one-body propagators of the form
$\hat{B} = \exp\!\left(\sum_{ij} \hat{a}_i^{\dagger} U_{ij} \hat{a}_j \right)$,
which, by Thouless' theorem, generate new SDs via orbital transformations induced by the one-body propagator. Consequently, to perform imaginary-time propagation while preserving the single-SD structure of the walkers, the two-body interaction $\hat{H}_2$ must be recast in terms of one-body operators. In AFQMC, this is achieved through a combination of a Trotter--Suzuki decomposition of the imaginary-time propagator, a factorization of the two-electron integrals (e.g., via Cholesky decomposition), and a Hubbard--Stratonovich transformation that introduces auxiliary fields. These steps transform the two-body propagator into a stochastic average over one-body propagators parameterized by a set of auxiliary fields.

As a result, the imaginary-time propagator can be written in the compact form:

\begin{equation}
    e^{-\Delta \tau \hat{H}} 
    = \int d^{N_{\mathrm{aux}}}\mathbf{x}\, p(\mathbf{x})\, \hat{B}(\mathbf{x}) + \mathcal{O}(\Delta \tau^2),
\label{eq: stochastic integral}
\end{equation}
where \(p(\mathbf{x})\) is a normalized Gaussian distribution and
\(\hat{B}(\mathbf{x})\) is a one-body propagator depending on the sampled
auxiliary fields \(\mathbf{x}\). This integral is evaluated
stochastically by sampling the auxiliary fields and propagating each walker
with the corresponding one-body propagator. A detailed derivation of this
procedure is provided in Appendix~\ref{app:ph-AFQMC}.

As already discussed, if no constraint were introduced, we would have effectively implemented fp-AFQMC. Propagation would be highly noise-prone due to the fermionic phase problem that worsens with increasing imaginary time $\tau$ and system size. To control this instability, Zhang and Krakauer introduced the phaseless approximation~\cite{zhang2003quantum}, described in detail in Appendix~\ref{app:ph-AFQMC}. In this formulation, propagation is guided by a trial wavefunction $\ket{\Psi_T}$ through importance sampling. The overlap between each walker and the trial state, $\langle \Psi_T | \phi \rangle$, enters the importance-sampling transformation that steers the random walk toward regions of Slater-determinant space where overlap with the trial state has large magnitude. At each imaginary-time step, the auxiliary-field sampling distribution $p(\mathbf{x})$ is shifted by an optimal force bias derived from this overlap, while the walker weights are updated according to the resulting importance function.

In addition, the phaseless constraint suppresses uncontrolled phase rotations by projecting the walker weights onto the real axis defined by the phase of the trial wavefunction. This procedure stabilizes the simulation while introducing a controllable bias whose magnitude depends on the quality of $\ket{\Psi_T}$. In particular, the trial wavefunction defines the reference phase against which the stochastic propagation is constrained, so the phaseless bias depends on how closely the phase structure of $\ket{\Psi_T}$ matches that of the true ground state. Because the exact phase structure of the ground state is not known in advance, identifying optimal trial wavefunctions for ph-AFQMC remains a nontrivial problem.

\section{Benchmark Results: Hydrogen Chains}
\label{sec:results}
\begin{figure*}[t]
    \centering

    \begin{subfigure}[t]{0.48\textwidth}
        \centering
        \includegraphics[width=\linewidth]{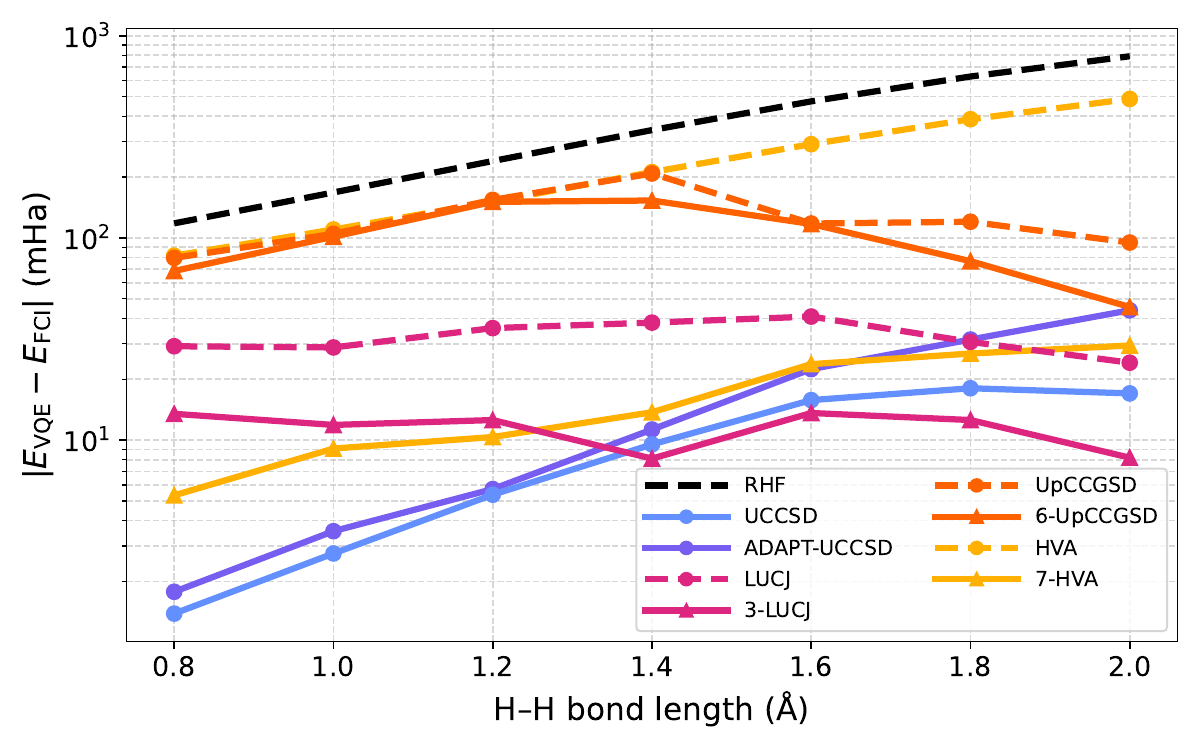}
        \caption{VQE H10 Energy Errors}
        \label{fig:VQE_H10_PES}
    \end{subfigure}
    \hfill
    \begin{subfigure}[t]{0.48\textwidth}
        \centering
        \includegraphics[width=\linewidth]{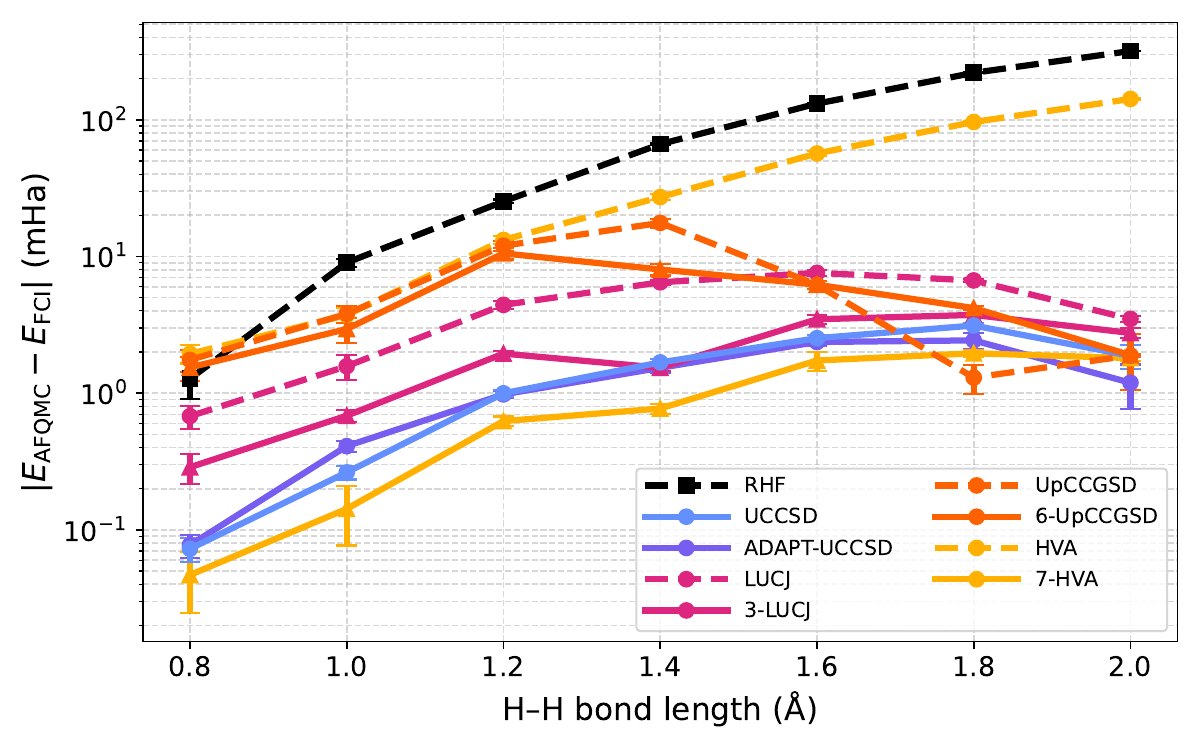}
        \caption{ph-AFQMC H10 Energy Errors}
        \label{fig:AFQMC H10 PES}
    \end{subfigure}

    \caption{
(a) Absolute energy errors $|E_{\mathrm{VQE}} - E_{\mathrm{FCI}}|$ (mHa) for the symmetric bond-stretching curve of \(\mathrm{H}_{10}\), and (b) the corresponding absolute ph-AFQMC errors $|E_{\mathrm{AFQMC}} - E_{\mathrm{FCI}}|$ obtained when each optimized state is used as the trial wavefunction. Both are shown on a logarithmic scale, with the RHF curve included as a classical single-determinant reference. Solid (dashed) lines indicate the higher- (lower-) depth variant of each ansatz family, while triangles (circles) denote multilayer (single-layer) variants; the depths of the multilayer ansatzes (3-LUCJ, 6-UpCCGSD, and 7-HVA) were chosen such that the total number of variational parameters is approximately equal to that of UCCSD.}
\label{fig:VQE Results}
\end{figure*}

To benchmark our quantum trial wavefunctions, we perform exact statevector simulations using optimized circuits with GPU accelerators. We implement all ansatzes using CUDA-Q \cite{cudaq} and perform ideal multi-node, multi-GPU statevector simulations. This enables efficient
evaluation of our quantum circuits, provides exact statevector representations
of the trial states, and allows seamless integration with our AFQMC pipeline.
All quantum circuit simulations were run on NVIDIA A100 GPUs on the Perlmutter
supercomputer at NERSC. We parallelize the wavefunction optimization by
distributing gradient evaluations, whether with respect to variational
parameters or operator pool elements, across multiple GPUs and compute nodes. 

As previously mentioned, each resulting statevector is expanded into an MSD trial wavefunction of the form given in Eq.~\ref{eq:MSD}, which is then passed directly to the AFQMC calculation. Detailed descriptions of ansatz implementations, wavefunction optimization procedure, AFQMC hyperparameters, and the associated statistical analysis are provided in Appendix~\ref{app: sim_details}.

We first use the STO-6G basis on a family of linear hydrogen chains under bond stretching. In this minimal basis, an $N$-atom chain corresponds to $N$ spatial orbitals (and thus $2N$ qubits), enabling controlled tests of our ansatz families as the system size increases. Near equilibrium geometries ($R \approx 1.0~\text{\AA}$), the electronic structure is dominated by dynamic correlation, and single-reference methods such as RHF and CCSD provide a qualitatively reasonable description of the ground state. In this regime, quantum ansatzes that closely resemble single-reference CC forms, such as UCCSD, perform competitively as both VQE solvers and as AFQMC trial wavefunctions. At stretched bond lengths ($R \gtrsim 1.2~\text{\AA}$), the system develops strong multi-reference character, with both static and dynamic correlation becoming significant. Here, conventional single-reference approaches like CCSD and MP2 begin to break down when built on a spin-restricted reference. CCSD based on an unrestricted determinant often remains qualitatively correct at stretched geometries, at the cost of spin contamination \cite{krylov2000spin}. The CCSD amplitudes used to initialize our ansatzes are unrestricted for this reason (Appendix~\ref{app: sim_details}). This regime is particularly challenging and serves as a critical test for more expressive ansatz families, as well as for the ability of AFQMC to recover correlation energy when supplied with quantum MSD trial wavefunctions. 
This makes the strongly stretched regime particularly well suited for assessing the robustness of quantum-derived trial states in AFQMC.

\subsection{VQE Results}
Since the UpCCGSD, LUCJ, and HVA ansatzes are relatively compact in their single-layer forms, they are most commonly employed in a multilayer construction, where their expressivity can be systematically enhanced with increasing depth. Consequently, we consider both a single layer ($k=1$) and a depth $k$ chosen such that the total number of variational parameters is approximately equal to that of UCCSD. This parameter-matching procedure enables a more direct comparison at roughly fixed expressivity. We note that the parameter $k$ is not directly commensurate across different ansatz families, as a single layer corresponds to distinct circuit structures and parameter counts in each case. In our plots and tables, the selected depth is prepended to the ansatz label (e.g., 7-HVA).
Resource requirements for all ansatz families, including circuit depth scaling, parameter counts, and number of optimization iterations, are summarized in Table~\ref{tab:depth_scaling} and Table~\ref{tab:VQE_AFQMC_Table}.
All VQE evaluations for \(\mathrm{H}_{10}\) were parallelized across 16 compute nodes totaling 64 GPUs. For a comprehensive study of the effect of increasing the number of layers for these multi-layer ansatzes on H$_8$, see Figure~\ref{fig:H8 iters} in Appendix~\ref{app:depth}.

\begin{figure}[t]
    \centering
    \includegraphics[width=\linewidth]{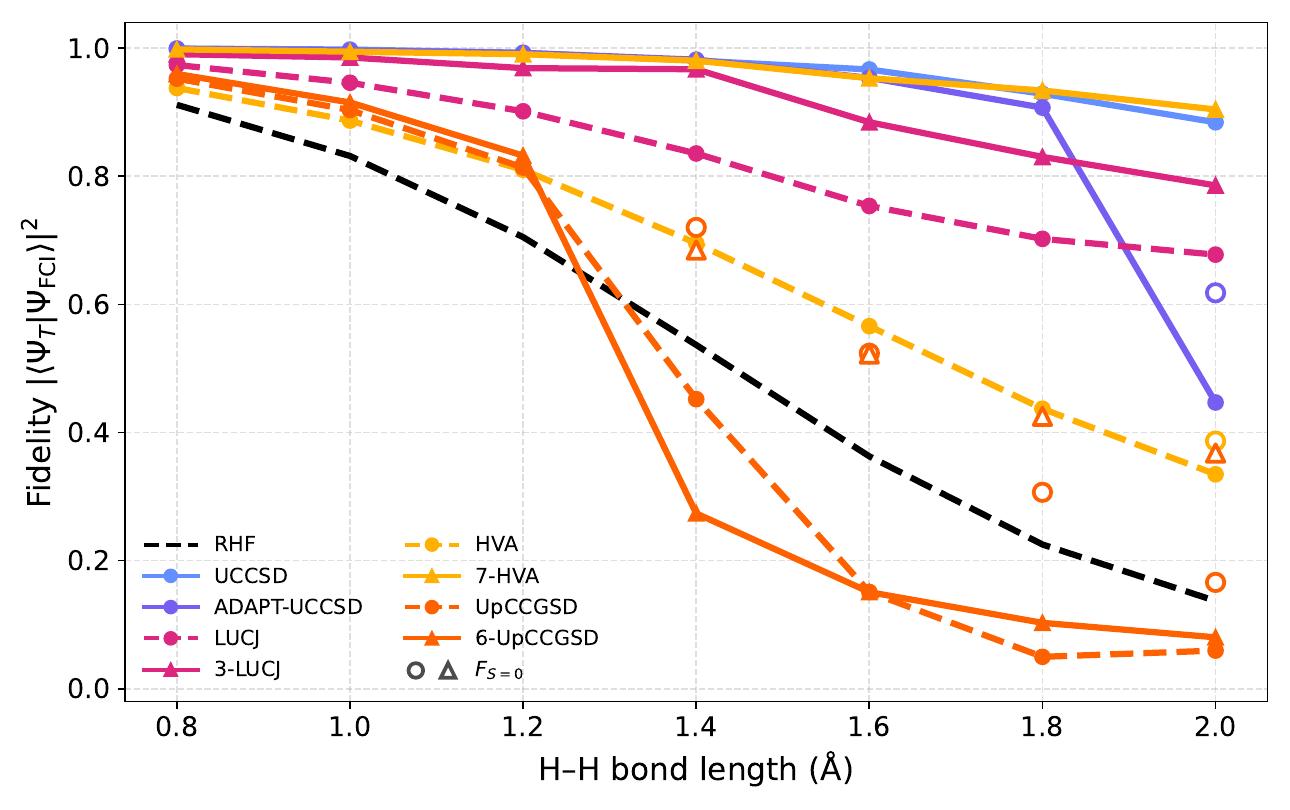}
    \caption{Fidelity of the optimized VQE states with respect to the FCI ground state, $F = |\langle \Psi_T | \Psi_{\mathrm{FCI}} \rangle|^2$, along the H$_{10}$ bond-stretching curve. Filled markers are unprojected fidelities. For trials where singlet projection changes the fidelity by more than 0.05, $F_{S=0}-F>0.05$, which happens when a trial is considerably spin-contaminated, an open marker of the same shape and color denotes the singlet-projected fidelity $F_{S=0}$ of Eq.~\eqref{eq:projected fidelity}. Only 10 of the 56 ansatz/geometry combinations satisfy this threshold.}
    \label{fig:VQE_fidelities}
\end{figure}

Figure~\ref{fig:VQE_H10_PES} presents the VQE energy errors in the potential energy surface (PES) for the symmetric stretch of the \(\mathrm{H}_{10}\) chain (20 qubits), across 7 bond lengths spanning from weakly to strongly correlated regimes. The absolute energy error \(|E_{\mathrm{VQE}} - E_{\mathrm{FCI}}|\) is reported on a logarithmic scale and includes the RHF error as a reference for the onset of strong correlation. Figure~\ref{fig:AFQMC H10 PES} shows the corresponding ph-AFQMC errors obtained from the same states, and is discussed in Section~\ref{sec:AFQMC results}. Figure~\ref{fig:VQE_fidelities} complements this analysis by showing the fidelity
$|\langle \Psi_{\mathrm{VQE}} | \Psi_{\mathrm{FCI}} \rangle|^{2}$ between the optimized VQE states and the exact ground state obtained from FCI. These fidelities, along with the corresponding $\langle \hat{S}^2 \rangle$ values, are summarized in Table~\ref{tab:VQE_AFQMC_Table}.

Across all bond lengths, UCCSD remains the most consistent and reliable ansatz overall, serving as the natural reference standard throughout this study. It achieves among the lowest energy errors at every geometry and maintains among the highest fidelities across the dissociation curve. Its strong and stable behavior is further aided by initialization with CCSD amplitudes, which significantly reduces the optimization cost relative to the other ansatzes. 

ADAPT-UCCSD remains a close second to UCCSD in both energy and fidelity across the entire PES. Its key advantage is circuit compactness, especially at large bond lengths where the number of selected operators decreases as correlation increases. 

The LUCJ and 3-LUCJ equilibrium accuracies are modest compared to UCCSD, but the LUCJ curves remain nearly flat across increasing correlation and 3-LUCJ even outperforms UCCSD in total energy error in the strongly correlated regime (\(R \gtrsim 1.4~\text{\AA}\)). Here, LUCJ achieves comparable accuracy with a significantly shallower circuit and requires only nearest-neighbor two-qubit gates. Its fidelity with the ground state also degrades more gently than those of the other low-depth alternatives. However, the lower overlap at larger bond lengths contrasts with its improved energy error, indicating a significant overlap with the low-lying spectrum of the Hamiltonian, but not necessarily the ground state. This indicates that judging the quality of the ansatz based on energy alone might be misleading.

Although single-layer HVA underperforms relative to the other ansatzes, it exhibits consistent and systematic improvement as the number of layers is increased. At sufficient depth (e.g., 7-HVA), HVA closely approaches the accuracy of UCCSD across all bond lengths. 

UpCCGSD underperforms relative to the other ansatzes across all bond lengths considered. Increasing the number of layers does not lead to consistent or systematic improvements in the energy, and both the single-layer and multilayer variants yield the highest VQE energy errors and lowest fidelities among the ansatzes studied.

Regarding optimization behavior, UCCSD was the most straightforward ansatz to train, largely due to the effectiveness of CCSD-based parameter initialization. In contrast, UpCCGSD was the most challenging ansatz to optimize. CCSD warm starts were less consistently beneficial, particularly for multilayer circuits, likely because the generalized operator pool introduces many additional excitation operators with vanishing CCSD amplitudes. Moreover, all-zero initialization occasionally resulted in premature convergence to suboptimal local minima. These optimization difficulties also made it hard to obtain systematic improvements by simply increasing the number of layers. Although ADAPT-UCCSD is expressive, compact, and flexible, these benefits come at the cost of expensive optimization, as the algorithm iteratively selects and re-optimizes operators. This results in a substantially larger number of energy evaluations during the optimization process than for the other methods considered, making it the most resource-demanding ansatz in our study. LUCJ was also comparatively resource-intensive to optimize, with iteration counts often exceeding those of the other ansatzes. Prior work has noted that quasi-Newton methods such as L-BFGS-B can struggle for this ansatz, and that stochastic reconfiguration or linear-method optimizers can substantially improve convergence~\cite{motta2024quantum}. We did not encounter major optimization issues for HVA during our hydrogen chain experiments, and its performance improved systematically with additional layers; nonetheless, standard HVA can exhibit barren plateaus, which can be mitigated through principled initialization schemes~\cite{park2024hamiltonian}. We did not employ these more advanced optimization or initialization strategies in order to maintain a fair and consistent comparison across all ansatzes. More generally, the VQE energy landscape is non-convex and a local optimizer guarantees only a local minimum \cite{bittel2021training, anschuetz2022quantum, wierichs2020avoiding}. All trial states reported here should therefore be understood as the outcome of a fixed, reproducible optimization protocol rather than as certified global minima. Global-search strategies such as genetic algorithms \cite{chivilikhin2020mog,wakaura2021evaluation} could reduce this uncertainty at considerably higher cost and are left for future work.

\subsection{AFQMC Results}
\label{sec:AFQMC results}
We employ these optimized quantum states as trial wavefunctions for AFQMC and compare their performance, as well as the simplest classical trial wavefunction, the RHF state. A comparison of the two panels of Figure~\ref{fig:VQE Results}, together with the fidelities in Figure~\ref{fig:VQE_fidelities}, shows that standard variational diagnostics, including VQE energy and state fidelity, are not perfect predictors of AFQMC performance. In particular, although UCCSD yields among the lowest VQE energies across the geometries, the deepest HVA circuit (7-HVA) produces the most accurate AFQMC energies despite having a higher variational energy. This highlights that variational optimality alone is not a sufficient indicator of trial quality for ph-AFQMC.

Fidelity provides a more reliable qualitative guide. ansatzes with higher fidelities generally produce more accurate AFQMC energies, and the ranking of ansatz families by fidelity in Figure~\ref{fig:VQE_fidelities} broadly mirrors their AFQMC performance. In particular, 7-HVA achieves among the highest fidelity trials and the most accurate projected energies. However, this relationship is not strictly monotonic. For example, the UpCCGSD ansatz exhibits relatively poor fidelity but yields moderately accurate AFQMC energies at larger bond lengths. To understand this behavior, we consider the role of spin symmetry. UpCCGSD exhibits significant spin contamination (Table~\ref{tab:VQE_AFQMC_Table}); however, because the Hamiltonian is spin symmetric and the walkers are initialized in RHF states, the walker ensemble remains in the correct spin sector, even when the trial wavefunction breaks this symmetry. As a result, AFQMC effectively samples a state closer to the spin-projected trial wavefunction.

This observation motivates a modified diagnostic. Since $S_z = 0$ throughout, we project the trial state onto the singlet subspace using the exact polynomial projector
\begin{equation}
\hat P_{S=0}
=
\prod_{s=1}^{S_{\max}}
\left(
I - \frac{\hat{S}^2}{s(s+1)}
\right),
\qquad
S_{\max} = \frac{N_\alpha+N_\beta}{2},
\label{eq:spin projector}
\end{equation}
in which each factor annihilates the component of total spin $S=s$ while leaving the singlet component unchanged. The normalized singlet-projected state is $|\Psi_{S=0}\rangle = \hat P_{S=0} |\Psi_T\rangle / \sqrt{\langle \Psi_T | \hat{P}_{S=0} | \Psi_T \rangle}$, and we define the spin-projected fidelity as
\begin{equation}
F_{S=0}
=
\left|
\langle \Psi_{\mathrm{FCI}} | \Psi_{S=0} \rangle
\right|^2 .
\label{eq:projected fidelity}
\end{equation}

By construction $F_{S=0} \geq F$, with the gap growing with the degree of spin contamination. Figure~\ref{fig:VQE_fidelities} shows both quantities on the same axes: open markers denote the spin-projected fidelities for points where spin projection produces an appreciable change. Because the walker ensemble is restricted to the spin-symmetric manifold, we therefore expect and observe $F_{S=0}$ to provide a more informative measure of trial quality whenever the ansatz breaks spin symmetry. We return to this point in Section~\ref{sec:molecules}, where we quantify how well each diagnostic predicts ph-AFQMC accuracy across the full benchmark set.

A key distinction among ansatzes is the smoothness and stability of the projected energy curves. The AFQMC energies obtained from UCCSD, ADAPT-UCCSD, 3-LUCJ, and 7-HVA trials are smooth and nearly indistinguishable across the potential energy surface, suggesting that sufficiently expressive ansatzes capture the phase structure relevant for the phaseless constraint. In contrast, the UpCCGSD trials produce visible cusps and irregular behavior in the strongly correlated regime. This behavior may reflect the more difficult optimization landscape, where convergence to different local minima at adjacent geometries can produce trial states with inconsistent phase structure along the PES. For comparison, the single-determinant RHF trial, despite having poor fidelity and variational energy, yields a smooth AFQMC curve that simply deviates systematically from the exact result.

Taken together, these observations suggest that the effectiveness of a trial wavefunction in ph-AFQMC is governed less by variational optimality and more by the accuracy and stability of its phase structure. ansatzes, such as UCCSD, ADAPT-UCCSD, LUCJ, and HVA appear to capture this structure consistently once sufficient expressivity is reached, resulting in smooth and nearly identical AFQMC energy curves, whereas optimization difficulties in UpCCGSD lead to inconsistencies in the resulting phase structure. We report our VQE and AFQMC results for each ansatz at representative bond lengths $R = 1.0, 1.4,\ \text{and}\ 1.8~\text{\AA}$ in Table~\ref{tab:VQE_AFQMC_Table}, providing numerical values corresponding to the data shown in Figures~\ref{fig:VQE Results} and \ref{fig:VQE_fidelities}. The table also includes analogous results for H$_{8}$, for which we observe that relative trends in energies and fidelities are consistent with those of H$_{10}$, indicating that our findings generalize across system size in hydrogen chains.

\begin{figure*}[t]
    \centering
    \includegraphics[width=\textwidth]{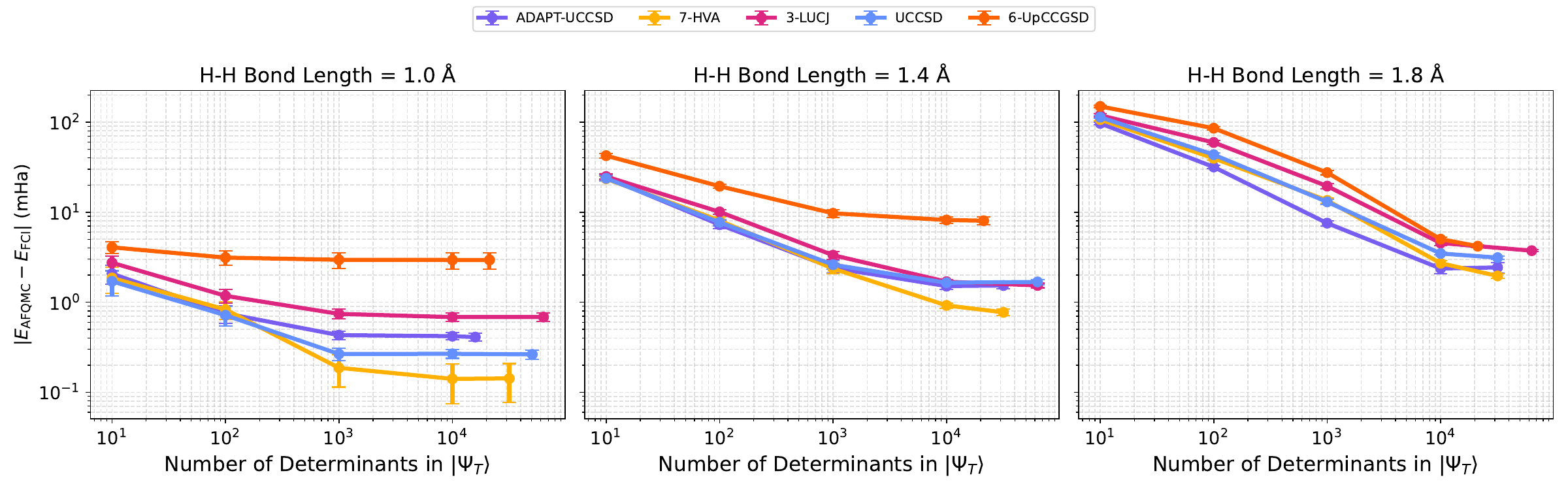}
    \caption{
AFQMC energy error as a function of the number of determinants retained in the MSD trial wavefunctions for H$_{10}$ at three bond lengths ($R$ = 1.0, 1.4, and 1.8~\AA). Trials are constructed by truncating the full quantum-prepared state, retaining the largest-magnitude coefficients, and renormalizing the resulting state. Note that on a logarithmic axis the drawn length of an error bar reflects the ratio $\sigma/|E_{\mathrm{AFQMC}} - E_{\mathrm{FCI}}|$ rather than $\sigma$ itself, so the most accurate trials display the longest bars at comparable statistical quality.}
    \label{fig:AFQMC vs numb dets}
\end{figure*}

In Figure~\ref{fig:AFQMC vs numb dets}, we present the AFQMC energy errors as a function of the number of determinants in each trial wavefunction. These results were obtained by truncating the full multi-determinant expansion, keeping the determinants with the largest coefficients, and renormalizing the truncated state. As expected, increasing the bond length enhances the multi-reference character of the system and, consequently, a larger number of determinants is required for the AFQMC energies to plateau. Also, increasing the number of determinants systematically improves the AFQMC accuracy in all cases. Overall, the different ansatzes exhibit qualitatively similar behavior as the number of determinants increases, and the observed differences appear to stem primarily from intrinsic trial quality rather than from a fundamentally more compact representation of correlation. In other words, no particular ansatz systematically captures significantly more correlation with fewer determinants than the others. Nevertheless, for all ansatzes considered, relatively compact approximations to the full trial wavefunction are sufficient to achieve converged AFQMC energies, especially near equilibrium geometries. This observation is particularly important in practice, as the trial overlap computation, which depends on the number of determinants, is the primary computational bottleneck of AFQMC. We note that these determinant-truncation results depend on the choice of single-particle basis. All results in Figure~\ref{fig:AFQMC vs numb dets} are obtained in the canonical RHF MO basis, whereas alternative choices, such as localized orbitals or natural orbitals obtained from a correlated wavefunction, can substantially alter the compactness of the MSD expansion and therefore the number of determinants required to reach converged ph-AFQMC energies \cite{huang2024gpu}.

\begin{figure}[t]
    \centering
    \includegraphics[width=\linewidth]{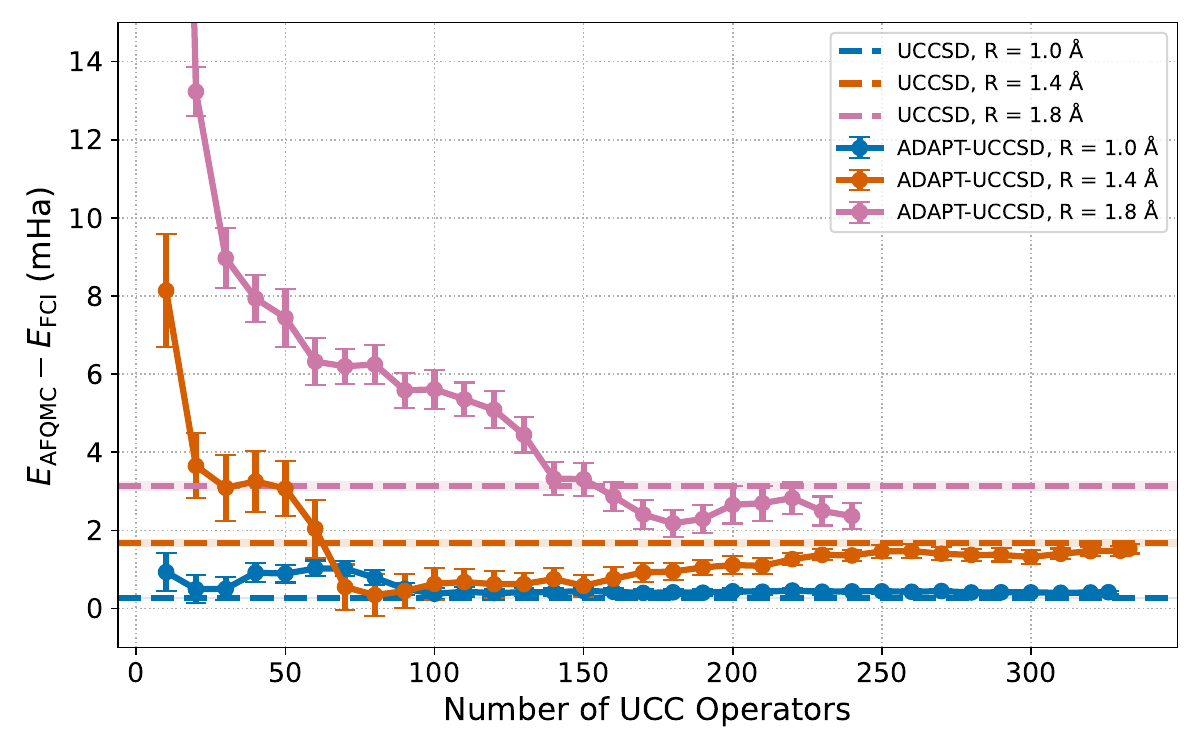}
    \caption{
AFQMC energy error as a function of the number of operators in the ADAPT-UCCSD ansatz for H$_{10}$ at three bond lengths ($R = 1.0$, 1.4, and 1.8~\AA). Each point corresponds to a trial wavefunction saved every ten ADAPT-VQE iterations and subsequently used in ph-AFQMC. Horizontal dashed lines indicate the corresponding AFQMC(UCCSD) reference energies.}    
\label{fig:ADAPT}
\end{figure}

In Figure~\ref{fig:ADAPT}, we focus specifically on the ADAPT-UCCSD trial by extracting the wavefunction after every ten iterations of the ADAPT-VQE algorithm and using each as a trial wavefunction for AFQMC. By construction, ADAPT-VQE monotonically lowers the variational energy as additional operators are appended to the ansatz until convergence. Consequently, as the number of operators increases in Figure~\ref{fig:ADAPT}, the trial energies decrease strictly monotonically.
As expected, we observe a general trend of improving AFQMC accuracy as the number of operators increases, followed by a plateau once the projected energy saturates. However, this improvement is not strictly monotonic. At the intermediate bond length ($R = 1.4$~\AA), after reaching its optimal AFQMC accuracy, the continued addition of operators leads to a gradual increase in AFQMC energy error, indicating ADAPT-VQE may over-optimize directions irrelevant (or harmful) for AFQMC.

This behavior does not appear to be solely attributable to statistical fluctuations and instead reflects a distortion in the phase structure of the trial wavefunction. We also note that the statistical uncertainty of the AFQMC simulations decreases as we increase the number of operators. 
This result indicates that systematic improvements in trial variational energy do not necessarily translate into improved AFQMC performance, further emphasizing the nontrivial relationship between variational optimality and the phase structure relevant to the phaseless constraint. This observation is consistent with similar recent findings that lower variational energies do not necessarily lead to more accurate ph-AFQMC energies \cite{danilov2026selecting, amsler2023classical}.

We also compare the AFQMC energies obtained from ADAPT-VQE trials to the corresponding AFQMC(UCCSD) results, shown as horizontal reference lines. This comparison reveals the operator count at which the adaptive trial surpasses the full UCCSD trial in AFQMC accuracy. Notably, although the UCCSD ansatz for H$_{10}$ contains 875 operators (Table~\ref{tab:VQE_AFQMC_Table}), in the strongly correlated regime a substantially smaller, adaptively selected subset of operators is sufficient to achieve superior AFQMC accuracy. This highlights the practical advantage of adaptive ansatz construction in generating compact yet high-quality AFQMC trial wavefunctions.

\subsection{UHF Initialization}
\label{UHF_init}
\begin{figure}[t]
    \centering
    \includegraphics[width=\linewidth]{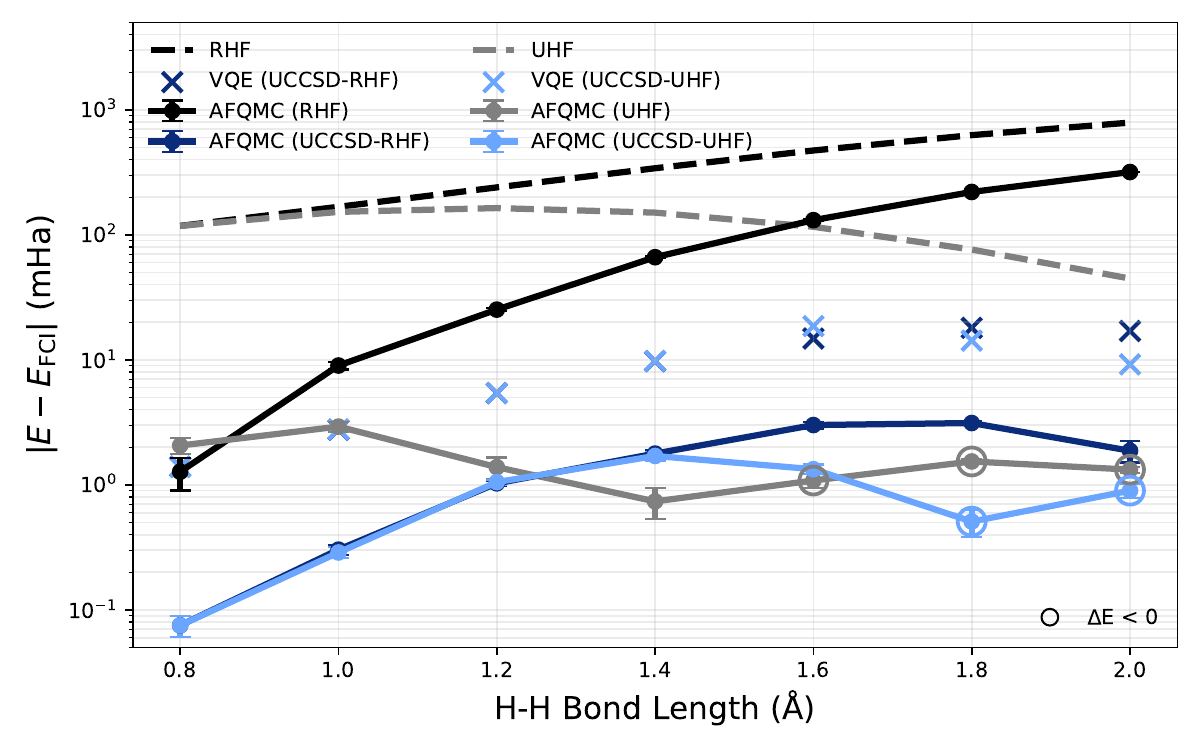}
    \caption{Comparison of RHF- and UHF-based references for VQE and ph-AFQMC along the H$_{10}$ bond-stretching curve. The vertical axis shows energy error in mHa on a logarithmic scale. Dashed lines denote mean-field RHF and UHF errors, crosses indicate VQE(UCCSD) results initialized on RHF and UHF references, and solid circles show the corresponding AFQMC energies using these trials. Encircled data points denote non-variational results for which $E - E_{\mathrm{FCI}} < 0$.}
\label{fig:UHF}
\end{figure}

We have performed a brief study comparing RHF- and UHF-based trials, the results of which are shown in Figure~\ref{fig:UHF}. Our implementation of AFQMC(UHF) trials follows the protocol of the Simons Foundation hydrogen chain benchmarking study~\cite{motta2017towards}. As a validation step, we first reproduce the reported energies for the corresponding geometries; our results are summarized in Table~\ref{tab:simons_comparison} and show excellent agreement.

AFQMC(RHF) and AFQMC(UHF) simulations for our set of geometries are performed using a projection time of $\tau = 200\,E_{\text{Ha}}^{-1}$. To obtain stable UHF solutions at stretched geometries, we explicitly break the spin symmetry of an RHF initial guess during the self-consistent field procedure to prevent convergence back to the restricted solution. As expected, the RHF and UHF mean-field solutions begin to diverge near $R \approx 1.0$~\AA, signaling the onset of static correlation, with UHF capturing additional correlation energy beyond this point. 

We next report VQE energies obtained with the UCCSD ansatz initialized with reference states $\ket{\Psi_{\mathrm{RHF}}}$ and $\ket{\Psi_{\mathrm{UHF}}}$, labeled UCCSD-RHF and UCCSD-UHF, respectively, with all variational parameters initialized to zero. We focus on UCCSD in this study because it is a widely used benchmark ansatz for VQE, allowing the effect of the reference state to be isolated while holding the variational form fixed. The circuits for the two initializations have the same variational structure; the difference arises from optimizing with respect to either the original Hamiltonian in the RHF spin-orbital basis or the classically rotated Hamiltonian in the UHF spin-orbital basis. Equivalently, this change of orbital basis can be represented in the original RHF basis by applying the corresponding fermionic orbital rotation to the UHF-based state. The resulting VQE energies are nearly identical at near-equilibrium bond lengths and begin to deviate slightly around $R = 1.6$~\AA, where correlation effects become more pronounced.

These optimized trial states are then supplied to AFQMC. For the UHF-initialized
circuits, the statevector is classically rotated back to the original RHF orbital
basis before AFQMC is performed. As shown in Figure~\ref{fig:UHF}, AFQMC(UCCSD-RHF) and AFQMC(UCCSD-UHF) yield nearly identical results near equilibrium, but begin to diverge around $R = 1.6$~\AA. In the intermediate stretched regime, the UHF-initialized trial produces systematically lower AFQMC errors, indicating improved phaseless constraint quality due to the enhanced static-correlation character of the UHF reference. At larger bond lengths ($R \gtrsim 1.8$~\AA), we observe non-variational AFQMC energies (i.e., $E < E_{\mathrm{FCI}}$). This behavior can be expected, as ph-AFQMC is not a strictly variational method~\cite{carlson1999issues,lee2019auxiliary,mahajan2025beyond}. Similar trends are observed when comparing single-determinant AFQMC(UHF) and AFQMC(RHF): UHF-based trials improve accuracy in the bond-stretched regime but can also introduce non-variational energies in the strongly stretched limit. 

\section{Benchmark Results: Molecular Test Set}
\label{sec:molecules}

Hydrogen chains isolate the crossover from dynamic to static correlation in a single, cleanly parameterized coordinate, but they are chemically homogeneous. To test whether the conclusions of Section~\ref{sec:results} transfer, we repeat the benchmark on nine closed-shell molecules at their experimental equilibrium geometries: BeH$_2$, BH$_3$, CH$_4$, NH$_3$, H$_2$O, N$_2$, C$_2$, Be$_2$, and C$_2$H$_2$. All calculations again use the STO-6G basis, so that FCI reference energies remain accessible, and the quantum calculations correspond to 14--20 qubits. For all systems except C$_2$H$_2$, both VQE state preparation and ph-AFQMC are performed in the full orbital space. For C$_2$H$_2$, the two carbon $1s$ orbitals are frozen during quantum state preparation, yielding a CAS(10e,10o) trial, which is subsequently embedded into the full orbital space for the ph-AFQMC calculation. The reported C$_2$H$_2$ VQE energy error is measured relative to the full-space FCI energy and therefore includes the 0.70 mHa by which the CASCI(10e,10o) energy lies above the full-space FCI energy, i.e., the core-correlation energy that no active-space ansatz can recover, whereas the trial fidelity is evaluated relative to the corresponding CASCI ground state. All reported ph-AFQMC energies are obtained in the full orbital space. For each molecule the five ansatzes are compared at matched parameter counts, with the repetition count $k$ of the multilayer ansatzes chosen per system to approximate the UCCSD parameter count. Unlike the hydrogen-chain calculations, for which each trial was obtained from a single optimization with deterministic initialization, ansatzes in the molecular benchmark exhibited greater sensitivity to initialization and were therefore optimized from multiple Gaussian-perturbed starting points, as detailed in Appendix~\ref{app: sim_details}. Full numerical results are collected in Table~\ref{tab:molecules}.

To place these systems on a common footing with the hydrogen chains we use the weight of the RHF determinant in the exact ground state, $w_{\mathrm{HF}} = |\langle \Phi_{\mathrm{RHF}} | \Psi_{\mathrm{FCI}} \rangle|^{2}$, as a simple proxy of multireference character: $w_{\mathrm{HF}} \rightarrow 1$ for a perfect single reference, and it falls as static correlation grows. Values for every system studied here are tabulated in Table~\ref{tab:corr_strength}. The molecular set spans $w_{\mathrm{HF}} = 0.97$ (H$_2$O) down to $0.66$ (C$_2$), and therefore interleaves with the hydrogen-chain series rather than duplicating it: the weakly correlated hydrides fill in the region above $w_{\mathrm{HF}} = 0.95$ that H$_{10}$ never reaches, while Be$_2$ ($0.77$) and C$_2$ ($0.66$) provide strong static correlation at equilibrium geometries. Taken together the two benchmarks cover $w_{\mathrm{HF}}$ from $0.97$ down to $0.14$ across sixteen systems.

\begin{figure*}[t]
    \centering
    \begin{subfigure}[t]{0.48\textwidth}
        \centering
        \includegraphics[width=\linewidth]{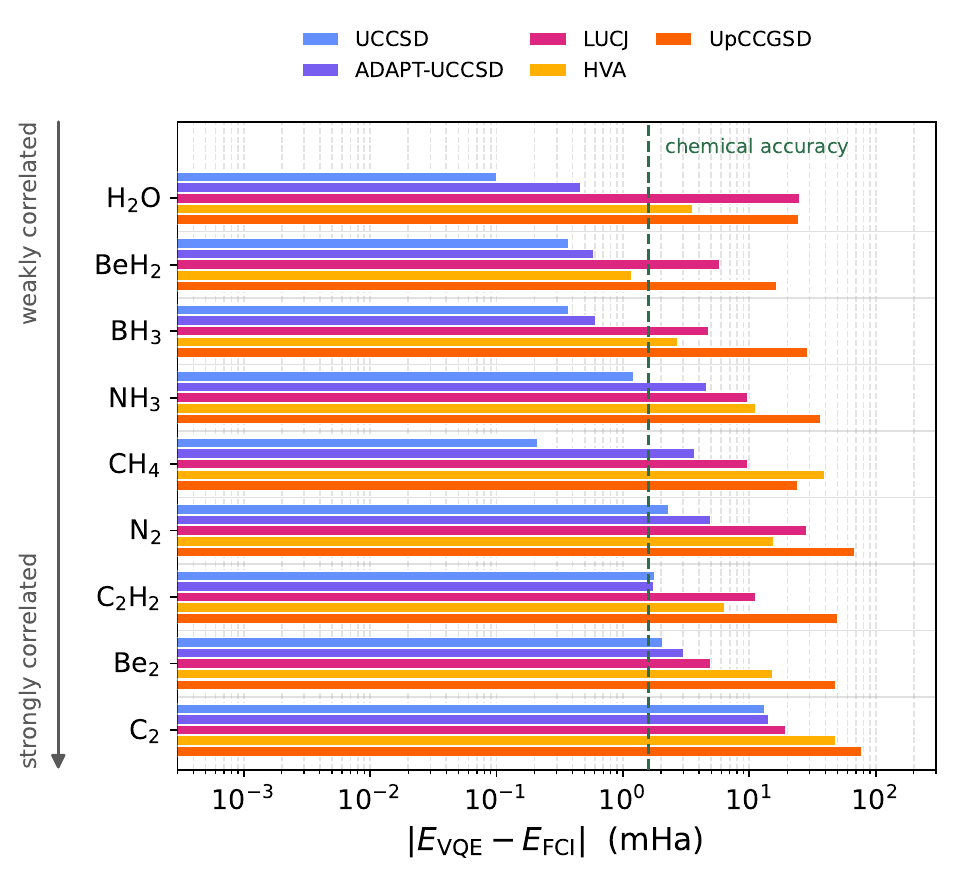}
        \caption{VQE energy errors}
        \label{fig:mol_vqe}
    \end{subfigure}
    \hfill
    \begin{subfigure}[t]{0.48\textwidth}
        \centering
        \includegraphics[width=\linewidth]{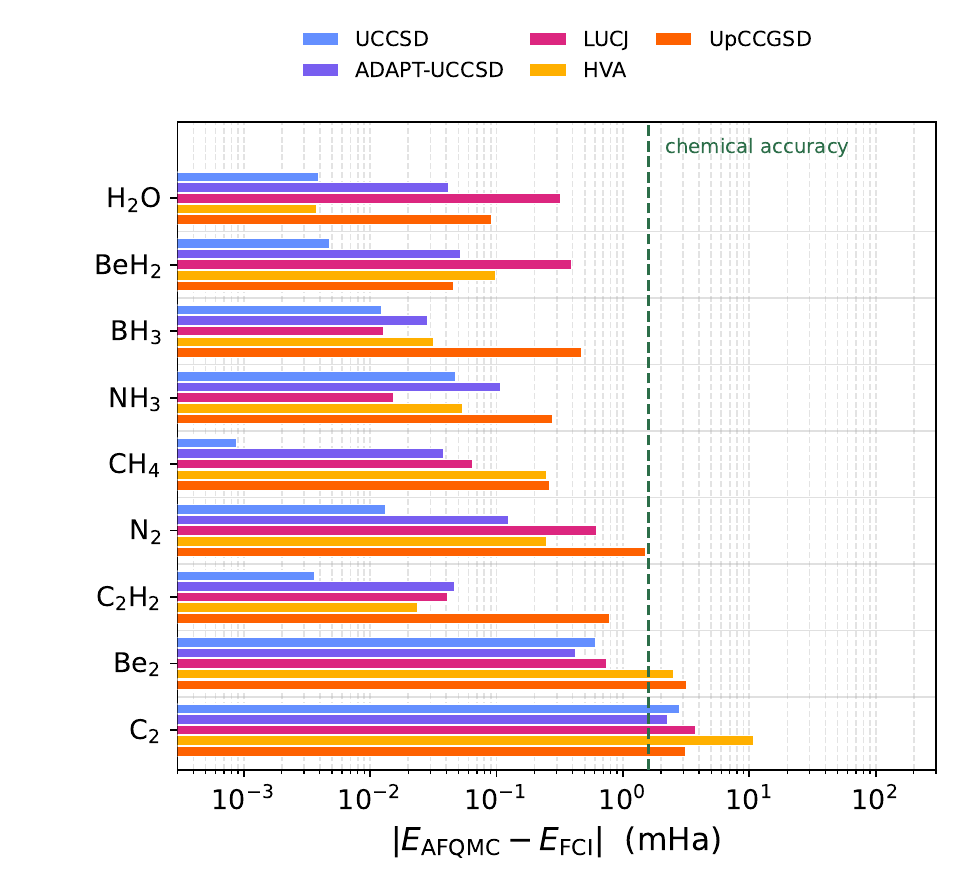}
        \caption{ph-AFQMC energy errors}
        \label{fig:mol_afqmc}
    \end{subfigure}
    \caption{(a) Absolute VQE energy error and (b) absolute ph-AFQMC energy error for the molecular test set in the STO-6G basis, ordered by decreasing weight $w_{\mathrm{HF}}$ of the Hartree--Fock determinant in the exact ground state (Table~\ref{tab:corr_strength}). For each molecule the five bars correspond to the five ansatzes at matched parameter counts. The dashed line marks chemical accuracy, $1.6$~mHa. Panel (b) plots the absolute deviation; some trials give projected energies slightly below FCI, and the signed values together with their statistical uncertainties are listed in Table~\ref{tab:molecules}}
    \label{fig:molecules}
\end{figure*}

Figure~\ref{fig:molecules} summarizes the results. The comparison between the two panels is the central observation. At the VQE level the ansatzes remain widely separated on every molecule as the errors span $0.10$--$76.7$~mHa, and only UCCSD and ADAPT-UCCSD are consistently around the boundary of chemical accuracy, with UpCCGSD the worst performer on most of the systems. After ph-AFQMC projection this spread largely collapses. For the seven weakly and moderately correlated molecules ($w_{\mathrm{HF}} \geq 0.85$) every trial, including the UpCCGSD states whose variational errors exceed 20~mHa, yields a projected energy within chemical accuracy; the largest error anywhere in that group is $1.5$~mHa. The choice of ansatz, which matters a great deal variationally, is almost irrelevant to the final projected energy. Only for Be$_2$ and C$_2$, the two strongly correlated members of the set, does the ansatz choice reassert itself. For these two molecules, just 30\% of the trials reach chemical accuracy and the errors span $0.43$--$10.8$~mHa. This reproduces, in a chemically diverse setting, the pattern seen along the H$_{10}$ curve, where the ansatzes are nearly indistinguishable near equilibrium and separate only under bond stretching.

The relative standing of the ansatz families is broadly consistent with the hydrogen-chain results, with one notable exception. UCCSD and ADAPT-UCCSD are again the most reliable, and ADAPT-UCCSD attains comparable accuracy with far fewer parameters, confirming that adaptive construction yields compact trials across chemically diverse systems and not only for hydrogen chains. LUCJ similarly retains the favorable tradeoff observed for H$_{10}$, providing competitive ph-AFQMC accuracy with a substantially shallower, locality-constrained circuit (Table~\ref{app:resources}). HVA, however, does not transfer as successfully. Although 7-HVA was among the most accurate trials along the H$_{10}$ dissociation curve, it is never cleaeryl the best-performing trial in the molecular test set and gives the largest ph-AFQMC errors for C$_2$.

The clearest molecular illustration of the spin-symmetry argument developed in Section~\ref{sec:results} is provided by C$_2$. The UpCCGSD trial for this molecule is heavily spin contaminated, $\langle \hat S^2 \rangle = 0.66$, and has by far the lowest raw fidelity of the five, $F = 0.652$ against $0.894$ for HVA. Raw fidelity would therefore rank it last. In fact its projected energy, $+3.2$~mHa, is markedly better than that of HVA, $+10.8$~mHa. The spin-projected fidelities resolve the discrepancy: $F_{S=0} = 0.932$ for UpCCGSD against $0.907$ for HVA, an ordering that matches the ph-AFQMC result. This is the same effect identified for stretched H$_{10}$, now observed at an equilibrium geometry of a genuine molecule, and it is the only system in the molecular set where the two fidelity measures differ appreciably, consistent with the fact that the remaining trials are nearly spin pure.

\begin{figure}[t]
    \centering
    \includegraphics[width=\linewidth]{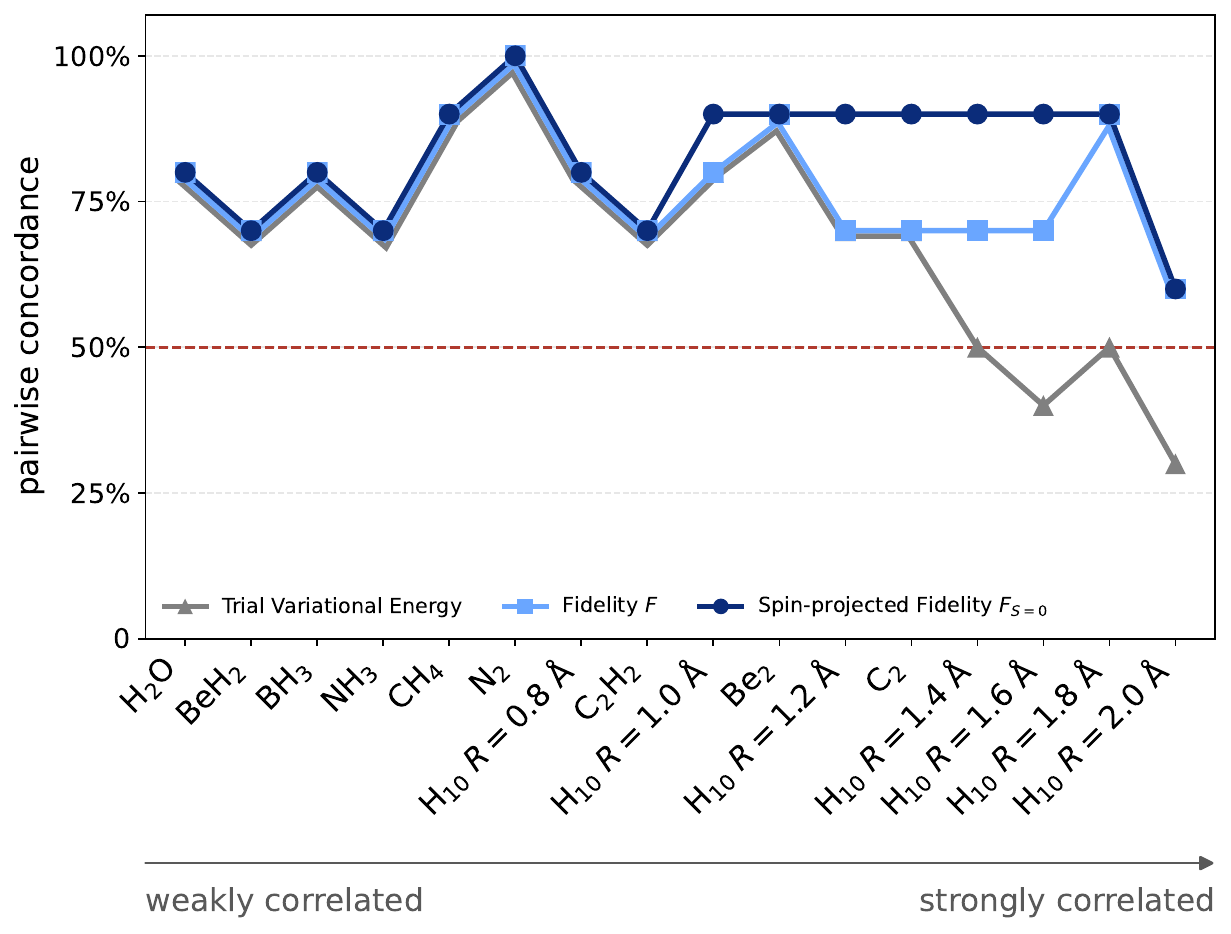}
    \caption{Pairwise concordance $C$ of three trial diagnostics with ph-AFQMC accuracy across the combined benchmark set. For each system, the five parameter-matched trials define $\binom{5}{2}=10$ pairs. Pairwise concordance is the fraction of these pairs for which the ordering predicted by a diagnostic agrees with the ordering of the corresponding absolute ph-AFQMC errors, $|E_{\mathrm{AFQMC}}-E_{\mathrm{FCI}}|$. A concordance of $0.5$ (dashed line) corresponds to the random-ordering baseline. Systems are ordered by decreasing $w_{\mathrm{HF}}$ (Table~\ref{tab:corr_strength}), corresponding to increasing multireference character.}
    \label{fig:concordance}
\end{figure}

Combining the two benchmarks allows the relative predictive value of these diagnostics to be compared quantitatively. Figure~\ref{fig:concordance} measures, for each of the sixteen systems studied, how often the ordering of a pair of trials according to a given diagnostic agrees with their ordering by absolute ph-AFQMC error. We quantify this agreement using pairwise concordance. For each system, we consider all $\binom{5}{2}=10$ pairs of parameter-matched trials and ask whether the diagnostic correctly identifies which trial has the smaller ph-AFQMC error. The concordance is then simply the fraction of trial pairs that are ordered correctly, $C = N_{\mathrm{concordant}}/{N_{\mathrm{pairs}}}$, where $N_{\mathrm{pairs}}=10$ for each system. Thus, $C=1$ indicates that the diagnostic correctly ranks every pair of trials, $C=0$ that it ranks every pair incorrectly, and $C=0.5$ is the value expected from random ordering. No ties occur in our data. In this case, the concordance is directly related to Kendall's rank correlation coefficient \cite{kendall1938new} by $C=(1+\tau)/2$. We base our conclusions on the consistency of the ordering across all sixteen systems rather than on any individual system. An uncertainty-aware version of this analysis, restricted to trial pairs whose ph-AFQMC energies are statistically resolvable and with bootstrap confidence intervals over systems, is given in Section~\ref{app:concordance} and yields the same ordering of the three diagnostics.

Three features stand out. Firstly, across the systems considered here, spin-projected fidelity is never less concordant with ph-AFQMC accuracy than unprojected fidelity, which in turn is never less concordant than variational energy; $F_{S=0}$ is strictly more concordant than $F$ in five systems, while $F$ is strictly more concordant than variational energy in four. Secondly, for the first half of the systems, which are weakly and moderately correlated systems, the three diagnostics generally give the same ordering, with a median concordance of 80\%. In this regime, essentially all trials already yield ph-AFQMC energies within chemical accuracy, limiting the practical significance of differences among their rankings. Thirdly, the diagnostics separate as the multireference character increases, with median concordances of 60\%, 70\%, and 90\% for variational energy, $F$, and $F_{S=0}$, respectively, among the strongly correlated systems. In the stretched hydrogen chains, the concordance of variational energy eventually falls to the random-ordering baseline, whereas $F_{S=0}$ remains at 90\% through $R=1.8$~\AA{}.

The improvement of $F_{S=0}$ over $F$ is concentrated in a relatively small number of spin-contaminated trials, primarily UpCCGSD in the stretched hydrogen chains and C$_2$. Spin projection should therefore be viewed as a correction that becomes important when spin symmetry is broken rather than as providing a uniform improvement across all trials. Moreover, at the most stretched geometry considered, H$_{10}$ at $R=2.0$~\AA{}, the concordance of all three diagnostics decreases substantially. Across the present benchmark, fidelity therefore provides a more reliable ranking diagnostic than variational energy, particularly when spin contamination is accounted for, but it is not a universal predictor of phaseless bias, consistent with recent reports that higher trial fidelity need not translate into improved ph-AFQMC energies \cite{kjonstad2025systematic, kjonstad2026can}.

\section{Discussion}
\label{sec:Discussion}

Our VQE results on H$_{10}$ show that UCCSD remains the most consistent and
reliable, general-purpose ansatz across the entire potential energy surface,
benefiting substantially from chemically informed parameter initialization.
In particular, CCSD warm-starts significantly reduced optimization difficulty
and yielded strong variational energies and fidelities at weak-to-moderate
correlation. ADAPT-UCCSD closely tracks UCCSD in both energy and fidelity
while producing systematically more compact circuits in the strongly correlated
regime. LUCJ exhibits comparatively modest equilibrium accuracy, but its error
remains nearly constant across bond lengths, demonstrating robustness to
increasing correlation. Additionally, the multilayer form (3-LUCJ) becomes
competitive with UCCSD in the strongly correlated regime while requiring only
nearest-neighbor two-qubit connectivity under the assumed square-topology qubit
layout. HVA shows the most systematic depth-driven improvement: although
single-layer HVA underperforms, at sufficient depth (e.g., 7-HVA) it approaches
the accuracy of UCCSD. UpCCGSD proves the most challenging ansatz to optimize
and generally underperforms across the geometries considered, with multilayer
variants not yielding consistent improvement. These VQE trends are consistent
with the broader observation that chemically structured ansatzes and informed
initialization strategies can substantially improve optimization behavior at
higher qubit counts, while more compact ansatzes may require careful depth
selection and specialized optimization techniques.

When these optimized states are then used as trials to ph-AFQMC, several central
conclusions emerge. Notably, the relationship between VQE optimality and AFQMC
performance is nontrivial: trial variational energy alone is not a reliable
predictor of ph-AFQMC accuracy. In contrast, fidelity with respect to the FCI
ground state generally provides a more informative qualitative indicator of
AFQMC performance, particularly away from extreme bond lengths. When the optimized trial exhibits appreciable spin contamination, as observed for several UpCCGSD and HVA trials, the relevant quantity is the fidelity of the singlet-projected trial, since that is the component the spin-restricted walker ensemble samples; across all sixteen systems it ranks candidate trials at least as reliably as the raw fidelity, which in turn is at least as reliable as the variational energy (Section~\ref{sec:molecules}). This reinforces the key point that the effectiveness of a trial in ph-AFQMC depends not only on
variational optimality or unprojected fidelity but also on the phase structure that enters the
phaseless constraint.

We observe that increasing ansatz expressivity generally improves the quality of the
phase structure relevant for ph-AFQMC. Across LUCJ, HVA, and UpCCGSD, moving
from single-layer to parameter-matched multilayer constructions typically
yields smoother projected-energy behavior and reduced AFQMC error. Importantly,
under approximately fixed parameter budgets, multiple ansatz families produce
comparable AFQMC performance, but the strength of this statement depends on
how strongly correlated the system is. For the weakly and moderately correlated
molecules of Section~\ref{sec:molecules} it is clear, as every ansatz, including
those with variational errors above 20~mHa, delivers a projected energy within
chemical accuracy, so the ansatz choice is effectively immaterial and the cheapest
adequate circuit should be preferred. In the strongly correlated regime the
families separate clearly, by an order of magnitude on C$_2$ and across the
stretched hydrogen chains, and the choice of ansatz once again matters. The
practical reading is therefore conditional: sufficient expressivity is what
matters, and near equilibrium the bar for ``sufficient'' is low enough that most
ansatzes clear it. This observation
is consistent with our determinant-truncation analysis, where all trial families
exhibited qualitatively similar convergence of AFQMC energies with the number
of retained determinants. Differences between methods therefore largely reflect
intrinsic trial quality rather than a systematically more compact MSD
representation.

Despite these broad similarities, clear tradeoffs among ansatzes emerge. HVA
achieves the strongest overall ph-AFQMC performance in our H$_{10}$ study,
including cases where it does not yield the lowest VQE energy, further
underscoring that AFQMC accuracy depends on more than variational optimality.
This advantage does not, however, transfer to the chemically
diverse molecular test set, where HVA is never clearly the best trial and gives among the largest
ph-AFQMC errors for the two most strongly correlated molecules, Be$_2$ and
C$_2$ (Section~\ref{sec:molecules}). This system dependence is consistent with
the construction of HVA, which is particularly natural for structured lattice
Hamiltonians, such as Hubbard- and Heisenberg-type models, that decompose into
a small number of physically meaningful local terms. Molecular
electronic-structure Hamiltonians mapped to qubits instead contain a much
larger and less local collection of Pauli terms, making the direct HVA
construction considerably less compact. In the implementation considered
here, where every Pauli term is exponentiated once per layer, the uncompiled
two-qubit gate count exceeds that of UCCSD by roughly a factor of five for
H$_{10}$ and four for C$_2$, and that of LUCJ by more than two orders of
magnitude (Table~\ref{tab:resources}). The classical optimization runtime cost of HVA is also substantial,
exceeding that of the cheapest ansatz for C$_2$ by more than a factor of
twenty (Table~\ref{tab:resources}). Thus, while HVA can provide highly accurate ph-AFQMC trials for the
regular hydrogen-chain problem, its accuracy does not transfer as consistently
to the broader molecular benchmark and comes with unfavorable resource costs
for molecular Hamiltonians.

LUCJ offers a compelling alternative: it is compact,
locality-constrained, and naturally compatible with hardware connectivity,
while remaining competitive in ph-AFQMC when given sufficient depth. UCCSD
remains a robust baseline with strong and consistent results, particularly when
paired with unrestricted CCSD warm-starts, although its gate-depth scaling and operator
count become limiting for larger systems. In contrast, UpCCGSD, while compact,
exhibits significant optimization difficulties; its ph-AFQMC curves are less
smooth and in some cases display unpredictable behavior, including instances
of non-variational projected energies. This highlights the practical risk of
using ansatzes whose optimization landscapes are difficult to navigate. Although analogous optimization and approximation issues can arise in constructing classical MSD trials, variational quantum circuits introduce circuit trainability as an additional consideration: the quality of the resulting phaseless constraint depends not only on the states that the ansatz can represent in principle, but also on whether those states can be reliably reached through circuit optimization.

Finally, ADAPT-UCCSD combines strong performance with compactness. Our analysis
in Figure~\ref{fig:ADAPT} reveals that the best ph-AFQMC performance is often
achieved before full ADAPT convergence, suggesting a promising strategy for
producing high-quality, resource-efficient trials by terminating the adaptive
procedure once the projected energy saturates.

We also examined the effect of the mean-field reference used to initialize the quantum trial using UCCSD along the H$_{10}$ bond-stretching curve.
Consistent with prior AFQMC studies, UHF-based
references become advantageous in the intermediate stretched regime, where
symmetry breaking provides a qualitatively improved description of static
correlation and can improve AFQMC accuracy. At the same time, in the
strongly stretched limit, we observe occasional
non-variational behavior, which emphasizes that UHF-based
trials can be beneficial but must be applied judiciously. Our findings are consistent with similar reports that unrestricted trials have led to non-variational energies \cite{danilov2026selecting}. The use of a UHF reference is not limited to UCCSD
and is, in principle, compatible with the other ansatzes considered here. For
LUCJ we primarily employ the spin-balanced formulation, in which
\( J^{\alpha\alpha} = J^{\beta\beta} \), but this restriction can be relaxed to
obtain a spin-unbalanced variant. In that case, one could use a UHF reference
state and apply the corresponding orbital rotation to the Hamiltonian in the
same manner as done for UCCSD. Similarly, for the HVA, the Pauli operators
may be constructed from the Hamiltonian expressed in the UHF-rotated basis.
The sensitivity to UHF initialization may depend on the structure and expressivity of the ansatz \cite{bertels2022symmetry,tsuchimochi2020spin}, and a systematic investigation of this dependence across different circuit families is left for future work.

Since all simulations in this work are performed in the noiseless setting, corresponding to ideal statevector-based implementations of the trial wavefunctions, our comparisons are most directly relevant to fault-tolerant quantum computing, where deep circuits and precise state preparation are feasible. Notably, prior quantum hardware implementations of QC-AFQMC have demonstrated a degree of inherent noise resilience \cite{huggins2022unbiasing,huang2024evaluating,zhao2025quantum}. Nevertheless, the relative performance of different ansatz families may be affected in the presence of noise, particularly for deeper or more highly entangled circuits. Despite this, many of the qualitative trends observed here, such as the reduced sensitivity of ph-AFQMC to variational energy accuracy and the comparable performance of different ansatz families at matched parameter counts, are expected to persist, as they primarily reflect properties of the AFQMC-projected state obtained via imaginary-time propagation rather than the specific details of the underlying circuit implementation. Understanding the interplay between noise, ansatz structure, and AFQMC performance remains an important direction for future work.

\section{Conclusion}
\label{sec:Conclusion}

In this work, we present a comprehensive benchmarking study of quantum trial
wavefunctions for ph-AFQMC within the QC-AFQMC framework. Building on recent
algorithmic progress that has brought the post-processing cost of matchgate-shadow
QC-AFQMC close to the scaling of classical AFQMC, our focus was on a complementary
and fundamental question: which families of quantum trial states provide the
most effective phaseless constraints in practice, and what tradeoffs arise among
expressivity, optimization cost, circuit depth, and AFQMC performance? To address
this, we benchmarked a broad set of chemically and physically motivated ansatzes,
including members of the UCC family (UCCSD and UpCCGSD), LUCJ, HVA, and adaptive
constructions such as the ADAPT-VQE algorithm based on the UCCSD operator pool.
Using GPU-accelerated statevector simulations distributed across multiple
nodes, we optimized each ansatz via VQE, expanded the resulting statevector into
an MSD trial wavefunction, and then performed standard
classical ph-AFQMC using these trials. Linear hydrogen chains in a minimal
STO-6G basis under bond stretching were selected as a controlled benchmark, along with a second benchmark of nine closed-shell molecules at their equilibrium geometries with varying degrees of static correlation.
Our results demonstrate that quantum-generated trial states can provide
reliable constraints for ph-AFQMC as several ansatz families achieve chemical
accuracy near equilibrium geometries, while the most expressive trials remain
close to this threshold even in the strongly correlated regime. From an
alternative perspective, ph-AFQMC may be viewed as an effective classical
post-processing step for quantum ground-state estimation \cite{danilov2025enhancing}.

This work provides several practical guidelines for QC-AFQMC trial design. Several of the trends identified here have analogues in the classical AFQMC literature. Our results establish how these principles manifest for quantum-prepared trials, for which trial quality is additionally shaped by the structure and symmetry properties of the variational circuit, its trainability, and the quantum and classical resources required for state preparation. The guidelines below therefore encompass both general principles of ph-AFQMC trial quality and the additional design tradeoffs introduced by quantum state preparation.

(i) Trial variational energy alone is not a reliable metric for
phaseless performance; diagnostics that better capture wavefunction quality,
such as fidelity or symmetry-projected fidelity in small systems or scalable
measures of state similarity, are more informative.
Quantifying this across all sixteen systems studied,
spin-projected fidelity is never less concordant with ph-AFQMC accuracy than
unprojected fidelity, which in turn is never less concordant than variational
energy when the diagnostics are used to rank candidate trials for a given
system. The three separate most clearly in the strongly correlated regime,
where trial choice is most consequential. When a trial breaks spin symmetry,
the singlet-projected fidelity provides the more relevant diagnostic because
the spin-restricted walker ensemble means that ph-AFQMC effectively samples
the projected component of the trial.

(ii) UCCSD initialized from unrestricted CCSD amplitudes
provides a strong and reliable baseline for molecular QC-AFQMC trials. Across
both benchmark sets it yields consistently accurate ph-AFQMC energies, while
the physically informed warm start makes the variational optimization
comparatively inexpensive, even for the strongly correlated systems considered
here. Its principal limitation is circuit depth: the number of excitation
operators and resulting gate count grow rapidly with system size. Thus, when
circuit depth is not the dominant constraint, warm-started UCCSD provides a
useful default against which more compact or hardware-oriented trial
constructions can be assessed.

(iii) Increasing the number of variational parameters, and thereby enhancing
expressivity, generally improves phaseless performance across ansatzes provided
that the enlarged parameter space remains tractable to optimize.
When parameter budgets are approximately matched, different
ansatz families exhibit comparable AFQMC behavior for weakly and moderately
correlated systems. Differences among ansatzes become more pronounced
as static correlation increases, emphasizing the need to consider trial
properties beyond variational energy, including information about symmetries and,
where available, measures of state similarity.

(iv) Hardware-aware constructions, such as LUCJ, can remain competitive in
ph-AFQMC while offering substantial advantages in circuit depth and
connectivity. On the two most strongly correlated systems
studied, LUCJ yields substantially shallower circuits than the other ansatzes considered (Table~\ref{tab:resources}), giving the most favorable accuracy--gate-count tradeoff observed here.

(v) Adaptive ansatzes offer a practical path to compact, high-quality trials,
and early stopping based on AFQMC-relevant saturation criteria may
substantially reduce quantum resources. ADAPT-UCCSD yields the
shortest circuit of the UCC family, with an uncompiled two-qubit gate count
approximately an order of magnitude below that of UCCSD, but this compactness
comes at additional classical cost because operator-pool gradient screening
must be performed before each operator is appended (Table~\ref{app:resources}). Where circuit
depth is the binding constraint this can represent a favorable exchange,
whereas the advantage is less clear when classical optimization cost is the
dominant limitation.

(vi) Orbital choices and symmetry breaking can provide an additional axis for
improving trials in strongly correlated and bond-stretched molecules without
increasing quantum resources.

(vii) Conclusions drawn from a single family of systems should
be treated with caution. HVA, which performs particularly well for the highly
regular hydrogen chains, does not retain the same advantage across the
chemically diverse molecular test set, illustrating the importance of
benchmarking trial constructions across systems with different correlation
and electronic-structure characteristics.

Several directions follow naturally from this study. A systematic exploration
of orbital optimization and spin-unbalanced generalizations across various ansatzes
could clarify when symmetry breaking is beneficial and how it should be
combined with variational circuit structure to minimize phaseless bias.
More scalable trial-quality diagnostics that correlate directly with
phaseless performance would enable principled early-stopping criteria and
more efficient trial selection at larger system sizes where fidelities are
unavailable. Extending these benchmarks beyond minimal basis sets, to larger active spaces and to systems such as transition-metal complexes and lattice models, remains important for establishing how far the present conclusions reach; the restriction to STO-6G here is set by the need for exact FCI references against which the phaseless bias can be measured. A systematic study of the single-particle basis, and of ansatzes built on symmetry-broken references such as GHF, are further natural extensions. As QC-AFQMC continues to mature algorithmically and
experimentally, the results presented here help map the practical design of
quantum trial wavefunctions and highlight concrete strategies for balancing
expressivity, hardware constraints, and ph-AFQMC accuracy in realistic
hybrid quantum--classical workflows.

\section*{Data and Software Availability}
The complete implementation used in this work, encompassing GPU-accelerated VQE and ph-AFQMC simulations with quantum trial wavefunctions, is publicly available at: \url{https://github.com/rr637/QC-AFQMC}.

\section*{Acknowledgements}
RR, KK, NM, and ER are supported by the U.S. Department of Energy (DOE) under Contract No. DE-AC02-05CH11231, through the National Energy Research Scientific Computing Center (NERSC), an Office of Science User Facility located at Lawrence Berkeley National Laboratory. RR and SS are supported by U.S. Department of Energy, Office of Science, National Quantum Information Science Research Centers, Quantum Science Center (QSC) and supported by PNNL’s Quantum Algorithms and Architecture
for Domain Science (QuAADS) Laboratory Directed Research and Development
(LDRD) Initiative. The Pacific Northwest National Laboratory is operated by Battelle for the U.S. Department of Energy under Contract
DE-AC05-76RL01830. JWM and NMT are supported by the U.S. Department of Energy, Office of Science, National Quantum Information Science Research Centers, Superconducting Quantum Materials and Systems Center (SQMS), under Contract No. 89243024CSC000002.
This research used resources of the National Energy Research
Scientific Computing Center, a DOE Office of Science User Facility
supported by the Office of Science of the U.S. Department of Energy
under Contract No. DE-AC02-05CH11231 using NERSC award
ASCR-ERCAP0037552 and ASCR-ERCAP0037527. RR thanks A. Mahajan for insightful discussions and comments on the manuscript.

\bibliography{main}

@article{huggins2022unbiasing,
   title={Unbiasing fermionic quantum Monte Carlo with a quantum computer},
   volume={603},
   ISSN={1476-4687},
   url={http://dx.doi.org/10.1038/s41586-021-04351-z},
   DOI={10.1038/s41586-021-04351-z},
   number={7901},
   journal={Nature},
   publisher={Springer Science and Business Media LLC},
   author={Huggins, William J. and O'Gorman, Bryan A. and Rubin, Nicholas C. and Reichman, David R. and Babbush, Ryan and Lee, Joonho},
   year={2022},
   month=Mar, pages={416--420} }

@article{wecker2015progress,
  title = {Progress towards practical quantum variational algorithms},
  author = {Wecker, Dave and Hastings, Matthew B. and Troyer, Matthias},
  journal = {Phys. Rev. A},
  volume = {92},
  issue = {4},
  pages = {042303},
  numpages = {10},
  year = {2015},
  month = {Oct},
  publisher = {American Physical Society},
  doi = {10.1103/PhysRevA.92.042303},
  url = {https://link.aps.org/doi/10.1103/PhysRevA.92.042303}
}

@inbook{Pan:2024,
   title={The sign problem in quantum Monte Carlo simulations},
   ISBN={9780323914086},
   url={http://dx.doi.org/10.1016/B978-0-323-90800-9.00095-0},
   DOI={10.1016/b978-0-323-90800-9.00095-0},
   booktitle={Encyclopedia of Condensed Matter Physics},
   publisher={Elsevier},
   author={Pan, Gaopei and Meng, Zi Yang},
   year={2024},
   pages={879--893} }

@article{Lee:2022,
    author = {Lee, Joonho and Pham, Hung Q. and Reichman, David R.},
    title = {Twenty Years of Auxiliary-Field Quantum Monte Carlo
in Quantum Chemistry: An Overview and Assessment on Main Group Chemistry
and Bond-Breaking},
    journal = {Journal of Chemical Theory and Computation},
    volume = {18},
    number = {12},
    pages = {7024-7042},
    year = {2022},
    month = {12},
    issn = {1549-9618},
    doi = {10.1021/acs.jctc.2c00802},
    url = {https://doi.org/10.1021/acs.jctc.2c00802},
}

@article{wiersema2020exploring,
  title = {Exploring Entanglement and Optimization within the Hamiltonian Variational Ansatz},
  author = {Wiersema, Roeland and Zhou, Cunlu and de Sereville, Yvette and Carrasquilla, Juan Felipe and Kim, Yong Baek and Yuen, Henry},
  journal = {PRX Quantum},
  volume = {1},
  issue = {2},
  pages = {020319},
  numpages = {14},
  year = {2020},
  month = {Dec},
  publisher = {American Physical Society},
  doi = {10.1103/PRXQuantum.1.020319},
  url = {https://link.aps.org/doi/10.1103/PRXQuantum.1.020319}
}

@article{zhao2025quantum,
  title = {Quantum-classical auxiliary field quantum Monte Carlo with matchgate shadows on trapped ion quantum computers},
  author = {Zhao, Luning and Goings, Joshua J. and Aboumrad, Willie and Arrasmith, Andrew and Calderin, Lazaro and Churchill, Spencer and Gabay, Dor and Harvey-Brown, Thea and Hiles, Melanie and Kaja, Magda and Keesan, Matthew and Kulesz, Karolina and Maksymov, Andrii and Maruo, Mei and Mu{\~n}oz, Mauricio and Nijholt, Bas and Schiller, Rebekah and de Sereville, Yvette and Smidutz, Amy and Tripier, Felix and Yao, Grace and Zaveri, Trishal and Collins, Coleman and Roetteler, Martin and Epifanovsky, Evgeny and Kovyrshin, Arseny and Tornberg, Lars and Broo, Anders and Hammond, Jeff R. and Chandani, Zohim and Khalate, Pradnya and Kyoseva, Elica and Chen, Yi-Ting and Kessler, Eric M. and Lin, Cedric Yen-Yu and Ramu, Gandhi and Shaffer, Ryan and Brett, Michael and Huang, Benchen and Hugues, Maxime R. and Takeshita, Tyler Y.},
  journal = {Phys. Rev. Res.},
  volume = {8},
  issue = {3},
  pages = {033061},
  numpages = {20},
  year = {2026},
  month = {Jul},
  publisher = {American Physical Society},
  doi = {10.1103/n1tf-8kr7},
  url = {https://link.aps.org/doi/10.1103/n1tf-8kr7}
}

@article{sugiyama1986auxiliary,
title = {Auxiliary field Monte-Carlo for quantum many-body ground states},
journal = {Annals of Physics},
volume = {168},
number = {1},
pages = {1-26},
year = {1986},
issn = {0003-4916},
doi = {10.1016/0003-4916(86)90107-7},
url = {https://www.sciencedirect.com/science/article/pii/0003491686901077},
author = {G Sugiyama and S.E Koonin}
}

@article{zhang1995constrained,
   title={Constrained Path Quantum Monte Carlo Method for Fermion Ground States},
   volume={74},
   ISSN={1079-7114},
   url={http://dx.doi.org/10.1103/PhysRevLett.74.3652},
   DOI={10.1103/physrevlett.74.3652},
   number={18},
   journal={Physical Review Letters},
   publisher={American Physical Society (APS)},
   author={Zhang, Shiwei and Carlson, J. and Gubernatis, J. E.},
   year={1995},
   month=May, pages={3652--3655} }

@article{zhang2003quantum,
  title = {Quantum Monte Carlo Method using Phase-Free Random Walks with Slater Determinants},
  author = {Zhang, Shiwei and Krakauer, Henry},
  journal = {Phys. Rev. Lett.},
  volume = {90},
  issue = {13},
  pages = {136401},
  numpages = {4},
  year = {2003},
  month = {Apr},
  publisher = {American Physical Society},
  doi = {10.1103/PhysRevLett.90.136401},
  url = {https://link.aps.org/doi/10.1103/PhysRevLett.90.136401}
}

@article{grimsley2019trotterized,
   title={Is the Trotterized UCCSD Ansatz Chemically Well-Defined?},
   volume={16},
   ISSN={1549-9626},
   url={http://dx.doi.org/10.1021/acs.jctc.9b01083},
   DOI={10.1021/acs.jctc.9b01083},
   number={1},
   journal={Journal of Chemical Theory and Computation},
   publisher={American Chemical Society (ACS)},
   author={Grimsley, Harper R. and Claudino, Daniel and Economou, Sophia E. and Barnes, Edwin and Mayhall, Nicholas J.},
   year={2019},
   month=Dec, pages={1--6} }

@article{landinez2019non,
    author = {Landinez Borda, Edgar Josu{\'e} and Gomez, John and Morales, Miguel A.},
    title = {Non-orthogonal multi-Slater determinant expansions in auxiliary field quantum Monte Carlo},
    journal = {The Journal of Chemical Physics},
    volume = {150},
    number = {7},
    pages = {074105},
    year = {2019},
    month = {02},
    issn = {0021-9606},
    doi = {10.1063/1.5049143},
    url = {https://doi.org/10.1063/1.5049143},
}

@article{morales2012multideterminant,
   title={Multideterminant Wave Functions in Quantum Monte Carlo},
   volume={8},
   ISSN={1549-9626},
   url={http://dx.doi.org/10.1021/ct3003404},
   DOI={10.1021/ct3003404},
   number={7},
   journal={Journal of Chemical Theory and Computation},
   publisher={American Chemical Society (ACS)},
   author={Morales, Miguel A. and McMinis, Jeremy and Clark, Bryan K. and Kim, Jeongnim and Scuseria, Gustavo E.},
   year={2012},
   month=June, pages={2181--2188} }

@article{clark2011computing,
    author = {Clark, Bryan K. and Morales, Miguel A. and McMinis, Jeremy and Kim, Jeongnim and Scuseria, Gustavo E.},
    title = {Computing the energy of a water molecule using multideterminants: A simple, efficient algorithm},
    journal = {The Journal of Chemical Physics},
    volume = {135},
    number = {24},
    pages = {244105},
    year = {2011},
    month = {12},
    issn = {0021-9606},
    doi = {10.1063/1.3665391},
    url = {https://doi.org/10.1063/1.3665391},}

@article{giner2013using,
author = {Giner, Emmanuel and Scemama, Anthony and Caffarel, Michel},
title = {Using perturbatively selected configuration interaction in quantum Monte Carlo calculations},
journal = {Canadian Journal of Chemistry},
volume = {91},
number = {9},
pages = {879-885},
year = {2013},
doi = {10.1139/cjc-2013-0017},
URL = {https://doi.org/10.1139/cjc-2013-0017}}

@article{mahajan2021taming,
    author = {Mahajan, Ankit and Sharma, Sandeep},
    title = {Taming the Sign Problem in Auxiliary-Field Quantum
Monte Carlo Using Accurate Wave Functions},
    journal = {Journal of Chemical Theory and Computation},
    volume = {17},
    number = {8},
    pages = {4786-4798},
    year = {2021},
    month = {08},
    issn = {1549-9618},
    doi = {10.1021/acs.jctc.1c00371},
    url = {https://doi.org/10.1021/acs.jctc.1c00371},
}

@article{roos1980complete,
title = {A complete active space SCF method (CASSCF) using a density matrix formulated super-CI approach},
journal = {Chemical Physics},
volume = {48},
number = {2},
pages = {157-173},
year = {1980},
issn = {0301-0104},
doi = {10.1016/0301-0104(80)80045-0},
url = {https://www.sciencedirect.com/science/article/pii/0301010480800450},
author = {Bj{\"o}rn O. Roos and Peter R. Taylor and Per E.M. Sigbahn}
}

@article{mahajan2020efficient,
    author = {Mahajan, Ankit and Sharma, Sandeep},
    title = {Efficient local energy evaluation for multi-Slater wave functions in orbital space quantum Monte Carlo},
    journal = {The Journal of Chemical Physics},
    volume = {153},
    number = {19},
    pages = {194108},
    year = {2020},
    month = {11},
    issn = {0021-9606},
    doi = {10.1063/5.0025055},
    url = {https://doi.org/10.1063/5.0025055}}

@misc{kjonstad2025systematic,
      title={Systematic improvement of trial states in phaseless auxiliary-field quantum Monte Carlo}, 
      author={Eirik F. Kj{\o}nstad and Yann Damour and Sandeep Sharma and Garnet Kin-Lic Chan},
      year={2025},
      eprint={2510.06486},
      archivePrefix={arXiv},
      primaryClass={physics.chem-ph}}

@article{jiang2025unbiasing,
  title = {Unbiasing fermionic auxiliary-field quantum Monte Carlo with matrix product state trial wavefunctions},
  author = {Jiang, Tong and O'Gorman, Bryan and Mahajan, Ankit and Lee, Joonho},
  journal = {Phys. Rev. Res.},
  volume = {7},
  issue = {1},
  pages = {013038},
  numpages = {24},
  year = {2025},
  month = {Jan},
  publisher = {American Physical Society},
  doi = {10.1103/PhysRevResearch.7.013038},
  url = {https://link.aps.org/doi/10.1103/PhysRevResearch.7.013038}
}

@article{huang2020predicting,
   title={Predicting many properties of a quantum system from very few measurements},
   volume={16},
   ISSN={1745-2481},
   url={http://dx.doi.org/10.1038/s41567-020-0932-7},
   DOI={10.1038/s41567-020-0932-7},
   number={10},
   journal={Nature Physics},
   publisher={Springer Science and Business Media LLC},
   author={Huang, Hsin-Yuan and Kueng, Richard and Preskill, John},
   year={2020},
   month=June, pages={1050--1057} }

@misc{mazzola2022exponential,
      title={Exponential challenges in unbiasing quantum Monte Carlo algorithms with quantum computers}, 
      author={Guglielmo Mazzola and Giuseppe Carleo},
      year={2022},
      eprint={2205.09203},
      archivePrefix={arXiv},
      primaryClass={quant-ph}}

@misc{lee2022response,
      title={Response to "Exponential challenges in unbiasing quantum Monte Carlo algorithms with quantum computers"}, 
      author={Joonho Lee and David R. Reichman and Ryan Babbush and Nicholas C. Rubin and Fionn D. Malone and Bryan O'Gorman and William J. Huggins},
      year={2022},
      eprint={2207.13776},
      archivePrefix={arXiv},
      primaryClass={quant-ph},}

@article{wan2023matchgate,
   title={Matchgate Shadows for Fermionic Quantum Simulation},
   volume={404},
   ISSN={1432-0916},
   url={http://dx.doi.org/10.1007/s00220-023-04844-0},
   DOI={10.1007/s00220-023-04844-0},
   number={2},
   journal={Communications in Mathematical Physics},
   publisher={Springer Science and Business Media LLC},
   author={Wan, Kianna and Huggins, William J. and Lee, Joonho and Babbush, Ryan},
   year={2023},
   month=Oct, pages={629--700} }

@article{huang2024evaluating,
  title = {Evaluating a quantum-classical quantum Monte Carlo algorithm with Matchgate shadows},
  author = {Huang, Benchen and Chen, Yi-Ting and Gupt, Brajesh and Suchara, Martin and Tran, Anh and McArdle, Sam and Galli, Giulia},
  journal = {Phys. Rev. Res.},
  volume = {6},
  issue = {4},
  pages = {043063},
  numpages = {25},
  year = {2024},
  month = {Oct},
  publisher = {American Physical Society},
  doi = {10.1103/PhysRevResearch.6.043063},
  url = {https://link.aps.org/doi/10.1103/PhysRevResearch.6.043063}
}

@article{kiser2024classical,
doi = {10.1088/1367-2630/ad2f67},
url = {https://doi.org/10.1088/1367-2630/ad2f67},
year = {2024},
month = {mar},
publisher = {IOP Publishing},
volume = {26},
number = {3},
pages = {033022},
author = {Kiser, Matthew and Schroeder, Anna and Anselmetti, Gian-Luca R and Kumar, Chandan and Moll, Nikolaj and Streif, Michael and Vodola, Davide},
title = {Classical and quantum cost of measurement strategies for quantum-enhanced auxiliary field quantum Monte Carlo},
journal = {New Journal of Physics}
}

@article{amsler2023classical,
    author = {Amsler, Maximilian and Deglmann, Peter and Degroote, Matthias and Kaicher, Michael P. and Kiser, Matthew and K{\"u}hn, Michael and Kumar, Chandan and Maier, Andreas and Samsonidze, Georgy and Schroeder, Anna and Streif, Michael and Vodola, Davide and Wever, Christopher and QUTAC Material Science Working Group},
    title = {Classical and quantum trial wave functions in auxiliary-field quantum Monte Carlo applied to oxygen allotropes and a CuBr2 model system},
    journal = {The Journal of Chemical Physics},
    volume = {159},
    number = {4},
    pages = {044119},
    year = {2023},
    month = {07},
    issn = {0021-9606},
    doi = {10.1063/5.0146934},
    url = {https://doi.org/10.1063/5.0146934},
}

@article{khinevich2025enhancing,
  title = {Enhancing quantum computations with the synergy of auxiliary field quantum Monte Carlo and computational basis tomography},
  author = {Khinevich, Viktor and Mizukami, Wataru},
  journal = {Phys. Rev. Res.},
  volume = {7},
  issue = {4},
  pages = {043344},
  numpages = {17},
  year = {2025},
  month = {Dec},
  publisher = {American Physical Society},
  doi = {10.1103/rbhc-dkh1},
  url = {https://link.aps.org/doi/10.1103/rbhc-dkh1}
}

@article{goings2025molecular,
    author = {Goings, Joshua J. and Shin, Kyujin and Noh, Seunghyo and Kyoung, Woomin and Kim, Donghwi and Baek, Jihye and Roetteler, Martin and Epifanovsky, Evgeny and Zhao, Luning},
    title = {Molecular Properties in Quantum-Classical Auxiliary-Field
Quantum Monte Carlo: Correlated Sampling with Application to Accurate
Nuclear Forces},
    journal = {Journal of Chemical Theory and Computation},
    volume = {21},
    number = {24},
    pages = {12573-12591},
    year = {2025},
    month = {12},
    issn = {1549-9618},
    doi = {10.1021/acs.jctc.5c01351},
    url = {https://doi.org/10.1021/acs.jctc.5c01351},
}

@article{blunt2025quantum,
   title={Quantum Computing Approach to Fixed-Node Monte Carlo Using Classical Shadows},
   volume={21},
   ISSN={1549-9626},
   url={http://dx.doi.org/10.1021/acs.jctc.4c01468},
   DOI={10.1021/acs.jctc.4c01468},
   number={4},
   journal={Journal of Chemical Theory and Computation},
   publisher={American Chemical Society (ACS)},
   author={Blunt, Nick S. and Caune, Laura and Quiroz-Fernandez, Javiera},
   year={2025},
   month=Feb, pages={1652--1666} }

@article{kiser2025contextual,
    author = {Kiser, Matthew and Beuerle, Matthias and {\v{S}}imkovic, Fedor, IV},
    title = {Contextual Subspace
Auxiliary-Field Quantum Monte
Carlo: Improved Bias with Reduced Quantum Resources},
    journal = {Journal of Chemical Theory and Computation},
    volume = {21},
    number = {5},
    pages = {2256-2271},
    year = {2025},
    month = {03},
    issn = {1549-9618},
    doi = {10.1021/acs.jctc.4c01280},
    url = {https://doi.org/10.1021/acs.jctc.4c01280},
}

@misc{yoshida2025auxiliary,
      title={Auxiliary-field quantum Monte Carlo method with quantum selected configuration interaction}, 
      author={Yuichiro Yoshida and Luca Erhart and Takuma Murokoshi and Rika Nakagawa and Chihiro Mori and Takafumi Miyanaga and Toshio Mori and Wataru Mizukami},
      year={2025},
      eprint={2502.21081},
      archivePrefix={arXiv},
      primaryClass={quant-ph},}

@article{zhao2023orbital,
  title={Orbital-optimized pair-correlated electron simulations on trapped-ion quantum computers},
  author={Zhao, Luning and Goings, Joshua and Shin, Kyujin and Kyoung, Woomin and Fuks, Johanna I. and Rhee, June-Koo Kevin and Rhee, Young Min and Wright, Kenneth and Nguyen, Jason and Kim, Jungsang and Johri, Sonika},
  journal={npj Quantum Information},
  volume={9},
  number={1},
  pages={60},
  year={2023},
  month={Jun},
  publisher={Springer Nature},
  doi={10.1038/s41534-023-00730-8},
  url={https://doi.org/10.1038/s41534-023-00730-8}
}

@book{pavarini2013emergent,
  title = {Emergent Phenomena in Correlated Matter},
  editor = {Pavarini, Eva and Koch, Erik and Schollw{\"o}ck, Ulrich},
  series = {Modeling and Simulation, Vol. 3},
  publisher = {Forschungszentrum J{\"u}lich},
  year = {2013}
}

@article{barkoutsos2018quantum,
  title = {Quantum algorithms for electronic structure calculations: Particle-hole Hamiltonian and optimized wave-function expansions},
  author = {Barkoutsos, Panagiotis Kl. and Gonthier, Jerome F. and Sokolov, Igor and Moll, Nikolaj and Salis, Gian and Fuhrer, Andreas and Ganzhorn, Marc and Egger, Daniel J. and Troyer, Matthias and Mezzacapo, Antonio and Filipp, Stefan and Tavernelli, Ivano},
  journal = {Phys. Rev. A},
  volume = {98},
  issue = {2},
  pages = {022322},
  numpages = {13},
  year = {2018},
  month = {Aug},
  publisher = {American Physical Society},
  doi = {10.1103/PhysRevA.98.022322},
  url = {https://link.aps.org/doi/10.1103/PhysRevA.98.022322}
}

@article{chen2021quantum,
    author = {Chen, Jia and Cheng, Hai-Ping and Freericks, James K.},
    title = {Quantum-Inspired Algorithm for the Factorized Form
of Unitary Coupled Cluster Theory},
    journal = {Journal of Chemical Theory and Computation},
    volume = {17},
    number = {2},
    pages = {841-847},
    year = {2021},
    month = {02},
    issn = {1549-9618},
    doi = {10.1021/acs.jctc.0c01052},
    url = {https://doi.org/10.1021/acs.jctc.0c01052},
}

@article{jordan1928paulische,
  title={{\"U}ber das {P}aulische {\"A}quivalenzverbot},
  author={Jordan, Pascual and Wigner, Eugene},
  journal={Zeitschrift f{\"u}r Physik},
  volume={47},
  number={9},
  pages={631--651},
  year={1928},
  publisher={Springer}
}

@article{bravyi2002fermionic,
   title={Fermionic Quantum Computation},
   volume={298},
   ISSN={0003-4916},
   url={http://dx.doi.org/10.1006/aphy.2002.6254},
   DOI={10.1006/aphy.2002.6254},
   number={1},
   journal={Annals of Physics},
   publisher={Elsevier BV},
   author={Bravyi, Sergey B. and Kitaev, Alexei Yu.},
   year={2002},
   month=May, pages={210--226} }

@misc{farhi2000quantum,
      title={Quantum Computation by Adiabatic Evolution}, 
      author={Edward Farhi and Jeffrey Goldstone and Sam Gutmann and Michael Sipser},
      year={2000},
      eprint={quant-ph/0001106},
      archivePrefix={arXiv},
      primaryClass={quant-ph},
}

@article{aspuru2005simulated,
   title={Simulated Quantum Computation of Molecular Energies},
   volume={309},
   ISSN={1095-9203},
   url={http://dx.doi.org/10.1126/science.1113479},
   DOI={10.1126/science.1113479},
   number={5741},
   journal={Science},
   publisher={American Association for the Advancement of Science (AAAS)},
   author={Aspuru-Guzik, Al{\'a}n and Dutoi, Anthony D. and Love, Peter J. and Head-Gordon, Martin},
   year={2005},
   month=Sept, pages={1704--1707} }

@article{abrams1999quantum,
   title={Quantum Algorithm Providing Exponential Speed Increase for Finding Eigenvalues and Eigenvectors},
   volume={83},
   ISSN={1079-7114},
   url={http://dx.doi.org/10.1103/PhysRevLett.83.5162},
   DOI={10.1103/physrevlett.83.5162},
   number={24},
   journal={Physical Review Letters},
   publisher={American Physical Society (APS)},
   author={Abrams, Daniel S. and Lloyd, Seth},
   year={1999},
   month=Dec, pages={5162--5165} }

@article{mcclean2018barren,
  title={Barren plateaus in quantum neural network training landscapes},
  author={McClean, Jarrod R. and Boixo, Sergio and Smelyanskiy, Vadim N. and Babbush, Ryan and Neven, Hartmut},
  journal={Nature Communications},
  volume={9},
  number={1},
  pages={4812},
  year={2018},
  month={Nov},
  publisher={Springer Nature},
  doi={10.1038/s41467-018-07090-4},
  url={https://doi.org/10.1038/s41467-018-07090-4}
}

@article{cerezo2021variational,
   title={Variational quantum algorithms},
   volume={3},
   ISSN={2522-5820},
   url={http://dx.doi.org/10.1038/s42254-021-00348-9},
   DOI={10.1038/s42254-021-00348-9},
   number={9},
   journal={Nature Reviews Physics},
   publisher={Springer Science and Business Media LLC},
   author={Cerezo, M. and Arrasmith, Andrew and Babbush, Ryan and Benjamin, Simon C. and Endo, Suguru and Fujii, Keisuke and McClean, Jarrod R. and Mitarai, Kosuke and Yuan, Xiao and Cincio, Lukasz and Coles, Patrick J.},
   year={2021},
   month=Aug, pages={625--644} }

@article{peruzzo2014variational,
  title={A variational eigenvalue solver on a photonic quantum processor},
  author={Peruzzo, Alberto and McClean, Jarrod and Shadbolt, Peter and Yung, Man-Hong and Zhou, Xiao-Qi and Love, Peter J. and Aspuru-Guzik, Al{\'a}n and O'Brien, Jeremy L.},
  journal={Nature Communications},
  volume={5},
  number={1},
  pages={4213},
  year={2014},
  month={Jul},
  publisher={Springer Nature},
  doi={10.1038/ncomms5213},
  url={https://doi.org/10.1038/ncomms5213}
}

@article{romero2018strategies,
doi = {10.1088/2058-9565/aad3e4},
url = {https://doi.org/10.1088/2058-9565/aad3e4},
year = {2018},
month = {oct},
publisher = {IOP Publishing},
volume = {4},
number = {1},
pages = {014008},
author = {Romero, Jonathan and Babbush, Ryan and McClean, Jarrod R and Hempel, Cornelius and Love, Peter J and Aspuru-Guzik, Al{\'a}n},
title = {Strategies for quantum computing molecular energies using the unitary coupled cluster ansatz},
journal = {Quantum Science and Technology}
}

@article{moller1934note,
  title = {Note on an Approximation Treatment for Many-Electron Systems},
  author = {M{\o{}}ller, Chr. and Plesset, M. S.},
  journal = {Phys. Rev.},
  volume = {46},
  issue = {7},
  pages = {618--622},
  numpages = {0},
  year = {1934},
  month = {Oct},
  publisher = {American Physical Society},
  doi = {10.1103/PhysRev.46.618},
  url = {https://link.aps.org/doi/10.1103/PhysRev.46.618}
}

@article{hirsbrunner2024beyond,
   title={Beyond MP2 initialization for unitary coupled cluster quantum circuits},
   volume={8},
   ISSN={2521-327X},
   url={http://dx.doi.org/10.22331/q-2024-11-26-1538},
   DOI={10.22331/q-2024-11-26-1538},
   journal={Quantum},
   publisher={Verein zur Forderung des Open Access Publizierens in den Quantenwissenschaften},
   author={Hirsbrunner, Mark R. and Chamaki, Diana and Mullinax, J. Wayne and Tubman, Norm M.},
   year={2024},
   month=Nov, pages={1538} }

@article{park2024hamiltonian,
   title={Hamiltonian variational ansatz without barren plateaus},
   volume={8},
   ISSN={2521-327X},
   url={http://dx.doi.org/10.22331/q-2024-02-01-1239},
   DOI={10.22331/q-2024-02-01-1239},
   journal={Quantum},
   publisher={Verein zur Forderung des Open Access Publizierens in den Quantenwissenschaften},
   author={Park, Chae-Yeun and Killoran, Nathan},
   year={2024},
   month=Feb, pages={1239} }

@article{matsuzawa2020jastrow,
   title={Jastrow-type Decomposition in Quantum Chemistry for Low-Depth Quantum Circuits},
   volume={16},
   ISSN={1549-9626},
   url={http://dx.doi.org/10.1021/acs.jctc.9b00963},
   DOI={10.1021/acs.jctc.9b00963},
   number={2},
   journal={Journal of Chemical Theory and Computation},
   publisher={American Chemical Society (ACS)},
   author={Matsuzawa, Yuta and Kurashige, Yuki},
   year={2020},
   month=Jan, pages={944--952} }

@article{cao2021towards,
  title = {Progress toward larger molecular simulation on a quantum computer: Simulating a system with up to 28 qubits accelerated by point-group symmetry},
  author = {Cao, Changsu and Hu, Jiaqi and Zhang, Wengang and Xu, Xusheng and Chen, Dechin and Yu, Fan and Li, Jun and Hu, Han-Shi and Lv, Dingshun and Yung, Man-Hong},
  journal = {Phys. Rev. A},
  volume = {105},
  issue = {6},
  pages = {062452},
  numpages = {12},
  year = {2022},
  month = {Jun},
  publisher = {American Physical Society},
  doi = {10.1103/PhysRevA.105.062452},
  url = {https://link.aps.org/doi/10.1103/PhysRevA.105.062452}
}

@article{lee2018generalized,
   title={Generalized Unitary Coupled Cluster Wave functions for Quantum Computation},
   volume={15},
   ISSN={1549-9626},
   url={http://dx.doi.org/10.1021/acs.jctc.8b01004},
   DOI={10.1021/acs.jctc.8b01004},
   number={1},
   journal={Journal of Chemical Theory and Computation},
   publisher={American Chemical Society (ACS)},
   author={Lee, Joonho and Huggins, William J. and Head-Gordon, Martin and Whaley, K. Birgitta},
   year={2018},
   month=Nov, pages={311--324} }

@article{motta2023bridging,
    author = {Motta, Mario and Sung, Kevin J. and Whaley, K. Birgitta and Head-Gordon, Martin and Shee, James},
    title = {Bridging physical intuition and hardware efficiency for correlated electronic states: the local unitary cluster Jastrow ansatz for electronic structure},
    journal = {Chemical Science},
    volume = {14},
    number = {40},
    pages = {11213-11227},
    year = {2023},
    month = {10},
    issn = {2041-6520},
    doi = {10.1039/d3sc02516k},
    url = {https://doi.org/10.1039/d3sc02516k},
}

@article{motta2024quantum,
    author = {Motta, Mario and Sung, Kevin J. and Shee, James},
    title = {Quantum Algorithms
for the Variational Optimization
of Correlated Electronic States with Stochastic Reconfiguration and
the Linear Method},
    journal = {The Journal of Physical Chemistry A},
    volume = {128},
    number = {40},
    pages = {8762-8776},
    year = {2024},
    month = {10},
    issn = {1089-5639},
    doi = {10.1021/acs.jpca.4c02847},
    url = {https://doi.org/10.1021/acs.jpca.4c02847}
}

@article{grimsley2019adaptive,
  title={An adaptive variational algorithm for exact molecular simulations on a quantum computer},
  author={Grimsley, Harper R. and Economou, Sophia E. and Barnes, Edwin and Mayhall, Nicholas J.},
  journal={Nature Communications},
  volume={10},
  number={1},
  pages={3007},
  year={2019},
  month={Jul},
  publisher={Springer Nature},
  doi={10.1038/s41467-019-10988-2},
  url={https://doi.org/10.1038/s41467-019-10988-2}
}

@article{sun2020recent,
    author = {Sun, Qiming and Zhang, Xing and Banerjee, Samragni and Bao, Peng and Barbry, Marc and Blunt, Nick S. and Bogdanov, Nikolay A. and Booth, George H. and Chen, Jia and Cui, Zhi-Hao and Eriksen, Janus J. and Gao, Yang and Guo, Sheng and Hermann, Jan and Hermes, Matthew R. and Koh, Kevin and Koval, Peter and Lehtola, Susi and Li, Zhendong and Liu, Junzi and Mardirossian, Narbe and McClain, James D. and Motta, Mario and Mussard, Bastien and Pham, Hung Q. and Pulkin, Artem and Purwanto, Wirawan and Robinson, Paul J. and Ronca, Enrico and Sayfutyarova, Elvira R. and Scheurer, Maximilian and Schurkus, Henry F. and Smith, James E. T. and Sun, Chong and Sun, Shi-Ning and Upadhyay, Shiv and Wagner, Lucas K. and Wang, Xiao and White, Alec and Whitfield, James Daniel and Williamson, Mark J. and Wouters, Sebastian and Yang, Jun and Yu, Jason M. and Zhu, Tianyu and Berkelbach, Timothy C. and Sharma, Sandeep and Sokolov, Alexander Yu. and Chan, Garnet Kin-Lic},
    title = {Recent developments in the PySCF program package},
    journal = {The Journal of Chemical Physics},
    volume = {153},
    number = {2},
    pages = {024109},
    year = {2020},
    month = {07},
    issn = {0021-9606},
    doi = {10.1063/5.0006074},
    url = {https://doi.org/10.1063/5.0006074},
}

@article{mcclean2020openfermion,
doi = {10.1088/2058-9565/ab8ebc},
url = {https://doi.org/10.1088/2058-9565/ab8ebc},
year = {2020},
month = {jun},
publisher = {IOP Publishing},
volume = {5},
number = {3},
pages = {034014},
author = {McClean, Jarrod R and Rubin, Nicholas C and Sung, Kevin J and Kivlichan, Ian D and Bonet-Monroig, Xavier and Cao, Yudong and Dai, Chengyu and Fried, E Schuyler and Gidney, Craig and Gimby, Brendan and Gokhale, Pranav and H{\"a}ner, Thomas and Hardikar, Tarini and Havl{\'i}{\v{c}}ek, Vojt{\v{e}}ch and Higgott, Oscar and Huang, Cupjin and Izaac, Josh and Jiang, Zhang and Liu, Xinle and McArdle, Sam and Neeley, Matthew and O'Brien, Thomas and O'Gorman, Bryan and Ozfidan, Isil and Radin, Maxwell D and Romero, Jhonathan and Sawaya, Nicolas P D and Senjean, Bruno and Setia, Kanav and Sim, Sukin and Steiger, Damian S and Steudtner, Mark and Sun, Qiming and Sun, Wei and Wang, Daochen and Zhang, Fang and Babbush, Ryan},
title = {OpenFermion: the electronic structure package for quantum computers},
journal = {Quantum Science and Technology}
}

@misc{cudaq,
  author = {{The CUDA-Q Development Team.}},
  title = {CUDA-Q},
  howpublished = {\url{https://github.com/NVIDIA/cuda-quantum} (accessed 2026-09-30)},
}

@misc{ffsim,
      title={ffsim: Faster simulation of fermionic quantum circuits}, 
      author={Kevin J. Sung and Inho Choi and Mirko Amico and Bartholomew Andrews and Esra Ayantuna and Yukio Kawashima and Wan-Hsuan Lin and David Omanovic and Samuele Piccinelli and Javier Robledo Moreno and Abdullah Ash Saki and James Shee and Soyoung Shin and Minh C. Tran and Kento Ueda and Haimeng Zhang and Mario Motta},
      year={2026},
      eprint={2605.03123},
      archivePrefix={arXiv},
      primaryClass={quant-ph},
}

@article{2020SciPy-NMeth,
   title={SciPy 1.0: fundamental algorithms for scientific computing in Python},
   volume={17},
   ISSN={1548-7105},
   url={http://dx.doi.org/10.1038/s41592-019-0686-2},
   DOI={10.1038/s41592-019-0686-2},
   number={3},
   journal={Nature Methods},
   publisher={Springer Science and Business Media LLC},
   author={Virtanen, Pauli and Gommers, Ralf and Oliphant, Travis E. and Haberland, Matt and Reddy, Tyler and Cournapeau, David and Burovski, Evgeni and Peterson, Pearu and Weckesser, Warren and Bright, Jonathan and van der Walt, St{\'e}fan J. and Brett, Matthew and Wilson, Joshua and Millman, K. Jarrod and Mayorov, Nikolay and Nelson, Andrew R. J. and Jones, Eric and Kern, Robert and Larson, Eric and Carey, C J and Polat, {\.I}lhan and Feng, Yu and Moore, Eric W. and VanderPlas, Jake and Laxalde, Denis and Perktold, Josef and Cimrman, Robert and Henriksen, Ian and Quintero, E. A. and Harris, Charles R. and Archibald, Anne M. and Ribeiro, Ant{\^o}nio H. and Pedregosa, Fabian and van Mulbregt, Paul and Vijaykumar, Aditya and Bardelli, Alessandro Pietro and Rothberg, Alex and Hilboll, Andreas and Kloeckner, Andreas and Scopatz, Anthony and Lee, Antony and Rokem, Ariel and Woods, C. Nathan and Fulton, Chad and Masson, Charles and H{\"a}ggstr{\"o}m, Christian and Fitzgerald, Clark and Nicholson, David A. and Hagen, David R. and Pasechnik, Dmitrii V. and Olivetti, Emanuele and Martin, Eric and Wieser, Eric and Silva, Fabrice and Lenders, Felix and Wilhelm, Florian and Young, G. and Price, Gavin A. and Ingold, Gert-Ludwig and Allen, Gregory E. and Lee, Gregory R. and Audren, Herv{\'e} and Probst, Irvin and Dietrich, J{\"o}rg P. and Silterra, Jacob and Webber, James T and Slavi{\v{c}}, Janko and Nothman, Joel and Buchner, Johannes and Kulick, Johannes and Sch{\"o}nberger, Johannes L. and de Miranda Cardoso, Jos{\'e} Vin{\'i}cius and Reimer, Joscha and Harrington, Joseph and Rodr{\'i}guez, Juan Luis Cano and Nunez-Iglesias, Juan and Kuczynski, Justin and Tritz, Kevin and Thoma, Martin and Newville, Matthew and K{\"u}mmerer, Matthias and Bolingbroke, Maximilian and Tartre, Michael and Pak, Mikhail and Smith, Nathaniel J. and Nowaczyk, Nikolai and Shebanov, Nikolay and Pavlyk, Oleksandr and Brodtkorb, Per A. and Lee, Perry and McGibbon, Robert T. and Feldbauer, Roman and Lewis, Sam and Tygier, Sam and Sievert, Scott and Vigna, Sebastiano and Peterson, Stefan and More, Surhud and Pudlik, Tadeusz and Oshima, Takuya and Pingel, Thomas J. and Robitaille, Thomas P. and Spura, Thomas and Jones, Thouis R. and Cera, Tim and Leslie, Tim and Zito, Tiziano and Krauss, Tom and Upadhyay, Utkarsh and Halchenko, Yaroslav O. and V{\'a}zquez-Baeza, Yoshiki},
   year={2020},
   month=Feb, pages={261--272} }

@article{malone2022ipie,
    author = {Malone, Fionn
D. and Mahajan, Ankit and Spencer, James S. and Lee, Joonho},
    title = {ipie: A Python-Based Auxiliary-Field
Quantum Monte Carlo Program with Flexibility and Efficiency on CPUs
and GPUs},
    journal = {Journal of Chemical Theory and Computation},
    volume = {19},
    number = {1},
    pages = {109-121},
    year = {2023},
    month = {01},
    issn = {1549-9618},
    doi = {10.1021/acs.jctc.2c00934},
    url = {https://doi.org/10.1021/acs.jctc.2c00934},
}

@article{motta2017towards,
  title = {Towards the Solution of the Many-Electron Problem in Real Materials: Equation of State of the Hydrogen Chain with State-of-the-Art Many-Body Methods},
  author = {Motta, Mario and Ceperley, David M. and Chan, Garnet Kin-Lic and Gomez, John A. and Gull, Emanuel and Guo, Sheng and Jim{\'e}nez-Hoyos, Carlos A. and Lan, Tran Nguyen and Li, Jia and Ma, Fengjie and Millis, Andrew J. and Prokof'ev, Nikolay V. and Ray, Ushnish and Scuseria, Gustavo E. and Sorella, Sandro and Stoudenmire, Edwin M. and Sun, Qiming and Tupitsyn, Igor S. and White, Steven R. and Zgid, Dominika and Zhang, Shiwei},
  collaboration = {Simons Collaboration on the Many-Electron Problem},
  journal = {Phys. Rev. X},
  volume = {7},
  issue = {3},
  pages = {031059},
  numpages = {28},
  year = {2017},
  month = {Sep},
  publisher = {American Physical Society},
  doi = {10.1103/PhysRevX.7.031059},
  url = {https://link.aps.org/doi/10.1103/PhysRevX.7.031059}
}

@article{PhysRevA.92.062318,
  title = {Solving strongly correlated electron models on a quantum computer},
  author = {Wecker, Dave and Hastings, Matthew B. and Wiebe, Nathan and Clark, Bryan K. and Nayak, Chetan and Troyer, Matthias},
  journal = {Phys. Rev. A},
  volume = {92},
  issue = {6},
  pages = {062318},
  numpages = {24},
  year = {2015},
  month = {Dec},
  publisher = {American Physical Society},
  doi = {10.1103/PhysRevA.92.062318},
  url = {https://link.aps.org/doi/10.1103/PhysRevA.92.062318}
}

@article{jiang2018quantum,
  title = {Quantum Algorithms to Simulate Many-Body Physics of Correlated Fermions},
  author = {Jiang, Zhang and Sung, Kevin J. and Kechedzhi, Kostyantyn and Smelyanskiy, Vadim N. and Boixo, Sergio},
  journal = {Phys. Rev. Appl.},
  volume = {9},
  issue = {4},
  pages = {044036},
  numpages = {23},
  year = {2018},
  month = {Apr},
  publisher = {American Physical Society},
  doi = {10.1103/PhysRevApplied.9.044036},
  url = {https://link.aps.org/doi/10.1103/PhysRevApplied.9.044036}
}

@misc{qiskit2024,
      title={Quantum computing with Qiskit}, 
      author={Ali Javadi-Abhari and Matthew Treinish and Kevin Krsulich and Christopher J. Wood and Jake Lishman and Julien Gacon and Simon Martiel and Paul D. Nation and Lev S. Bishop and Andrew W. Cross and Blake R. Johnson and Jay M. Gambetta},
      year={2024},
      eprint={2405.08810},
      archivePrefix={arXiv},
      primaryClass={quant-ph},
}

@article{purwanto2008eliminating,
    author = {Purwanto, Wirawan and Al-Saidi, W. A. and Krakauer, Henry and Zhang, Shiwei},
    title = {Eliminating spin contamination in auxiliary-field quantum Monte Carlo: Realistic potential energy curve of F2},
    journal = {The Journal of Chemical Physics},
    volume = {128},
    number = {11},
    pages = {114309},
    year = {2008},
    month = {03},
    issn = {0021-9606},
    doi = {10.1063/1.2838983},
    url = {https://doi.org/10.1063/1.2838983},
}

@article{lee2019auxiliary,
    author = {Lee, Joonho and Malone, Fionn D. and Morales, Miguel A.},
    title = {An auxiliary-Field quantum Monte Carlo perspective on the ground state of the dense uniform electron gas: An investigation with Hartree-Fock trial wavefunctions},
    journal = {The Journal of Chemical Physics},
    volume = {151},
    number = {6},
    pages = {064122},
    year = {2019},
    month = {08},
    issn = {0021-9606},
    doi = {10.1063/1.5109572},
    url = {https://doi.org/10.1063/1.5109572},
}

@article{al2007bond,
    author = {Al-Saidi, W. A. and Zhang, Shiwei and Krakauer, Henry},
    title = {Bond breaking with auxiliary-field quantum Monte Carlo},
    journal = {The Journal of Chemical Physics},
    volume = {127},
    number = {14},
    pages = {144101},
    year = {2007},
    month = {10},
    issn = {0021-9606},
    doi = {10.1063/1.2770707},
    url = {https://doi.org/10.1063/1.2770707},
}

@article{carlson1999issues,
   title={Issues and observations on applications of the constrained-path Monte Carlo method to many-fermion systems},
   volume={59},
   ISSN={1095-3795},
   url={http://dx.doi.org/10.1103/PhysRevB.59.12788},
   DOI={10.1103/physrevb.59.12788},
   number={20},
   journal={Physical Review B},
   publisher={American Physical Society (APS)},
   author={Carlson, J. and Gubernatis, J. E. and Ortiz, G. and Zhang, Shiwei},
   year={1999},
   month=May, pages={12788--12798} }

@article{mahajan2025beyond,
    author = {Mahajan, Ankit and Thorpe, James H. and Kurian, Jo S. and Reichman, David R. and Matthews, Devin A. and Sharma, Sandeep},
    title = {Beyond CCSD(T)
Accuracy at Lower Scaling with Auxiliary
Field Quantum Monte Carlo},
    journal = {Journal of Chemical Theory and Computation},
    volume = {21},
    number = {4},
    pages = {1626-1642},
    year = {2025},
    month = {02},
    issn = {1549-9618},
    doi = {10.1021/acs.jctc.4c01314},
    url = {https://doi.org/10.1021/acs.jctc.4c01314},
}

@article{xiao2025implementing,
    author = {Xiao, Zhi-Yu and Xiang, Tao and Lu, Zixiang and Chen, Yixiao and Zhang, Shiwei},
    title = {Implementing advanced trial wave functions in fermion quantum Monte Carlo via stochastic sampling},
    journal = {The Journal of Chemical Physics},
    volume = {163},
    number = {16},
    pages = {164109},
    year = {2025},
    month = {10},
    issn = {0021-9606},
    doi = {10.1063/5.0296390},
    url = {https://doi.org/10.1063/5.0296390}}

@article{hachmann2006multireference,
    author = {Hachmann, Johannes and Cardoen, Wim and Chan, Garnet Kin-Lic},
    title = {Multireference correlation in long molecules with the quadratic scaling density matrix renormalization group},
    journal = {The Journal of Chemical Physics},
    volume = {125},
    number = {14},
    pages = {144101},
    year = {2006},
    month = {10},
    issn = {0021-9606},
    doi = {10.1063/1.2345196},
    url = {https://doi.org/10.1063/1.2345196},
}

@article{huggins2021efficient,
  title={Efficient and noise resilient measurements for quantum chemistry on near-term quantum computers},
  author={Huggins, William J. and McClean, Jarrod R. and Rubin, Nicholas C. and Jiang, Zhang and Wiebe, Nathan and Whaley, K. Birgitta and Babbush, Ryan},
  journal={npj Quantum Information},
  volume={7},
  number={1},
  pages={23},
  year={2021},
  month={Feb},
  publisher={Springer Nature},
  doi={10.1038/s41534-020-00341-7},
  url={https://doi.org/10.1038/s41534-020-00341-7}
}

@misc{lin2025improved,
      title={Improved parameter initialization for the (local) unitary cluster Jastrow ansatz}, 
      author={Wan-Hsuan Lin and Fangchun Liang and Mario Motta and Haimeng Zhang and Kenneth M. Merz Jr. and Kevin J. Sung},
      year={2025},
      eprint={2511.22476},
      archivePrefix={arXiv},
      primaryClass={quant-ph},
}

@article{zhang2026size,
    author = {Zhang, Yichi and Mahajan, Ankit and Damour, Yann and Sharma, Sandeep},
    title = {Size-extensive auxiliary-field quantum Monte Carlo with perturbative coupled cluster trial wavefunctions},
    journal = {The Journal of Chemical Physics},
    volume = {165},
    number = {9},
    pages = {094107},
    year = {2026},
    month = {09},
    issn = {0021-9606},
    doi = {10.1063/5.0341498},
    url = {https://doi.org/10.1063/5.0341498},
}

@article{danilov2025capturing,
    author = {Danilov, Don and Ganoe, Brad and Munyi, Mark and Shee, James},
    title = {Capturing Strong
Correlation in Molecules with Phaseless
Auxiliary-Field Quantum Monte Carlo Using Generalized Hartree–Fock
Trial Wave Functions},
    journal = {Journal of Chemical Theory and Computation},
    volume = {21},
    number = {3},
    pages = {1136-1152},
    year = {2025},
    month = {02},
    issn = {1549-9618},
    doi = {10.1021/acs.jctc.4c01251},
    url = {https://doi.org/10.1021/acs.jctc.4c01251},
}

@article{danilov2025enhancing,
    author = {Danilov, Don and Robledo-Moreno, Javier and Sung, Kevin J. and Motta, Mario and Shee, James},
    title = {Enhancing the Accuracy
and Efficiency of Sample-Based
Quantum Diagonalization with Phaseless Auxiliary-Field Quantum Monte
Carlo},
    journal = {Journal of Chemical Theory and Computation},
    volume = {21},
    number = {22},
    pages = {11585-11594},
    year = {2025},
    month = {11},
    issn = {1549-9618},
    doi = {10.1021/acs.jctc.5c01407},
    url = {https://doi.org/10.1021/acs.jctc.5c01407},
}

@article{danilov2026selecting,
    author = {Danilov, Don and Ganoe, Brad and Otis, Leon and Gong, Zhi and Lu, Zixiang and Shee, James},
    title = {Selecting Optimal
Unrestricted Hartree–Fock
Trial Wave Functions for Phaseless Auxiliary-Field Quantum Monte Carlo:
Accuracy and Limitations in Modeling Three Iron–Sulfur Clusters},
    journal = {Journal of Chemical Theory and Computation},
    volume = {22},
    number = {16},
    pages = {8274-8287},
    year = {2026},
    month = {08},
    issn = {1549-9618},
    doi = {10.1021/acs.jctc.6c00899},
    url = {https://doi.org/10.1021/acs.jctc.6c00899},
}

@misc{huang2024gpu,
      title={GPU-accelerated Auxiliary-field quantum Monte Carlo with multi-Slater determinant trial states}, 
      author={Yifei Huang and Zhen Guo and Hung Q. Pham and Dingshun Lv},
      year={2024},
      eprint={2406.08314},
      archivePrefix={arXiv},
      primaryClass={physics.chem-ph},
}

@article{mullinax2025classical,
   title={Classical Preoptimization Approach for ADAPT-VQE: Maximizing the Potential of High-Performance Computing Resources to Improve Quantum Simulation of Chemical Applications},
   volume={21},
   ISSN={1549-9626},
   url={http://dx.doi.org/10.1021/acs.jctc.5c00150},
   DOI={10.1021/acs.jctc.5c00150},
   number={8},
   journal={Journal of Chemical Theory and Computation},
   publisher={American Chemical Society (ACS)},
   author={Mullinax, J. Wayne and Anastasiou, Panagiotis G. and Larson, Jeffrey and Economou, Sophia E. and Tubman, Norm M.},
   year={2025},
   month=Apr, pages={4006--4015} }

@article{kendall1938new,
    author = {Kend, M. G.},
    title = {A NEW MEASURE OF RANK CORRELATION},
    journal = {Biometrika},
    volume = {30},
    number = {1-2},
    pages = {81-93},
    year = {1938},
    month = {06},
    issn = {0006-3444},
    doi = {10.1093/biomet/30.1-2.81},
    url = {https://doi.org/10.1093/biomet/30.1-2.81},
}

@article{stanisic2022observing,
  title={Observing ground-state properties of the Fermi-Hubbard model using a scalable algorithm on a quantum computer},
  author={Stanisic, Stasja and Bosse, Jan Lukas and Gambetta, Filippo Maria and Santos, Raul A. and Mruczkiewicz, Wojciech and O'Brien, Thomas E. and Ostby, Eric and Montanaro, Ashley},
  journal={Nature Communications},
  volume={13},
  number={1},
  pages={5743},
  year={2022},
  month={Oct},
  publisher={Springer Nature},
  doi={10.1038/s41467-022-33335-4},
  url={https://doi.org/10.1038/s41467-022-33335-4}
}

@article{hempel2018quantum,
  title = {Quantum Chemistry Calculations on a Trapped-Ion Quantum Simulator},
  author = {Hempel, Cornelius and Maier, Christine and Romero, Jonathan and McClean, Jarrod and Monz, Thomas and Shen, Heng and Jurcevic, Petar and Lanyon, Ben P. and Love, Peter and Babbush, Ryan and Aspuru-Guzik, Al{\'a}n and Blatt, Rainer and Roos, Christian F.},
  journal = {Phys. Rev. X},
  volume = {8},
  issue = {3},
  pages = {031022},
  numpages = {22},
  year = {2018},
  month = {Jul},
  publisher = {American Physical Society},
  doi = {10.1103/PhysRevX.8.031022},
  url = {https://link.aps.org/doi/10.1103/PhysRevX.8.031022}
}

@article{nam2020ground,
  title={Ground-state energy estimation of the water molecule on a trapped-ion quantum computer},
  author={Nam, Yunseong and Chen, Jwo-Sy and Pisenti, Neal C. and Wright, Kenneth and Delaney, Conor and Maslov, Dmitri and Brown, Kenneth R. and Allen, Stewart and Amini, Jason M. and Apisdorf, Joel and Beck, Kristin M. and Blinov, Aleksey and Chaplin, Vandiver and Chmielewski, Mika and Collins, Coleman and Debnath, Shantanu and Hudek, Kai M. and Ducore, Andrew M. and Keesan, Matthew and Kreikemeier, Sarah M. and Mizrahi, Jonathan and Solomon, Phil and Williams, Mike and Wong-Campos, Jaime David and Moehring, David and Monroe, Christopher and Kim, Jungsang},
  journal={npj Quantum Information},
  volume={6},
  number={1},
  pages={33},
  year={2020},
  publisher={Nature Publishing Group},
  doi={10.1038/s41534-020-0259-3},
  url={https://doi.org/10.1038/s41534-020-0259-3}
}

@article{robledo2025chemistry,
author = {Javier Robledo-Moreno  and Mario Motta  and Holger Haas  and Ali Javadi-Abhari  and Petar Jurcevic  and William Kirby  and Simon Martiel  and Kunal Sharma  and Sandeep Sharma  and Tomonori Shirakawa  and Iskandar Sitdikov  and Rong-Yang Sun  and Kevin J. Sung  and Maika Takita  and Minh C. Tran  and Seiji Yunoki  and Antonio Mezzacapo },
title = {Chemistry beyond the scale of exact diagonalization on a quantum-centric supercomputer},
journal = {Science Advances},
volume = {11},
number = {25},
pages = {eadu9991},
year = {2025},
doi = {10.1126/sciadv.adu9991},
URL = {https://www.science.org/doi/abs/10.1126/sciadv.adu9991}}

@article{tang2021qubit,
  title = {Qubit-ADAPT-VQE: An Adaptive Algorithm for Constructing Hardware-Efficient Ans\"atze on a Quantum Processor},
  author = {Tang, Ho Lun and Shkolnikov, V.O. and Barron, George S. and Grimsley, Harper R. and Mayhall, Nicholas J. and Barnes, Edwin and Economou, Sophia E.},
  journal = {PRX Quantum},
  volume = {2},
  issue = {2},
  pages = {020310},
  numpages = {16},
  year = {2021},
  month = {Apr},
  publisher = {American Physical Society},
  doi = {10.1103/PRXQuantum.2.020310},
  url = {https://link.aps.org/doi/10.1103/PRXQuantum.2.020310}
}

@article{bauman2025coupled,
  title = {Coupled cluster downfolding theory in simulations of correlated systems on quantum hardware},
  author = {Bauman, Nicholas P. and Zheng, Muqing and Liu, Chenxu and Myers, Nathan M. and Panyala, Ajay and Peng, Bo and Li, Ang and Kowalski, Karol},
  journal = {Phys. Rev. Res.},
  volume = {8},
  issue = {1},
  pages = {013072},
  numpages = {16},
  year = {2026},
  month = {Jan},
  publisher = {American Physical Society},
  doi = {10.1103/b1t6-ln6v},
  url = {https://link.aps.org/doi/10.1103/b1t6-ln6v}
}

@article{guo2024experimental,
   title={Experimental quantum computational chemistry with optimized unitary coupled cluster ansatz},
   volume={20},
   ISSN={1745-2481},
   url={http://dx.doi.org/10.1038/s41567-024-02530-z},
   DOI={10.1038/s41567-024-02530-z},
   number={8},
   journal={Nature Physics},
   publisher={Springer Science and Business Media LLC},
   author={Guo, Shaojun and Sun, Jinzhao and Qian, Haoran and Gong, Ming and Zhang, Yukun and Chen, Fusheng and Ye, Yangsen and Wu, Yulin and Cao, Sirui and Liu, Kun and Zha, Chen and Ying, Chong and Zhu, Qingling and Huang, He-Liang and Zhao, Youwei and Li, Shaowei and Wang, Shiyu and Yu, Jiale and Fan, Daojin and Wu, Dachao and Su, Hong and Deng, Hui and Rong, Hao and Li, Yuan and Zhang, Kaili and Chung, Tung-Hsun and Liang, Futian and Lin, Jin and Xu, Yu and Sun, Lihua and Guo, Cheng and Li, Na and Huo, Yong-Heng and Peng, Cheng-Zhi and Lu, Chao-Yang and Yuan, Xiao and Zhu, Xiaobo and Pan, Jian-Wei},
   year={2024},
   month=June, pages={1240--1246} }

@article{bittel2021training,
  title = {Training Variational Quantum Algorithms Is NP-Hard},
  author = {Bittel, Lennart and Kliesch, Martin},
  journal = {Phys. Rev. Lett.},
  volume = {127},
  issue = {12},
  pages = {120502},
  numpages = {6},
  year = {2021},
  month = {Sep},
  publisher = {American Physical Society},
  doi = {10.1103/PhysRevLett.127.120502},
  url = {https://link.aps.org/doi/10.1103/PhysRevLett.127.120502}
}

@article{anschuetz2022quantum,
  title={Quantum variational algorithms are swamped with traps},
  author={Anschuetz, Eric R. and Kiani, Bobak T.},
  journal={Nature Communications},
  volume={13},
  number={1},
  pages={7760},
  year={2022},
  month={Dec},
  publisher={Springer Nature},
  doi={10.1038/s41467-022-35364-5},
  url={https://doi.org/10.1038/s41467-022-35364-5}
}

@article{wierichs2020avoiding,
  title = {Avoiding local minima in variational quantum eigensolvers with the natural gradient optimizer},
  author = {Wierichs, David and Gogolin, Christian and Kastoryano, Michael},
  journal = {Phys. Rev. Res.},
  volume = {2},
  issue = {4},
  pages = {043246},
  numpages = {18},
  year = {2020},
  month = {Nov},
  publisher = {American Physical Society},
  doi = {10.1103/PhysRevResearch.2.043246},
  url = {https://link.aps.org/doi/10.1103/PhysRevResearch.2.043246}
}

@misc{chivilikhin2020mog,
      title={MoG-VQE: Multiobjective genetic variational quantum eigensolver}, 
      author={D. Chivilikhin and A. Samarin and V. Ulyantsev and I. Iorsh and A. R. Oganov and O. Kyriienko},
      year={2020},
      eprint={2007.04424},
      archivePrefix={arXiv},
      primaryClass={quant-ph},
}

@misc{wakaura2021evaluation,
      title={Evaluation on Genetic Algorithms as an optimizer of Variational Quantum Eigensolver(VQE) method}, 
      author={Hikaru Wakaura and Takao Tomono and Shoya Yasuda},
      year={2021},
      eprint={2110.07441},
      archivePrefix={arXiv},
      primaryClass={quant-ph},
}

@article{krylov2000spin,
    author = {Krylov, Anna I.},
    title = {Spin-contamination of coupled-cluster wave functions},
    journal = {The Journal of Chemical Physics},
    volume = {113},
    number = {15},
    pages = {6052-6062},
    year = {2000},
    month = {10},
    issn = {0021-9606},
    doi = {10.1063/1.1308557},
    url = {https://doi.org/10.1063/1.1308557},
}

@article{bertels2022symmetry,
    author = {Bertels, Luke W. and Grimsley, Harper R. and Economou, Sophia E. and Barnes, Edwin and Mayhall, Nicholas J.},
    title = {Symmetry Breaking
Slows Convergence of the ADAPT Variational
Quantum Eigensolver},
    journal = {Journal of Chemical Theory and Computation},
    volume = {18},
    number = {11},
    pages = {6656-6669},
    year = {2022},
    month = {11},
    issn = {1549-9618},
    doi = {10.1021/acs.jctc.2c00709},
    url = {https://doi.org/10.1021/acs.jctc.2c00709},
}

@article{tsuchimochi2020spin,
  title = {Spin-projection for quantum computation: A low-depth approach to strong correlation},
  author = {Tsuchimochi, Takashi and Mori, Yuto and Ten-no, Seiichiro L.},
  journal = {Phys. Rev. Res.},
  volume = {2},
  issue = {4},
  pages = {043142},
  numpages = {8},
  year = {2020},
  month = {Oct},
  publisher = {American Physical Society},
  doi = {10.1103/PhysRevResearch.2.043142},
  url = {https://link.aps.org/doi/10.1103/PhysRevResearch.2.043142}
}

@article{shi2014symmetry,
  title = {Symmetry-projected wave functions in quantum Monte Carlo calculations},
  author = {Shi, Hao and Jim{\'e}nez-Hoyos, Carlos A. and Rodr{\'{\i}}guez-Guzm{\'a}n, R. and Scuseria, Gustavo E. and Zhang, Shiwei},
  journal = {Phys. Rev. B},
  volume = {89},
  issue = {12},
  pages = {125129},
  numpages = {8},
  year = {2014},
  month = {Mar},
  publisher = {American Physical Society},
  doi = {10.1103/PhysRevB.89.125129},
  url = {https://link.aps.org/doi/10.1103/PhysRevB.89.125129}
}

@article{rosenberg2017ultracold,
  title = {Ultracold Atoms in a Square Lattice with Spin-Orbit Coupling: Charge Order, Superfluidity, and Topological Signatures},
  author = {Rosenberg, Peter and Shi, Hao and Zhang, Shiwei},
  journal = {Phys. Rev. Lett.},
  volume = {119},
  issue = {26},
  pages = {265301},
  numpages = {5},
  year = {2017},
  month = {Dec},
  publisher = {American Physical Society},
  doi = {10.1103/PhysRevLett.119.265301},
  url = {https://link.aps.org/doi/10.1103/PhysRevLett.119.265301}
}

@misc{chang2017multi,
      title={Multi-determinant generalized Hartree-Fock wave functions in Monte Carlo calculations}, 
      author={Chia-Chen Chang and Miguel A. Morales},
      year={2018},
      eprint={1711.02154},
      archivePrefix={arXiv},
      primaryClass={cond-mat.str-el},
}

@article{qin2016benchmark,
  title = {Benchmark study of the two-dimensional Hubbard model with auxiliary-field quantum Monte Carlo method},
  author = {Qin, Mingpu and Shi, Hao and Zhang, Shiwei},
  journal = {Phys. Rev. B},
  volume = {94},
  issue = {8},
  pages = {085103},
  numpages = {12},
  year = {2016},
  month = {Aug},
  publisher = {American Physical Society},
  doi = {10.1103/PhysRevB.94.085103},
  url = {https://link.aps.org/doi/10.1103/PhysRevB.94.085103}
}

@article{ganoe2024notion,
    author = {Ganoe, Brad and Shee, James},
    title = {On the notion of strong correlation in electronic structure theory},
    journal = {Faraday Discussions},
    volume = {254},
    pages = {53-75},
    year = {2024},
    month = {11},
    issn = {1359-6640},
    doi = {10.1039/d4fd00066h},
    url = {https://doi.org/10.1039/d4fd00066h},
}

@article{motta2018ab,
  title={Ab initio computations of molecular systems by the auxiliary-field quantum Monte Carlo method},
  author={Motta, Mario and Zhang, Shiwei},
  journal={WIREs Computational Molecular Science},
  volume={8},
  number={5},
  pages={e1364},
  year={2018},
  month={May},
  publisher={Wiley},
  doi={10.1002/wcms.1364},
  url={https://doi.org/10.1002/wcms.1364}
}

@misc{kjonstad2026can,
      title={Can phaseless auxiliary-field quantum Monte Carlo with broken symmetry trials describe iron-sulfur clusters?}, 
      author={Eirik F. Kj{\o}nstad and Huanchen Zhai and James Shee and Sandeep Sharma and Garnet Kin-Lic Chan},
      year={2026},
      eprint={2605.03270},
      archivePrefix={arXiv},
      primaryClass={physics.chem-ph},
}

@misc{johnson2022cccbdb,
  author       = {{Johnson III}, Russell D.},
  title        = {{NIST} Computational Chemistry Comparison and Benchmark Database},
  howpublished = {{NIST} Standard Reference Database Number 101, Release 22},
  year         = {2022},
  month        = {May},
  publisher    = {National Institute of Standards and Technology},
  doi          = {10.18434/T47C7Z},
  url          = {https://cccbdb.nist.gov/},
  note         = {Editor: Russell D. Johnson III. Accessed September 30, 2026}
}

\appendix

\clearpage

\section{Simulation Details}
\label{app: sim_details}
All Hamiltonian integrals, SCF, CCSD, and FCI calculations were performed using
the \texttt{PySCF} \cite{sun2020recent} quantum chemistry package. wave function parameters were
optimized with the L-BFGS-B optimizer from the \texttt{SciPy} library \cite{2020SciPy-NMeth}, using an energy tolerance of $10^{-6}$ Ha. For ADAPT-VQE, we used a gradient magnitude convergence threshold of $10^{-3}$. Gradients were estimated using central finite-difference methods with a parameter shift of $\epsilon = 10^{-3}$. The \texttt{OpenFermion} package \cite{mcclean2020openfermion} was used to construct UCC excitation operators, while the \texttt{ffsim} library \cite{ffsim} was used to generate the matrix
representations required for the LUCJ ansatz, which were then decomposed into quantum gates.

Our UCCSD and UpCCGSD ansatzes were constructed using non--spin-adapted and 
non--spin-complemented excitation operators. Consequently, the ansatzes are built 
from spin-orbital excitations that conserve the spin projection $S_z$, but do not explicitly enforce total spin symmetry, as 
excitation operators are assigned independent variational parameters. We use warm-start initialization of these parameters with unrestricted CCSD $t_1$ and $t_2$ values. In cases where CCSD did not converge (e.g., for $R \gtrsim 1.8\,\text{\AA}$ in H$_{10}$), we initialized all variational parameters to zero so that the initial circuit reduces to the identity acting on $\lvert \Psi_0 \rangle$. The ordering of operators in the UCCSD ansatz is not uniquely defined \cite{grimsley2019trotterized,grimsley2019adaptive}. Following 
Ref.~\cite{hirsbrunner2024beyond}, we therefore employ magnitude ordering, 
in which operators are ordered according to the absolute values of their initial CCSD
 parameters $|\theta|$, with operators having larger magnitudes placed 
to the right in Eq.~4 of the main text to reduce Trotter error. Operators with zero initial CCSD parameter values, such as generalized excitation operators, were placed at the end of the ansatz and ordered lexicographically by spin-orbital indices. We found that multilayer UpCCGSD was difficult to optimize when all variational parameters were updated simultaneously. To address this, we employed a layered optimization strategy in which the first layer was optimized independently, followed by the first two layers, and so on up to $k$ layers. In the multilayer setting, we observed that zero initialization of the parameters was more reliable than the standard CCSD initialization for the non-generalized excitations; therefore, all-zeros initialization was used for the final results. The ADAPT-VQE operator pool we used is the set of UCCSD operators described above. For each hydrogen-chain geometry, the VQE optimization was performed independently so that every point on the curve is an independent state-preparation task.

The LUCJ ansatz was initialized using CCSD amplitudes, and the interactions were constrained to nearest neighbors on a square lattice topology (Eq.~7 of the main text). Although the double factorization of the $T_2$ operators in Eq.~8 of the main text can approximate UCCSD reasonably well, imposing locality constraints on the interactions can significantly degrade the quality of the approximation. For this reason, we enabled compressed double factorization and mitigated Trotter errors by regularizing the norm of the diagonal Coulomb matrix \cite{lin2025improved, ffsim}.

For the HVA, we employed a simple all-ones initialization of the variational parameters. The Hamiltonian was partitioned into commuting groups following the default implementation in Qiskit \cite{qiskit2024}, without attempting to optimize for minimal sets of commuting terms. 

In Sec.~4.3 of the main text, we explore optimizing circuits initialized to a UHF reference $\lvert \Psi_{\mathrm{UHF}} \rangle$. Rather than
modifying the circuit structure, we implement UHF initialization through an orbital-basis transformation of the one- and two-electron integrals in Eq.~1 of the main text. Concretely, we obtain
spin-dependent UHF MO coefficient matrices $C^{\alpha}$ and $C^{\beta}$ from the SCF calculation and form the corresponding change-of-basis matrices from the RHF MO basis to
the UHF MO basis. Using these matrices, we rotate the one- and two-body integrals $T_{pq}$ and $V_{pqrs}$ into the UHF orbital basis. With the Hamiltonian expressed in this
rotated basis, the standard occupation-state preparation circuit that would initialize $\lvert \Psi_{\mathrm{RHF}} \rangle$ in the RHF basis instead initializes $\lvert \Psi_{\mathrm{UHF}} \rangle$ in the UHF basis, and all VQE optimization is carried
out with respect to this rotated Hamiltonian. This approach is computationally efficient, as the orbital rotation is performed entirely at the classical integral level and does not
introduce any additional quantum circuit depth.

Once our circuits were optimized and the final statevector was obtained, we converted it into an MSD trial of the form given in Eq.~13 of the main text. This was done by masking all non--particle-number-conserving amplitudes as well as any particle-number-conserving amplitudes whose magnitudes were below double-precision threshold. We then performed ph-AFQMC using this trial state with the \texttt{ipie} package~\cite{malone2022ipie}. The initial walker states were chosen to be the RHF determinant. In combination with a spin-symmetric Hamiltonian, this ensures that the walker ensemble remains within the spin-restricted manifold throughout the imaginary-time projection, such that $\langle \hat{S}^2 \rangle = 0$, even when the trial wave function itself contains spin contamination. In the rare case that the trial state is orthogonal to the RHF determinant, the walkers are instead initialized from the determinant with the largest coefficient in the MSD expansion of the trial. 

The molecular test set of Section~5 of the main text were treated with the following specifics. Geometries are the experimental equilibrium structures tabulated in the NIST Computational Chemistry Comparison and Benchmark Database \cite{johnson2022cccbdb}, and all calculations use the STO-6G basis. Every molecule is treated in the full minimal-basis space except C$_2$H$_2$, for which the two carbon $1s$ orbitals are frozen and the circuit spans a CAS(10e,10o) constructed with CASCI in the RHF orbital basis, so that the active integrals are an exact frozen-core reduction of the full-space ones and the trial embeds consistently in the full-space ph-AFQMC Hamiltonian. As for the hydrogen chains, UCCSD and LUCJ were initialized from unrestricted CCSD amplitudes and UpCCGSD, HVA and ADAPT-UCCSD from zero or all-ones parameters. The repetition count $k$ of each multilayer ansatz was chosen per molecule so that its parameter count approximately matches that of UCCSD for that molecule, which is why $k$ varies across the set (Table~\ref{tab:molecules}).

For the hydrogen chains, each trial was obtained from a single optimization using the deterministic initialization described above; the multi-start protocol was introduced for the molecular test set, where initialization sensitivity was more pronounced, particularly for UpCCGSD, HVA, and multilayer LUCJ. For these cases, we optimized from multiple starting points: in addition to the deterministic warm start, we performed restarts in which each initial parameter was perturbed by Gaussian noise with standard deviation $\sigma = 0.05$, using independent random seeds. Where multiple optimizations of the same molecule and ansatz were available, the one reaching the lowest variational energy was retained as the trial wave function, and it is that trial whose properties are reported in Table~\ref{tab:molecules}. We note that this restart procedure does not remove the underlying difficulty: for UpCCGSD in particular, independent optimizations from equally reasonable starting points can converge to distinct local minima with substantially different variational energies and, consequently, different projected energies. This is discussed further in Section~6 of the main text.

All AFQMC simulations were run on Perlmutter CPU nodes equipped with AMD EPYC 7763 processors, with the population of single-SD walkers parallelized across many CPU threads. We used a population of 1024 walkers and the total imaginary projection time used for the quantum trial wave functions was $\tau = 40 E_{\text{Ha}}^{-1}$, which we found was approximately the projection time to achieve within $\sim 1\,\mathrm{mHa}$ statistical error bars for all ansatzes, molecules, and geometries, with a few exceptions in which case we increase to $\tau = 80 E_{\text{Ha}}^{-1}$. For the classical benchmarks using single-determinant RHF and UHF trial wave functions, a longer imaginary-time projection was required to achieve statistical error bars below 1 mHa. Accordingly, we employed a projection time of $\tau = 200\,E_{\text{Ha}}^{-1}$ for these calculations. In all AFQMC simulations, the initial portion of the imaginary-time evolution was treated as equilibration (burn-in). Specifically, the first 25\% of the total projection time was discarded to ensure that the walker population had equilibrated and lost memory of the initial determinant before measurements were accumulated. Additionally, after every five propagation steps of $\Delta\tau = 0.01\,E_{\text{Ha}}^{-1}$, each walker is reorthonormalized to maintain numerical stability. At the same frequency, population control is performed: walkers with larger weights are stochastically cloned, while those with smaller weights are removed, in order to stabilize the ensemble and reduce statistical variance.

After imaginary-time evolution, we accounted for the autocorrelation present in successive local-energy measurements. To obtain statistically reliable error estimates, we performed reblocking analysis, in which correlated samples are grouped into blocks whose size is increased until the block-averaged energies become uncorrelated. The resulting integrated autocorrelation time was then used to compute accurate error bars for all reported AFQMC energies.

\section{ph-AFQMC: Algorithmic Details}
\label{app:ph-AFQMC}

To perform imaginary-time propagation of the Hamiltonian in Eq.~1 of the main text, the two-body term must be recast in terms of one-body operators through the introduction of auxiliary fields. This can be done by first choosing a small imaginary-time step \(\Delta \tau > 0\) and applying the Trotter–Suzuki approximation:
\begin{equation}
    e^{-\Delta \tau \hat{H}} \approx e^{-\Delta \tau \hat{H}_1} e^{-\Delta \tau \hat{H}_2} + \mathcal{O}(\Delta \tau^2).
\label{eq:Trotter}
\end{equation}

Next, we factorize the two-electron integrals \(V_{pqrs}\) using the Cholesky decomposition:
\begin{equation}
    V_{pqrs} = \frac{1}{2} \sum_{\gamma=1}^{N_{\mathrm{aux}}} L^\gamma_{pr} L^\gamma_{qs},
\label{eq: Cholesky}
\end{equation}
which allows the two-body part of the Hamiltonian, \(\hat{H}_2\), to be written as a sum of squares of one-body operators \(\hat{v}_\gamma\):
\begin{equation}
    \hat{H}_2 = \frac{1}{2} \sum_{\gamma=1}^{N_{\mathrm{aux}}} \lambda_\gamma \hat{v}_\gamma^2.
\label{eq: one-body operators}
\end{equation}

Having expressed \(\hat{H}_2\) in this quadratic form, we can apply the Hubbard–Stratonovich transformation to express its imaginary-time propagator as an integral over \(N_{\mathrm{aux}}\) auxiliary fields:
\begin{equation}
    e^{-\Delta \tau \hat{H}_2}
    = \prod_{\gamma=1}^{N_{\mathrm{aux}}} \int dx_\gamma \, e^{-x_\gamma^2/2} 
      e^{x_\gamma \sqrt{-\Delta \tau \, \lambda_\gamma}\, \hat{v}_\gamma}
      + \mathcal{O}(\Delta \tau^2).
\label{eq:prod stochastic integral}
\end{equation}

By incorporating the one-body term \(\hat{H}_1\) and collecting constants, the full imaginary-time propagator of the Hamiltonian can be written in the compact form of Eq.~14 of the main text.

The phaseless approximation incorporates importance sampling to constrain the walkers to remain within the phase sector defined by the trial wave function $\ket{\Psi_T}$. This constraint suppresses destructive phase cancellations arising from random fluctuations, thereby stabilizing the simulation at the cost of a controllable bias, the accuracy of which is determined by the quality of $\ket{\Psi_T}$.

With importance sampling, we assign a manipulable weight, $w_k(\tau)$, to each walker such that the weighted average over all $N_W$ walkers represents the many-body wave function:
\begin{equation}
    |\Psi(\tau)\rangle = \sum_{k=1}^{N_W} 
    w_k(\tau)\,
    \frac{|\phi_k(\tau)\rangle}
         {\langle \Psi_T | \phi_k(\tau)\rangle}.
\label{eq:AFQMC wave function}
\end{equation}

To constrain the propagation of the walkers toward the trial state $|\Psi_T\rangle$, we modify the distribution $p(\mathbf{x})$ of auxiliary fields to minimize fluctuations in the overlap ratio $\langle \Psi_T|\phi_k'\rangle / \langle \Psi_T|\phi_k\rangle$, where $|\phi_k'\rangle = \hat{B}(\mathbf{x})|\phi_k\rangle$. The optimal shift of the auxiliary fields, known as the force bias, is:
\begin{equation}
\bar{\mathbf{x}}_k(\tau)
= -\sqrt{\Delta\tau}\,
  \frac{\langle \Psi_T|\hat{\mathbf{v}}|\phi_k(\tau)\rangle}
       {\langle \Psi_T|\phi_k(\tau)\rangle},
\label{eq: force bias}
\end{equation}
where $\hat{\mathbf{v}}$ denotes the collection of one-body operators introduced in Eq.~\ref{eq: one-body operators}.

We then use a shifted distribution $p(\mathbf{x}-\bar{\mathbf{x}}_k)$ to propagate the walkers. Under this shift, the overlap ratio becomes:
\begin{equation}
O_k(\tau)
= \frac{\langle \Psi_T|
\hat{B}(\mathbf{x}_k - \bar{\mathbf{x}}_k)
|\phi_k(\tau)\rangle}
       {\langle \Psi_T|\phi_k(\tau)\rangle}.
\label{eq:overlap_ratio_rewrite}
\end{equation}

To assign larger weights to walkers with larger overlaps, we define the importance function:
\begin{equation}
I(\mathbf{x}_k,\bar{\mathbf{x}}_k,\tau)
= O_k(\tau)\,
\exp\!\left(
\mathbf{x}_k\!\cdot\!\bar{\mathbf{x}}_k 
- \tfrac{1}{2}\bar{\mathbf{x}}_k\!\cdot\!\bar{\mathbf{x}}_k
\right),
\label{eq: importance function}
\end{equation} 
where the exponential prefactor arises from the force-bias shift.

The phaseless approximation projects each walker’s complex weight onto the real axis by modifying the importance function:
\begin{equation}
I_{\mathrm{ph}}(\mathbf{x}_k,\bar{\mathbf{x}}_k,\tau)
= \big| I(\mathbf{x}_k,\bar{\mathbf{x}}_k,\tau) \big|\,
  \max\!\left[0, \cos\!\big(\theta_k(\tau)\big)\right],
\label{eq:phaseless_mod_rewrite}
\end{equation}
where the phase angle that is determined by the trial state is:
\begin{equation}
\theta_k(\tau) = \arg\!\big(O_k(\tau)\big).
\label{eq:phase_rewrite}
\end{equation}

With this constraint, the walker weights and SDs are updated as:
\begin{align}
w_k(\tau + \Delta\tau) &= 
I_{\mathrm{ph}}(\mathbf{x}_k,\bar{\mathbf{x}}_k,\tau)\,
w_k(\tau),\\[3pt]
|\phi_k(\tau + \Delta \tau)\rangle &= 
\hat{B}(\mathbf{x}_k - \bar{\mathbf{x}}_k)\,
|\phi_k(\tau)\rangle.
\label{eq:update_rewrite}
\end{align}

To summarize the importance sampling procedure, at each time step and for each walker, the auxiliary-field distribution 
$p(\mathbf{x})$ is shifted by the optimal force bias $\bar{\mathbf{x}}_k$, while the importance function 
$I_{ph}$ reweights the walkers to 
minimize fluctuations in overlap ratio, thereby steering the population toward physically meaningful regions of SD space guided by $\ket{\Psi_T}$ \cite{pavarini2013emergent,malone2022ipie}.

\section{Tabulated Benchmark Results}
\label{app:supp_results}

Table~\ref{tab:VQE_AFQMC_Table} provides a detailed summary of the VQE and ph-AFQMC results for H$_8$ and H$_{10}$ at multiple bond lengths spanning weakly to strongly correlated regimes. For each ansatz, we report the number of variational parameters, the number of optimization iterations required for convergence, the VQE correlation energy and fidelity with respect to FCI, the $\langle \hat{S^2}\rangle$ of the trial states, the singlet-projected fidelity $F_{S=0}$ of Eq.~16 of the main text, the number of determinants in the multideterminant expansion obtained from the optimized statevector, and the corresponding AFQMC correlation energy with statistical uncertainties. The H$_{10}$ results reported here are a subset of the results used to construct the potential energy surface and fidelity comparisons shown in Figs.~1 and~2 of the main text. We include results for H$_8$ to demonstrate that the trends observed for H$_{10}$ persist across other hydrogen chains. 

Table~\ref{tab:simons_comparison} reports a direct comparison of our AFQMC(UHF) total energies for H$_{10}$ with the Simons Foundation hydrogen chain benchmark data. The agreement is within statistical uncertainty at all reported geometries, validating both our AFQMC implementation and our UHF-based trial protocol.

\begin{table*}
    \centering
    \begingroup
    \renewcommand{\arraystretch}{1.1}
    \resizebox{\textwidth}{!}{%
    \begin{tabular}{cc|cc|ccccc|cc}
        \hline
        System\rule{0pt}{3.0ex} &
        FCI (mHa) &
        Ansatz &
        \makecell{\# of\\Parameters} &
        \makecell{\# of\\Iterations$^{*}$} &
        \makecell{VQE Energy\\(mHa)} &
        \makecell{Fidelity} &
        \makecell{$ \log_{10}\langle \hat{S^2} \rangle$} &
        \makecell{$F_{S=0}$} &
        \makecell{\# of\\Determinants} &
        \makecell{AFQMC Energy\\(mHa)} \\
        \hline

        \multirow{8}{*}{\makecell{H$_{8}$\\($R = 1.0$~\AA)}} & \multirow{8}{*}{134.68}
            & UCCSD  & 360  &  5   & 133.11 & 0.999 & -6.3 & 0.999 & 4,526 & 134.60(1)\\
        & & ADAPT-UCCSD  & 153  &  153 & 132.79 & 0.998 & -3.9 & 0.998 & 1,252  & 134.60(2)\\
        & & LUCJ  & 200  &  75  & 114.57 & 0.968 &-2.1 & 0.969 & 4,900  & 134.16(8)\\
        & & 2-LUCJ  & 336  &  163  & 126.26 & 0.991 & -1.8& 0.993 & 4,900  & 134.39(6)\\

        & & HVA  & 54  &  18 & 75.61 & 0.939 & -1.6& 0.945 & 2,468  & 133.7(3)\\
        & & 6-HVA  & 324  &  67 & 131.21 & 0.998 & -2.4& 0.999 & 2,468  & 134.60(3)\\

        & & UpCCGSD  & 84  & 57   & 52.70  & 0.933 & -15.6 & 0.933 & 4,174  & 133.3(4)\\
        & & 4-UpCCGSD  & 336  & 73   & 54.01  & 0.937 & -18.0 & 0.937 & 1,810 & 133.5(3)\\

        \hline

        \multirow{8}{*}{\makecell{H$_{8}$\\($R = 1.4$~\AA)}} & \multirow{8}{*}{275.24}
            & UCCSD  & 360  &  10  & 268.93 & 0.990 & -4.2& 0.990 & 4,899  & 274.00(7)\\
        & & ADAPT-UCCSD  & 163  &  163 & 268.55 & 0.990 & -3.1& 0.991 & 2,468  & 274.04(7)\\
        & & LUCJ  & 200  &  146  & 251.05 & 0.898 & -2.1 & 0.899 & 4,900  & 270.2(2)\\
        & & 2-LUCJ  & 336  &  121  & 261.04 & 0.938 & -1.8 & 0.940 & 4,900  & 272.34(7)\\

        & & HVA  & 54  &  17 & 151.56 & 0.811 & -1.0 & 0.829 & 2,468  & 255.9(5)\\
        & & 6-HVA  & 324  &  59 & 267.70 & 0.989 & -1.7& 0.994 & 2,468  & 274.45(4)\\

        & & UpCCGSD  & 84  &  46  & 90.69 & 0.753 &-16.8 & 0.753 & 4,178  & 256(1)\\
        & & 4-UpCCGSD  & 336  &  603  & 204.77 & 0.846 & -1.6 & 0.882 & 1,810  & 270.7(3)\\

        \hline

        \multirow{8}{*}{\makecell{H$_{8}$\\($R = 1.8$~\AA)}} & \multirow{8}{*}{504.77}
            & UCCSD  & 360  &  17   & 491.71 & 0.958 & -3.5 & 0.958 & 2,468 & 503.0(2)\\
        & & ADAPT-UCCSD  & 127  &  127 & 485.96 & 0.951 & -1.3& 0.959 & 2,468 & 503.2(2)\\
        & & LUCJ  & 200  &  123 & 486.48 & 0.793 &-2.9 & 0.793 & 4,900 & 500.7(3)\\
        & & 2-LUCJ  & 336  &  258 & 497.58 & 0.895 &-1.9 & 0.897 & 4,900 & 502.56(8)\\

        & & HVA  & 54 &  18 & 289.03 & 0.627 & -0.3 & 0.692 &2,468 & 464.4(9)\\
        & & 6-HVA  & 324 &  44 & 492.39 & 0.950 & -1.0 & 0.966 & 2,468 & 503.4(1)\\

        & & UpCCGSD  & 84   &  65  & 415.83  & 0.107   & 0.4 & 0.347 & 4,900 & 496(1)\\
        & & 4-UpCCGSD  & 336   &  74  & 416.05  & 0.108 &0.4 & 0.351 & 4,900 & 496(1)\\

        \hline

        \multirow{8}{*}{\makecell{H$_{10}$\\($R = 1.0$~\AA)}} & \multirow{8}{*}{167.78}
            & UCCSD  & 875  &  5  & 165.02 & 0.998 & -5.7 & 0.998 & 50,586 &  167.51(3)\\
        & & ADAPT-UCCSD & 320   &  320  & 164.23 & 0.996 & -3.3 & 0.996 & 15,912 & 167.37(4)\\
        & & LUCJ  & 310  &  74  & 139.04 & 0.946 & -1.9& 0.948 & 63,504 & 166.2(3)\\
        & & 3-LUCJ  & 730  &  93  & 155.88 & 0.985 & -1.6 & 0.989 & 63,504 & 167.09(7)\\

        & & HVA & 120 &  13 & 57.69 & 0.887 & -1.3 & 0.897 & 31,752 & 164.0(5)\\
        & & 7-HVA & 840 &  32 & 158.68 & 0.994 &-2.1& 0.997 & 31,752 & 167.63(7)\\

        & & UpCCGSD & 135 &  90 & 63.01 & 0.903 & -10.4 & 0.903 & 50,964 & 164.0(5)\\
        & & 6-UpCCGSD & 810 &  124 & 66.38 & 0.915 & -17.1 & 0.915 & 21,252 & 164.8(6)\\

        \hline

        \multirow{8}{*}{\makecell{H$_{10}$\\($R = 1.4$~\AA)}} & \multirow{8}{*}{341.36}
            & UCCSD  & 875  &  10  & 331.83 & 0.981 & -4.0 & 0.981 & 63,468 & 339.7(1)\\
        & & ADAPT-UCCSD & 333   &  333  & 330.08 & 0.982 &-2.4& 0.983 & 31,752 & 339.8(1)\\
        & & LUCJ  & 310  &  154  & 303.23 & 0.836 & -2.3 & 0.836 & 63,504 & 334.9(3)\\
        & & 3-LUCJ  & 730  &  222  & 333.26 & 0.967 & -1.8& 0.970 & 63,504 & 339.83(9)\\

        & & HVA & 120 &  12 & 129.74 & 0.696 &-0.8& 0.723 & 31,752 & 314(1)\\
        & & 7-HVA & 840 &  30 & 327.62 & 0.980 & -1.6& 0.988 & 31,752 & 340.59(6)\\

        & & UpCCGSD & 135 &  92 & 133.11 & 0.452 & -0.1& 0.720 & 55,079 & 324(1)\\
        & & 6-UpCCGSD & 810 &  118 & 188.24 & 0.274 & 0.3& 0.685 & 21,252 & 333.3(8)\\

        \hline

        \multirow{8}{*}{\makecell{H$_{10}$\\($R = 1.8$~\AA)}} & \multirow{8}{*}{628.17}
            & UCCSD  & 875  &  14  & 610.11 & 0.929 & -3.3& 0.929 & 31,752 & 625.0(1)\\
        & & ADAPT-UCCSD & 244   &  244  & 596.71 & 0.907 &-0.9& 0.925 & 31,752 & 625.7(3)\\
        & & LUCJ  & 310  &  85  & 597.54 & 0.702& -4.1& 0.702 & 63,504 & 621.5(2)\\
        & & 3-LUCJ  & 730  &  142  & 615.59 & 0.830 & -1.6& 0.834 & 63,504 & 624.4(1)\\

        & & HVA & 120 &  14 & 241.78 & 0.437 &-0.4 & 0.483 & 31,752 & 532(1)\\
        & & 7-HVA & 840 &  25 & 601.35 & 0.934 & -1.1 & 0.954 & 31,752 & 626.2(1)\\

        & & UpCCGSD & 135 &  74 & 507.92 & 0.050 &0.5& 0.307 & 63,504 & 626.9(3)\\
        & & 6-UpCCGSD & 810 &  88 & 551.42 & 0.102 & 0.5 & 0.425 & 21,252 & 624.0(1)\\

        \hline
    \end{tabular}%
    }
    \endgroup
    \caption{Summary of VQE and AFQMC results for H$_8$ and H$_{10}$ at multiple bond lengths. Reported energies are correlation energies, defined as $|E - E_{\mathrm{RHF}}|$ in mHa. Thus, larger values indicate improved recovery of correlation energy. For each ansatz, we list the number of variational parameters, optimization iterations, VQE correlation energy and fidelity with respect to FCI, the $\langle \hat{S^2}\rangle$ of the trial states, the singlet-projected fidelity $F_{S=0}$ of Eq.~16 of the main text, the number of determinants in the multideterminant expansion (i.e., the number of nonzero, particle-number-conserving amplitudes in the VQE statevector), and the corresponding AFQMC correlation energy (with statistical uncertainties in parentheses). The H$_{10}$ results reported here are a subset of the results used to construct the potential energy surface and fidelity comparisons shown in Figs.~1 and~2 of the main text.}
    \begin{flushleft}
    \footnotesize
    
    $^{*}$For ADAPT-UCCSD, the reported iteration count denotes iterations of the adaptive ansatz construction, with one operator added per iteration, rather than iterations of the classical optimizer.
    \end{flushleft}
    
    \label{tab:VQE_AFQMC_Table}
\end{table*}

\begin{table*}
    \centering
    \begingroup
    \renewcommand{\arraystretch}{1.1}
    \resizebox{\textwidth}{!}{%
    \begin{tabular}{cc|cc|ccccc|cc}
        \hline
        Molecule\rule{0pt}{3.0ex} &
        FCI (mHa) & Ansatz & \makecell{\# of\\Parameters} & \makecell{\# of\\Iterations$^{*}$} &
        \makecell{VQE Energy\\(mHa)} & \makecell{Fidelity} & \makecell{$ \log_{10}\langle \hat{S^2} \rangle$} &
        \makecell{$F_{S=0}$} & \makecell{\# of\\Determinants} & \makecell{AFQMC Energy\\(mHa)} \\
        \hline

        \multirow{5}{*}{H$_2$O} & \multirow{5}{*}{50.09}
            & UCCSD  & 140  & 3  & 49.99 & 1.000 & -6.5 & 1.000 & 169 & 50.085(3) \\
        & & ADAPT-UCCSD  & 30  & 30  & 49.63 & 1.000 & -4.9 & 1.000 & 63 & 50.13(2) \\
        & & 1-LUCJ  & 154  & 24  & 25.28 & 0.985 & -2.9 & 0.986 & 441 & 50.4(2) \\
        & & 3-HVA  & 132  & 39  & 46.50 & 0.998 & -2.9 & 0.999 & 133 & 50.09(4) \\
        & & 2-UpCCGSD  & 126  & 12  & 25.36 & 0.987 & -20.9 & 0.987 & 121 & 50.2(2) \\
        \hline

        \multirow{5}{*}{BeH$_2$} & \multirow{5}{*}{35.01}
            & UCCSD  & 204  & 3  & 34.64 & 1.000 & -7.9 & 1.000 & 178 & 35.009(5) \\
        & & ADAPT-UCCSD  & 21  & 21  & 34.43 & 0.999 & -5.7 & 0.999 & 88 & 35.07(1) \\
        & & 1-LUCJ  & 154  & 69  & 29.24 & 0.993 & -3.6 & 0.993 & 1,225 & 35.40(5) \\
        & & 6-HVA  & 204  & 29  & 33.83 & 0.999 & -3.8 & 0.999 & 169 & 35.11(1) \\
        & & 3-UpCCGSD  & 189  & 18  & 18.64 & 0.991 & -21.2 & 0.991 & 151 & 35.0(1) \\
        \hline

        \multirow{5}{*}{BH$_3$} & \multirow{5}{*}{53.97}
            & UCCSD  & 360  & 3  & 53.60 & 1.000 & -7.7 & 1.000 & 2,271 & 53.960(5) \\
        & & ADAPT-UCCSD  & 61  & 61  & 53.36 & 1.000 & -4.5 & 1.000 & 634 & 54.001(8) \\
        & & 2-LUCJ  & 336  & 35  & 49.20 & 0.997 & -3.5 & 0.997 & 4,900 & 53.98(4) \\
        & & 4-HVA  & 324  & 63  & 51.28 & 0.999 & -3.0 & 0.999 & 1,250 & 53.94(2) \\
        & & 4-UpCCGSD  & 336  & 18  & 24.78 & 0.984 & -21.6 & 0.984 & 1,350 & 53.5(1) \\
        \hline

        \multirow{5}{*}{NH$_3$} & \multirow{5}{*}{69.10}
            & UCCSD  & 315  & 3  & 67.89 & 0.999 & -7.6 & 0.999 & 1,656 & 69.05(2) \\
        & & ADAPT-UCCSD  & 75  & 75  & 64.49 & 0.996 & -2.4 & 0.997 & 590 & 69.21(2) \\
        & & 2-LUCJ  & 336  & 14  & 59.34 & 0.994 & -2.1 & 0.995 & 3,136 & 69.12(6) \\
        & & 3-HVA  & 282  & 57  & 57.81 & 0.993 & -2.3 & 0.995 & 1,576 & 69.16(5) \\
        & & 4-UpCCGSD  & 336  & 18  & 32.32 & 0.979 & -19.9 & 0.979 & 1,336 & 69.4(2) \\
        \hline

        \multirow{5}{*}{CH$_4$} & \multirow{5}{*}{80.16}
            & UCCSD  & 560  & 11  & 79.95 & 1.000 & -7.7 & 1.000 & 15,876 & 80.164(4) \\
        & & ADAPT-UCCSD  & 117  & 117  & 76.46 & 0.998 & -2.9 & 0.998 & 3,635 & 80.12(3) \\
        & & 3-LUCJ  & 594  & 121  & 70.43 & 0.995 & -2.8 & 0.995 & 15,876 & 80.10(8) \\
        & & 3-HVA  & 507  & 92  & 40.91 & 0.980 & -2.3 & 0.981 & 15,876 & 79.9(2) \\
        & & 5-UpCCGSD  & 540  & 171  & 56.12 & 0.988 & -3.4 & 0.988 & 15,876 & 79.9(2) \\
        \hline

        \multirow{5}{*}{N$_2$} & \multirow{5}{*}{158.68}
            & UCCSD  & 609  & 3  & 156.36 & 0.999 & -5.4 & 0.999 & 4,495 & 158.69(6) \\
        & & ADAPT-UCCSD  & 77  & 77  & 153.73 & 0.997 & -2.5 & 0.997 & 689 & 158.80(4) \\
        & & 2-LUCJ  & 520  & 25  & 130.00 & 0.980 & -1.6 & 0.984 & 14,400 & 159.3(1) \\
        & & 8-HVA  & 616  & 38  & 143.17 & 0.992 & -2.1 & 0.994 & 3,616 & 158.93(9) \\
        & & 5-UpCCGSD  & 675  & 36  & 90.48 & 0.958 & -19.2 & 0.958 & 1,288 & 160.2(4) \\
        \hline

        \multirow{5}{*}{C$_2$H$_2$} & \multirow{5}{*}{174.21}
            & UCCSD  & 875  & 3  & 172.41 & 0.999 & -5.6 & 0.999 & 33,018 & 174.21(5) \\
        & & ADAPT-UCCSD  & 198  & 198  & 172.46 & 1.000 & -4.0 & 1.000 & 15,912 & 174.25(4) \\
        & & 3-LUCJ  & 730  & 123  & 162.93 & 0.992 & -1.8 & 0.995 & 63,504 & 174.16(7) \\
        & & 12-HVA  & 816  & 44  & 167.77 & 0.996 & -2.6 & 0.997 & 15,912 & 174.18(4) \\
        & & 6-UpCCGSD  & 810  & 199  & 124.30 & 0.970 & -1.7 & 0.973 & 63,492 & 173.4(3) \\
        \hline

        \multirow{5}{*}{Be$_2$} & \multirow{5}{*}{107.79}
            & UCCSD  & 804  & 9  & 105.72 & 0.997 & -5.2 & 0.997 & 44,022 & 107.18(3) \\
        & & ADAPT-UCCSD  & 62  & 62  & 104.75 & 0.994 & -2.3 & 0.995 & 2,165 & 107.37(4) \\
        & & 3-LUCJ  & 730  & 43  & 102.84 & 0.991 & -2.0 & 0.992 & 44,100 & 107.05(9) \\
        & & 10-HVA  & 780  & 61  & 92.51 & 0.962 & -2.5 & 0.962 & 11,076 & 105.3(2) \\
        & & 6-UpCCGSD  & 810  & 34  & 59.99 & 0.884 & -20.8 & 0.884 & 1,481 & 104.6(4) \\
        \hline

        \multirow{5}{*}{C$_2$} & \multirow{5}{*}{271.88}
            & UCCSD  & 804  & 3  & 258.54 & 0.990 & -3.8 & 0.990 & 32,185 & 269.1(3) \\
        & & ADAPT-UCCSD  & 101  & 101  & 257.76 & 0.989 & -2.6 & 0.990 & 4,898 & 269.6(2) \\
        & & 3-LUCJ  & 730  & 68  & 252.38 & 0.977 & -2.1 & 0.978 & 44,100 & 268.2(3) \\
        & & 10-HVA  & 770  & 59  & 223.26 & 0.894 & -1.1 & 0.907 & 11,076 & 261.1(6) \\
        & & 6-UpCCGSD  & 810  & 130  & 195.13 & 0.652 & -0.2 & 0.932 & 44,100 & 268.7(6) \\
        \hline
    \end{tabular}%
    }
    \endgroup
    \caption{Summary of VQE and ph-AFQMC results for the molecular test set, all in the STO-6G basis and ordered by decreasing $w_{\mathrm{HF}}$ (Table~\ref{tab:corr_strength}). Reported energies are correlation energies, defined as $|E - E_{\mathrm{RHF}}|$ in mHa, so that larger values indicate improved recovery of correlation energy. Columns are as in Table~\ref{tab:VQE_AFQMC_Table}. Every system is treated in the full minimal-basis space except C$_2$H$_2$, for which the two carbon $1s$ orbitals are frozen and the circuit spans a CAS(10e,10o); the fidelity and $F_{S=0}$ of C$_2$H$_2$ are therefore evaluated against the CASCI state that the circuit can represent, whereas its energies are referenced to the full-space FCI value. The multilayer ansatzes (LUCJ, HVA, UpCCGSD) are evaluated at the repetition count $k$ that approximately matches the UCCSD parameter count for each molecule, so $k$ varies between systems. These data correspond directly to Figure~6 of the main text.}
    \begin{flushleft}
    \footnotesize
    
    $^{*}$For ADAPT-UCCSD, the reported iteration count denotes iterations of the adaptive ansatz construction, with one operator added per iteration, rather than iterations of the classical optimizer.
    \end{flushleft}
    \label{tab:molecules}
\end{table*}

\begin{table}[t]
\centering
\renewcommand{\arraystretch}{1.15}
\begin{tabular}{lccc}
\toprule
System & $N_{\mathrm{orb}}$ & $E_{\mathrm{corr}}$ (mHa) & $w_{\mathrm{HF}}$ \\
\midrule
H$_2$O & 7 & 50.1 & 0.973 \\
BeH$_2$ & 7 & 35.0 & 0.972 \\
BH$_3$ & 8 & 54.0 & 0.967 \\
NH$_3$ & 8 & 69.1 & 0.961 \\
CH$_4$ & 9 & 80.2 & 0.957 \\
N$_2$ & 10 & 158.7 & 0.916 \\
H$_{10}$, $R = 0.8$~\AA & 10 & 118.1 & 0.912 \\
C$_2$H$_2$ & 12 & 174.2 & 0.888 \\
H$_{10}$, $R = 1.0$~\AA & 10 & 167.8 & 0.832 \\
Be$_2$ & 10 & 107.8 & 0.766 \\
H$_{10}$, $R = 1.2$~\AA & 10 & 239.9 & 0.705 \\
C$_2$ & 10 & 271.9 & 0.659 \\
H$_{10}$, $R = 1.4$~\AA & 10 & 341.4 & 0.537 \\
H$_{10}$, $R = 1.6$~\AA & 10 & 473.4 & 0.363 \\
H$_{10}$, $R = 1.8$~\AA & 10 & 628.2 & 0.225 \\
H$_{10}$, $R = 2.0$~\AA & 10 & 791.2 & 0.137 \\
\bottomrule
\end{tabular}
\caption{Correlation strength across the combined benchmark set, all in the STO-6G basis, ordered by decreasing $w_{\mathrm{HF}}$. Here $w_{\mathrm{HF}} = |\langle \Phi_{\mathrm{RHF}} | \Psi_{\mathrm{FCI}} \rangle|^{2}$ is the weight of the HF determinant in the exact ground state, and $E_{\mathrm{corr}} = |E_{\mathrm{FCI}} - E_{\mathrm{RHF}}|$. Both are evaluated in the full minimal-basis space for every system, C$_2$H$_2$ included. A value $w_{\mathrm{HF}} \rightarrow 1$ indicates a well-described single reference; decreasing $w_{\mathrm{HF}}$ indicates increasing multireference character.}
\label{tab:corr_strength}
\end{table}

\begin{table}[t]
\centering
\renewcommand{\arraystretch}{1.15}
\begin{tabular}{c c c}
\toprule
$R$ (Bohr) & This work (Ha) & Simons benchmark (Ha) \\
\midrule
1.2 & $-4.7661(1)$ & $-4.7666(3)$ \\
1.6 & $-5.3818(3)$ & $-5.3819(6)$ \\
1.8 & $-5.4219(3)$ & $-5.4218(6)$ \\
2.0 & $-5.3878(2)$ & $-5.3874(5)$ \\
2.4 & $-5.2268(2)$ & $-5.2271(8)$ \\
2.8 & $-5.0503(1)$ & $-5.0507(5)$ \\
\bottomrule
\end{tabular}
\caption{Comparison of AFQMC(UHF) energies for the H$_{10}$ chain in the STO-6G basis versus Simons Foundation benchmark data \cite{motta2017towards}.}
\label{tab:simons_comparison}
\end{table}

\section{Ansatz Depth Study}
\label{app:depth}

\begin{figure*}[t]
    \centering
    \includegraphics[width=\textwidth]{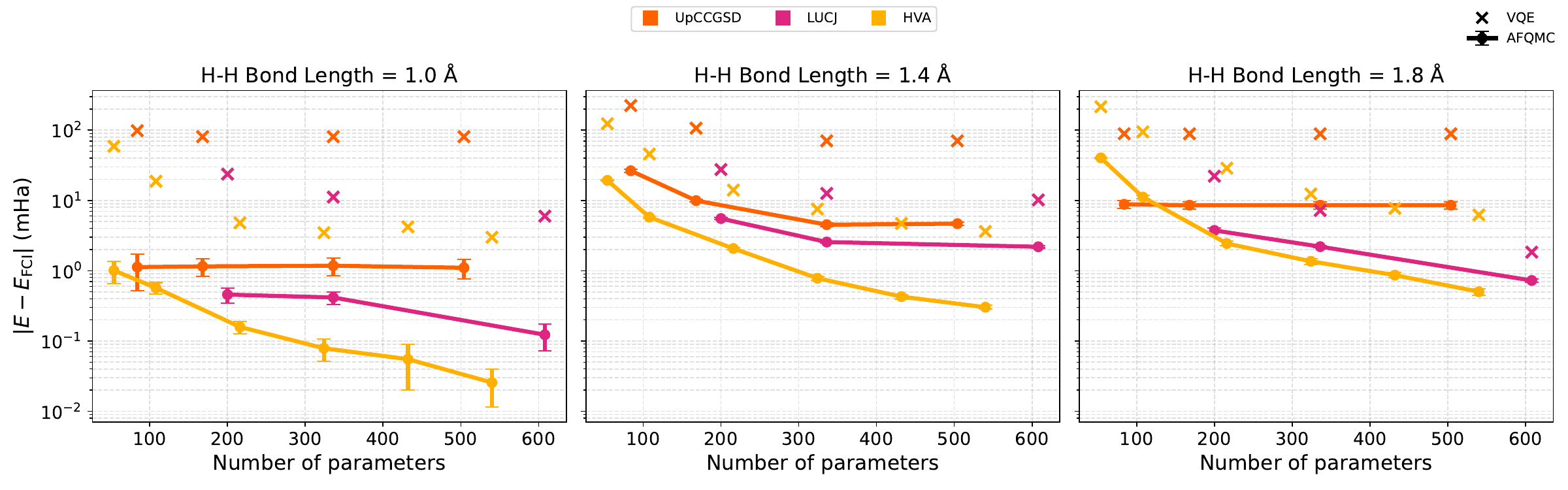}
\caption{
AFQMC and VQE energy errors for H$_8$ as a function of the number of variational parameters at three bond lengths ($R = 1.0$, 1.4, and 1.8$~\text{\AA}$). Successive points along a curve correspond to increasing repetition count $k$ for these layered ansatzes. The vertical axis shows the absolute deviation from FCI, $|E - E_{\mathrm{FCI}}|$, in mHa on a logarithmic scale. Cross markers denote VQE results, while solid circles with error bars correspond to ph-AFQMC energies using the optimized states as trial wave functions.} 
\label{fig:H8 iters}
\end{figure*}

Fig.~\ref{fig:H8 iters} presents a systematic study of ansatz expressivity for H$_8$ at three representative bond lengths, with both the VQE and the ph-AFQMC errors shown as functions of the number of variational parameters. Successive points along each curve correspond to increasing repetition count $k$, and the range shown extends to roughly 1.5 times the parameter count of UCCSD for H$_8$ (360, Table~\ref{tab:VQE_AFQMC_Table}), which brackets the parameter-matched depths used for the comparisons in the main text. Within this range, LUCJ and HVA both improve systematically in the variational and in the projected energy as parameters are added, at every bond length, with the largest gains occurring in the intermediate and strongly correlated regimes. At matched parameter counts, HVA yields the more accurate projected energy at all three bond lengths, and at $R = 1.8$~\AA{} it does so despite a higher variational energy than LUCJ, suggesting that HVA circuits provide a phase structure better suited to the phaseless constraint for these systems than their variational energy alone would indicate. This is consistent with the H$_{10}$ results in the main text. In contrast, UpCCGSD plateaus earliest in both the variational and the projected energy, and additional layers do not yield further improvement. These results reinforce the central theme of this work, with an important qualification: adequate expressivity, achieved either through increased depth or adaptive construction, improves phaseless performance, but only insofar as the optimization can realize it, and the optimal balance between parameter count and optimization tractability is ansatz dependent.

\section{Optimization Cost and Circuit Resources}
\label{app:resources}

Table~\ref{tab:resources} reports the cost of preparing each trial wave function for the two most strongly correlated systems in the benchmark set, H$_{10}$ at $R = 2.0$~\AA{} and C$_2$, along two independent axes: the classical cost of carrying out the variational optimization, and the size of the circuit that would have to be executed on hardware. The two do not track each other, and neither tracks accuracy.

The first is the measured wall-clock cost of the optimization on GPU nodes, expressed relative to the cheapest ansatz for that system so that the comparison is not tied to a particular machine or node count. HVA is the most expensive optimization for C$_2$ by more than an order of magnitude, and among the most expensive for H$_{10}$, while being one of the more inaccurate ansatz on both systems (Table~\ref{tab:molecules}). The cost follows from the structure of the ansatz: the number of variational parameters per HVA layer is the number of mutually commuting groups of the qubit Hamiltonian, 120 for H$_{10}$ and 77 for C$_2$, which grows quickly with the complexity of the electronic structure, and every one of those parameters requires its own gradient evaluation at every optimization step. ADAPT-UCCSD is expensive for the opposite reason: it produces the shortest circuit of the UCC family, but the operator-pool gradient screening performed before each operator is appended dominates its total cost. The low optimization cost of UCCSD relative to the other ansatzes should also be understood in light of its initialization. Its variational parameters are initialized from amplitudes obtained from an unrestricted CCSD calculation, providing a physically informed starting point that substantially reduces the number of optimization iterations required compared with an uninformed initialization. Although this initialization is most effective in the weakly correlated regime, it remains a useful starting point for the strongly correlated systems considered here.

The second is an estimate of the two-qubit gate count of the corresponding circuit. These counts are structural rather than compiled. All simulations in this work were performed with GPU state-vector simulators, so the circuits are never transpiled to a hardware gate set and no compiled counts exist for them. For the ansatzes built from Pauli exponentials, UCCSD, ADAPT-UCCSD, UpCCGSD and HVA, each factor $e^{-i\theta P}$ acting on a Pauli word $P$ of weight $w$ is counted in the textbook CNOT-ladder form: a staircase of $w-1$ \textsc{cnot}s, one single-qubit rotation, and the reversed staircase, for $2(w-1)$ two-qubit gates. The counts are summed over every Pauli word of every generator in the circuit and multiplied by the number of layers. LUCJ is counted directly from its own gate structure, as Givens-rotation networks for the orbital rotations together with the paired $XX+YY$ interactions and the diagonal-Coulomb $ZZ$ phases. No circuit optimization of any kind is applied: adjacent \textsc{cnot}s between neighboring Pauli terms are not cancelled, JW strings are not shared between terms, and all-to-all qubit connectivity is assumed so that no \textsc{swap} overhead is incurred. A real compiler targeting a specific device would reduce these numbers substantially for the UCC-type circuits, while restricted connectivity would increase them. They should therefore be read as an order-of-magnitude proxy for relative circuit depth across ansatzes, not as a resource requirement for any particular device.

Read that way, the gate counts separate the ansatzes far more sharply than the optimization cost does, and they do not order them the same way. LUCJ yields substantially shallower circuits than the other ansatzes considered, by construction: its two-body correlations are restricted to a fixed interaction pattern rather than to the full set of excitations. HVA is the deepest by a wide margin, roughly $5\times$ UCCSD for H$_{10}$ and $4\times$ for C$_2$, because it exponentiates every Pauli term of the JW Hamiltonian once per layer, including the long parity strings, which reach weight 20 on 20 qubits. The combination of LUCJ's shallow circuits with its accuracy at these geometries (Figures~1(b) and~6 of the main text) is the most favorable trade-off in the set.

\begin{table}[t]
\centering
\renewcommand{\arraystretch}{1.2}
\begin{tabular}{lcccc}
\toprule
 & \multicolumn{2}{c}{H$_{10}$, $R = 2.0$~\AA{}} & \multicolumn{2}{c}{C$_2$} \\
\cmidrule(lr){2-3}\cmidrule(lr){4-5}
Ansatz & rel.\ cost & 2-qubit gates & rel.\ cost & 2-qubit gates \\
\midrule
UCCSD           & 1.0            & $1.3 \times 10^{5}$ & 1.0  & $1.2 \times 10^{5}$ \\
ADAPT-UCCSD     & 33.6           & $1.1 \times 10^{4}$ & 5.4  & $1.5 \times 10^{4}$ \\
LUCJ            & 5.8            & $2.7 \times 10^{3}$ & 1.1  & $2.5 \times 10^{3}$ \\
HVA             & 11.8           & $6.7 \times 10^{5}$ & 23.5 & $4.6 \times 10^{5}$ \\
UpCCGSD         & 6.2            & $4.4 \times 10^{4}$ & 1.0  & $4.4 \times 10^{4}$ \\
\bottomrule
\end{tabular}
\caption{Cost of preparing each trial wave function for the two most strongly correlated systems in the benchmark set, at the parameter-matched depths used throughout ($k = 3$ for LUCJ on both; $k = 7$ and $k = 10$ for HVA; $k = 6$ for UpCCGSD). Relative cost is the measured GPU wall-clock cost of the variational optimization multiplied by the number of compute nodes, normalized within each system to the cheapest ansatz for that system. 2-qubit gates are uncompiled structural estimates: every Pauli exponential is counted in the CNOT-ladder form, $2(w-1)$ two-qubit gates for a Pauli word of weight $w$, summed over all terms and layers, with no gate cancellation, no sharing of JW strings between terms, and all-to-all connectivity assumed. Because all simulations were run on GPU state-vector simulators the circuits were never transpiled, so these numbers are a proxy for relative circuit depth and should not be read as compiled resource counts for a specific device.}
\label{tab:resources}
\end{table}

\section{Concordance Analysis with Statistical Uncertainty}
\label{app:concordance}

\begin{figure}[!htbp]
    \centering
    \includegraphics[width=0.75\linewidth]{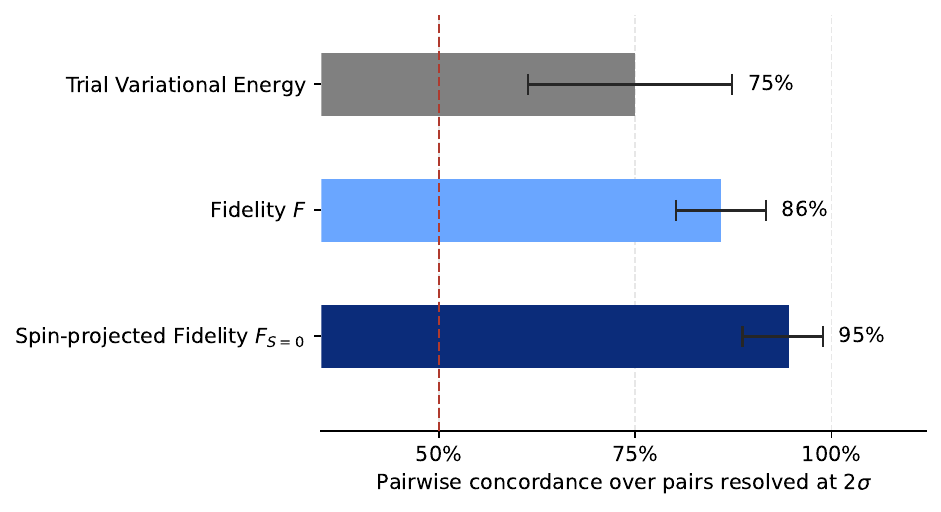}
    \caption{Fraction of statistically resolvable same-system trial pairs (92 pairs across 15 systems) for which the ordering by each diagnostic agrees with the ordering by $|E_{\mathrm{AFQMC}}-E_{\mathrm{FCI}}|$. Error bars show 95\% confidence intervals obtained from a cluster bootstrap over systems and the dashed line indicates the 50\% agreement expected from random ordering.}
    \label{fig:concordance_pooled}
\end{figure}

The per-system concordances in Figure~7 of the main text treat each ph-AFQMC energy as exact. To account for statistical uncertainty in the projected energies, we also compute concordance pooled across systems, restricting the analysis to trial pairs $(A,B)$ from the same system whose absolute ph-AFQMC errors are resolvable according to $\big||\Delta E_A|-|\Delta E_B|\big|
>
2\sqrt{\sigma_A^2+\sigma_B^2}$, where $\Delta E_i = E_{\mathrm{AFQMC},i}-E_{\mathrm{FCI}}$ is the ph-AFQMC energy error for trial $i$ and $\sigma_i$ is its statistical uncertainty. Because trial pairs within a system are not independent, we estimate 95\% confidence intervals using a cluster bootstrap over systems. Specifically, we generate 4000 resampled datasets by randomly selecting whole systems with replacement and recompute the concordance for each dataset. The resulting variation in concordance provides an estimate of its statistical uncertainty.

As shown in Figure~\ref{fig:concordance_pooled}, the ordering by variational energy, fidelity $F$, and spin-projected fidelity $F_{S=0}$ agrees with the ordering by absolute ph-AFQMC error for 75\%, 86\%, and 95\% of the resolvable pairs, respectively. To compare the diagnostics directly, their differences in concordance are computed within each of the same bootstrap resamples. The paired differences are $+10$\% (95\% CI: $+2$ to $+21$\%) for $F$ relative to variational energy, $+19$\% ($+8$ to $+31$\%) for $F_{S=0}$ relative to variational energy, and $+8$\% ($+3$ to $+15$\%) for $F_{S=0}$ relative to $F$. The same ordering of the three diagnostics is obtained for resolvability thresholds ranging from $0\sigma$ to $3\sigma$.

\end{document}